hep-th/9607016

\def\bbC           {\mathbb{C}}
\def\bbM           {\mathbb{M}}
\def\bbN           {\mathbb{N}}
\def\bbNo          {\mathbb{N}_0}
\def\bbT           {\mathbb{T}}
\newcommand\BBe[3] {B^{#1,#2}_{#3}}
\newcommand\bbe[2] {b^{#1}_{#2}}
\def\be            {\begin{equation}}
\def\bearl         {\begin{array}{l}}
\def\bearll        {\begin{array}{ll}}
\def\bearlll       {\begin{array}{lll}}
\def\bea           {\begin{eqnarray}}
\def\beaa          {\begin{eqnarray*}}
\newcommand\bet[3] {\beta^{#1,#2}_{#3}}
\def\bfe           {{\bf1}}
\def\BK            {\mathfrak{B}(\mathcal{K})}
\def\bP            {\overline{P}}
\newcommand\bproof {\noindent {\it Proof. }}
\newcommand\brae[2]{\langle e^{#1}_{#2}|}
\def\CKG           {\mathcal{C}(\mathcal{K},\Gamma)}
\def\CKIG          {\mathcal{C}(\mathcal{K}(I),\Gamma)}
\def\CKGh          {\mathcal{C}(\hat{\mathcal{K}},\hat{\Gamma})}
\def\cA            {\mathcal{A}}
\def\cAloc         {\mathcal{A}}
\def\cCA           {\mathfrak{C}}
\def\cG            {\mathcal{G}}
\def\cH            {\mathcal{H}}
\def\cK            {\mathcal{K}}
\def\cO            {\mathcal{O}}
\def\Cos           {\mathfrak{Cos}}
\def\cR            {\mathcal{R}}

\def\D             {\mathrm{d}}
\def\DiffS         {\mathrm{Diff}S^1}

\def\dimg          {{\mathrm{dim}\,\g}}
\newcommand\del[2] {\delta_{{#1},{#2}}}
\def\DK            {\mathcal{DK}}
\def\dQP           {\mathrm{d}Q_P}
\def\E             {\mathrm{e}}
\newcommand\EE[2]  {\mathcal{E}^{#1}_{#2}}
\def\ee            {\end{equation}}
\def\eear          {\end{array}}
\def\eea           {\end{eqnarray}}
\def\eeaa          {\end{eqnarray*}}
\newcommand\ef[2]  {e^{#1}_{#2}}
\def\eps           {\epsilon}
\newcommand\eproof {\hspace*{\fill}\nolinebreak\hspace*{\fill}
                   $\Box$\par\vspace{3mm}}
\newcommand\EO[2]  {E^{#1}_{#2}}
\def\fA            {\mathfrak{A}}
\def\fAg           {\mathfrak{A}}
\def\fB            {\mathfrak{B}}
\def\fF            {\mathfrak{F}}
\def\fFg           {\mathfrak{F}}
\def\fh            {\mathfrak{h}}
\def\fhh           {\hat{\mathfrak{h}}}
\def\fn            {\mathfrak{n}}
\def\fnh           {\hat{\mathfrak{n}}}
\def\forzl         {{\rm for}\ N=2\ell}
\def\forzle        {{\rm for}\ N=2\ell+1}
\def\fsqz          {\mbox{\large$\frac1{\sqrt2}$}\,}

\def\fW            {\mathfrak{W}}
\def\fWb           {\overline{\mathfrak{W}}}
\def\g             {{\mathfrak g}}
\def\gh            {\hat{\mathfrak g}}
\def\GG            {\Gamma}
\def\GGh           {\hat{\Gamma}}
\def\GLZ           {\mathit{GL}\,(2;\bbC)}
\def\Gr            {\mathit{G}}
\def\h             {{1/2}}
\def\half          {\mbox{\large$\frac12$}\,}
\def\halfi         {\mbox{\large$\frac\I2$}\,}
\def\Hl            {\mathcal{H}_\Lambda}
\def\HNS           {\mathcal{H}_{\rm NS}}
\def\HNSf          {\mathcal{H}_{\rm NS}^{\rm fin}}
\def\HNSh          {\hat{\mathcal{H}}_{\rm NS}}
\def\HNShf         {\hat{\mathcal{H}}_{\rm NS}^{\rm fin}}
\def\Hph           {\mathcal{H}_\mathrm{phys}}
\def\HR            {\mathcal{H}_{\rm R}}
\def\HRf           {\mathcal{H}_{\rm R}^{\rm fin}}
\newcommand\hsp[1] {\mbox{\hspace{#1em}}}
\def\id            {\mbox{\sl id}}
\def\I             {{\rm i}}
\def\IKG           {\mathcal{I}(\mathcal{K},\Gamma)}

\def\IoKG          {\mathcal{U}(\mathcal{K},\Gamma)}
\def\IPoKG         {\mathcal{U}_P(\mathcal{K},\Gamma)}
\def\iPoKG         {\mathfrak{u}_P^\mathrm{b}(\mathcal{K},\Gamma)}
\def\iPNSoKG       {\mathfrak{u}_{\PNS}^\mathrm{b}(\mathcal{K},\Gamma)}
\newcommand\J[3]   {J^{#1,#2}_{#3}}
\newcommand\Jm[2]  {J_{m}^{#1,#2}}
\newcommand\Jmt[4] {J_{m}(\te{#1}{#2}{#3}{#4})}
\newcommand\Jmtee[2]{J_{m}(\ttee{#1}{#2})}
\newcommand\Jmte[1]{J_m(t^{#1}_{\eps})}
\newcommand\Jo[2]  {J_{0}^{#1,#2}}
\newcommand\Jot[4] {J_{0}(\te{#1}{#2}{#3}{#4})}
\newcommand\Jt[5]  {J_{#1}(\te{#2}{#3}{#4}{#5})}

\def\JHS           {\mathfrak{J}_2(\mathcal{K})}
\def\Jz            {\mathcal{J}_\zeta}
\newcommand\kete[2]{|e^{#1}_{#2}\rangle}
\def\KK            {\mathcal{K}}
\def\KKh           {\hat{\mathcal{K}}}

\newcommand\Lamf[1]{\Lambda_{(#1)}}

\newcommand\labl[1]{\label{#1}\ee}
\newcommand\lablth[1]{\label{#1}}
\def\lc            {\Lambda_{\rmsm}}
\def\LG            {\mathit{LG}}
\def\LGer          {\mathcal{L}\mathit{G}}
\def\LIN           {L^2(I;\mathbb{C}^N)}
\def\LIG           {\mathit{L}_I\mathit{G}}
\newcommand\lj[1]  {\Lambda_{[#1]}}
\newcommand\Lj[1]  {\Lambda_{(#1)}}
\def\ls            {\Lambda_{\rmsp}}
\def\LSE           {L^2(S^1)}
\def\LSN           {L^2(S^1;\mathbb{C}^N)}
\def\LSUN          {\mathit{LSU}(N)}
\def\lv            {\Lambda_{\rmv}}
\def\mh            {{-1/2}}
\def\mnh           {{-n-1/2}}

\def\natnum        {\bbN}
\def\natnumo       {\bbNo}
\def\nh            {{n+1/2}}
\def\nmh           {{n-1/2}}
\def\Oml           {|\Omega_\Lambda\rangle}
\def\Omd           {|\Omega_\Delta\rangle}
\def\Omego         {|\Omega_0\rangle}
\def\OmP           {|\Omega_P\rangle}
\def\onehalf       {\mbox{$\frac12$}}
\def\onetol        {1,2,...\,,\ell}
\def\onetolme      {1,2,...\,,\ell-1}
\def\onetolmz      {1,2,...\,,\ell-2}
\def\onetoN        {1,2,...\,,N}
\def\ONS           {|\Omega_{\rm NS}\rangle}
\def\ONSh          {|\hat\Omega_{\rm NS}\rangle}
\def\OR            {|\Omega_{\rm R}\rangle}
\def\otol          {0,1,...\,,\ell}
\def\otolme        {0,1,...\,,\ell-1}

\def\PiNS          {\pi_{\rm NS}}
\def\PiNSh         {\hat\pi_{\rm NS}}
\def\PiR           {\pi_{\rm R}}
\def\PNS           {P_{\rm NS}}
\def\PNSh          {\hat{P}_{\rm NS}}
\def\Poin          {\mathcal{P}^\uparrow_+}
\def\Point         {\tilde{\mathcal{P}}^\uparrow_+}
\newcommand\prodni[1]{\prod_{n={#1}}^\infty}
\def\PSLZ          {\mathit{PSL}(2;\zet)}
\def\PSU           {\mathit{PSU}(1,1)}
\def\Psik          {|\Psi\rangle}

\def\QKG           {\mathcal{Q}(\KK,\GG)}
\def\QKGh          {\mathcal{Q}(\KKh,\GGh)}
\def\qmh           {q^{m+1/2}}
\def\qzme          {q^{2m+1}}
\def\reals         {\mathbb{R}}
\def\rmc           {{\mathrm{c}}}
\def\rms           {{\mathrm{s}}}
\def\rmv           {{\mathrm{v}}}

\def\rseta         {\varrho_{U(\eta)}}
\def\rsgamt        {\varrho_{U(\gamma_t)}}
\newcommand\sfrac[2]{\mbox{\large$\frac{#1}{#2}$}\,}
\def\seta          {\eta}
\def\sgamt         {\gamma_t}

\def\son           {\mathfrak{so}(N)}
\def\SON           {\mathit{SO}(N)}
\def\sonh          {\widehat{\mathfrak{so}}(N)}
\def\sonhe         {\widehat{\mathfrak{so}}(N)_1}
\def\sonhz         {\widehat{\mathfrak{so}}(N)_2}
\def\SR            {S_\mathrm{R}}
\def\sumnZ         {\sum_{n\in\mathbb{Z}}}
\def\sumnN         {\sum_{n\in\mathbb{N}}}
\def\sumrZp        {\sum_{r\in\mathbb{Z}+1/2}}
\def\sumrNp        {\sum_{r\in\mathbb{N}_0+1/2}}
\def\SU            {\mathit{SU}(1,1)}
\def\SUN           {\mathit{SU}(N)}

\newcommand\T[2]   {T^{#1,#2}}
\newcommand\te[4]  {t^{#1,#2}_{#3,#4}}
\newcommand\tee[2] {t^{#1}_{#2}}
\def\tr            {\mathrm{tr}\,}
\newcommand\ttau[3]{\tau^{#1,#2}_{#3}}
\def\ue            {\mathfrak{u}(1)}
\def\Vir           {\mathfrak{Vir}}
\def\Virb          {\overline{\mathfrak{Vir}}}
\def\zb            {\bar{z}}
\def\zet           {\mathbb{Z}}

\def\Acos          {\mathfrak{A}_{\Cos}}
\def\aff           {affine Lie algebra}

\def\alg           {algebra}

\def\Avirc         {\mathfrak{A}_{{\Vir\coset}}}

\def\AW            {\mathfrak{A}_{\rm WZW}}
\renewcommand\b[3] {b^{#1;#2}_{#3}}
\newcommand\BX[4]  {B^{#1,#2;#3}_{#4}}

\def\be            {\begin{equation}}

\def\bearl         {\begin{array}{l}}
\def\bearll        {\begin{array}{ll}}
\def\bearlll       {\begin{array}{lll}}
\def\bfe           {{\bf1}}

\newcommand\cbm[2] {\bar c^{#1,-}_{#2}}
\newcommand\cbp[2] {\bar c^{#1,+}_{#2}}
\newcommand\cbpm[2]{\bar c^{#1,\pm}_{#2}}

\def\cft           {conformal field theory}
\def\cfts          {conformal field theories}

\def\CKG           {{\cal C}({\cal K},\Gamma)}
\def\CKGh          {{\cal C}(\hat{\cal K},\hat{\Gamma})}

\newcommand\ceps[2]{c^{#1,\eps}_{#2}}
\newcommand\ceta[2]{c^{#1,\eta}_{#2}}
\def\chiCc         {\raisebox{.15em}{$\chi$}\Coset_\rmc} 
 
\def\chicjl        {\raisebox{.15em}{$\chi$}^{{\rm c};\scriptstyle \el}_{[j]}} 
\newcommand\chicjzN[1]{\raisebox{.15em}{$\chi$}^{{\rm c};
                   \scriptstyle 2N}_{[#1]}} 
\def\chiCj         {\raisebox{.15em}{$\chi$}\Coset_{[j]}} 
\def\chiCm         {\raisebox{.15em}{$\chi$}\Coset_{[M]}} 
\def\chicol        {\raisebox{.15em}{$\chi$}^{{\rm c};\scriptstyle \el}\Null} 
\def\chicozN       {\raisebox{.15em}{$\chi$}^{{\rm c};\scriptstyle 2N}\Null} 
\def\chiCo         {\raisebox{.15em}{$\chi$}\Coset\Null} 
\def\chiCoM        {\raisebox{.15em}{$\chi$}^{{\rm c};\scriptstyle M}\Null} 
\def\chiCs         {\raisebox{.15em}{$\chi$}\Coset_\rms} 
 
\def\chicvl        {\raisebox{.15em}{$\chi$}^{{\rm c};\scriptstyle \el}_\rmv} 
\def\chicvzN       {\raisebox{.15em}{$\chi$}^{{\rm c};\scriptstyle 2N}_\rmv}  
\def\chiCv         {\raisebox{.15em}{$\chi$}\Coset_\rmv}   
\def\chieo         {\raisebox{.15em}{$\chi$}\ke\Null} 
\def\chiev         {\raisebox{.15em}{$\chi$}\ke_\rmv}  

\def\chiJ          {\raisebox{.15em}{$\chi$}\ns_J}
\newcommand\chivir[1]{\raisebox{.15em}{$\chi$}^{\rm Vir}_{#1}}
\newcommand\chim[1]{\raisebox{.15em}{$\chi$}\ns_{[#1]}}
\def\chio          {\raisebox{.15em}{$\chi$}\ns_0}
\def\chizc         {\raisebox{.15em}{$\chi$}\kz_\rmc} 
\def\chizj         {\raisebox{.15em}{$\chi$}\kz_{[j]}} 
\def\chizl         {\raisebox{.15em}{$\chi$}\kz_{[\el]}} 
\def\chizo         {\raisebox{.15em}{$\chi$}\kz\Null} 
\def\chizs         {\raisebox{.15em}{$\chi$}\kz_\rms} 
\def\chizv         {\raisebox{.15em}{$\chi$}\kz_\rmv}

\def\CKGIh         {{\cal C}(\hat{\cal K}(I),\hat{\Gamma})}
\newcommand\cm[2]  {c^{#1,-}_{#2}}

\def\complex       {{\mathbb C}}
\def\coset         {^{\rm c}}
\def\Coset         {^{{\rm c};\scriptscriptstyle M}}

\def\cost          {\cos(t)\,}

\newcommand\cp[2]  {c^{#1,+}_{#2}}
\newcommand\cpm[2] {c^{#1,\pm}_{#2}}
\def\csa           {Cartan subalgebra}

\newcommand\Delns[1]{\Delta\Ns_{n;#1}}
\newcommand\Delnsb[1]{\bar \Delta\Ns_{n;#1}}
\newcommand\Delc[1]{\Delta^{\!\rm c}_{n;#1}}
\newcommand\Delcb[1]{\bar \Delta^{\!\rm c}_{n;#1}}
\def\dh            {\Delta}
\def\Dn            {\mbox{${\cal D}_N$}}

\let\dstyle=\displaystyle

\def\ee            {\end{equation}}
\def\eit           {{\rm e}^{{\rm i}t}}
\def\el            {\ell}

\def\emit          {{\rm e}^{-{\rm i}t}}

\let\eps=\varepsilon
\newcommand\erf[1] {(\ref{#1})}
\def\findim        {finite-dimensional}

\def\forzl         {{\rm for}\ N=2\el}
\def\forzle        {{\rm for}\ N=2\el+1}
\newcommand\Frac[2]{\mbox{\large$\frac{#1}{#2}$}}
\def\fsqz          {\mbox{\large$\frac1{\sqrt2}$}\,}

\def\futnot#1      {\ifnum\draftcontrol=1%
                   \footnote{~{\sc internal footnote:} #1}\ \fi}
\def\futnote#1     {\footnote{~#1}\ }
\def\g             {\mathfrak{g}}

\def\gv            {g_{}^{\scriptscriptstyle\vee}}
\def\h             {{1/2}}
\def\half          {\mbox{\large$\frac12$}\,}
\def\halfi         {\mbox{\large$\frac\I 2$}\,}
\newcommand\Hh[1]  {{\cal H}_{#1}}
\newcommand\HH[1]  {{\cal H}^{#1}_{}}
\def\Hheo          {{\cal H}_\circ\ke}
\def\Hhev          {{\cal H}_\rmv\ke}

\def\Hhzc          {{\cal H}_\rmc\kz}
\def\Hhzj          {{\cal H}_{[j]}\kz}

\def\Hhzo          {{\cal H}_\circ\kz}
\def\Hhzs          {{\cal H}_\rms\kz}
\def\Hhzv          {{\cal H}_\rmv\kz}
\newcommand\Hhh[1] {{\cal H}_{[#1]}}
\def\Hhj           {{\cal H}_J}

\def\Hhm           {{\cal H}_{[m]}}
\def\Hho           {{\cal H}_0}
\def\HNS           {{\cal H}_{\rm NS}}
\def\HNSh          {\hat{\cal H}_{\rm NS}}
\def\hsa           {horizontal subalgebra}
\def\hw            {highest weight}
\def\hwm           {highest weight module}

\def\hws           {highest weight state}
\def\hwv           {highest weight vector}
\def\hy            {$\mbox{-\hspace{-.66 mm}-}$}
\def\id            {\mbox{\sl id}}

\def\ihwm          {irreducible highest weight module}
\def\iN            {\!\in\!}

\def\Intt          {\Frac1{2\pi}\dstyle\int_0^{2\pi}\!\!{\rm d}t\,}
\def\IntT          {\Frac1{4\pi}\dstyle\int_0^{2\pi}\!\!{\rm d}t\,}
\def\intt          {\Frac1{2\pi}\int_0^{2\pi}\!\!{\rm d}t\,}
\def\intT          {\Frac1{4\pi}\int_0^{2\pi}\!\!{\rm d}t\,}
\def\irmod         {irreducible module}
\def\irrep         {irreducible representation}
\newcommand\Jmttm[1]{J_m(t^{#1}_-)}
\newcommand\Jmttp[1]{J_m(t^{#1}_+)}
\newcommand\Jmttpm[1]{J_m(t^{#1}_{\pm})}
\def\ke            {^{\scriptscriptstyle(1)}}

\def\kv            {k_{}^{\scriptscriptstyle\vee}}

\def\kz            {^{\scriptscriptstyle(2)}}
\def\lc            {\Lambda_{{\rm c}}}

\def\lie           {Lie algebra}

\def\llb           {\mbox{\large[}}
\def\lLb           {\mbox{\large(}}
\def\Llb           {\mbox{\Large\{}}
\def\LLb           {\mbox{\Large[}}
\newcommand\LNS[1] {\mbox{$L_{#1}\ns$}}
\def\lo            {\Lambda\Null}
\def\lrb           {\mbox{\large]}}
\def\lRb           {\mbox{\large)}}
\def\LRb           {\mbox{\Large]}}
\def\Lrb           {\mbox{\Large\}}}
\def\ls            {\Lambda_{{\rm s}}}
\def\lv            {\Lambda_{{\rm v}}}
 
\def\mh            {{-1/2}}
\def\mnh           {{-n-1/2}}

\def\nh            {{n+1/2}}
\newcommand\nline[1] {\\{}\\[-.#1em]}
\def\nmh           {{n-1/2}}

\newcommand\normord[1] {\,{\bf:}#1{\bf:}\,}
\def\ns            {^{\scriptscriptstyle\rm NS}}
\def\Ns            {^{\!\scriptscriptstyle\rm NS}}

\def\NS            {Neveu-Schwarz }
\def\Null          {_\circ}

\newcommand\Ocnpm[1] {|\Omega^{#1,\pm}_{\rm c}\rangle}
\newcommand\Ocnmp[1] {|\Omega^{#1,\mp}_{\rm c}\rangle}

\def\Ok            {\mathit{O}(\kv)}

\def\Omo           {|\Omega\Null\rangle}

\newcommand\Ombnmp[2]{|\overline\Omega^{#2,\mp}_{[#1]}\rangle}

\newcommand\Ombnpm[2]{|\overline\Omega^{#2,\pm}_{[#1]}\rangle}

\newcommand\OmfJo[1]{|\Omega^{J,#1}\Null\rangle}
\newcommand\OmfJv[1]{|\Omega^{J,#1}_\rmv\rangle}
\newcommand\OmfOo[1]{|\Omega^{0,#1}\Null\rangle}
\newcommand\OmfOv[1]{|\Omega^{0,#1}_\rmv\rangle}

\newcommand\Omp[1] {|\Omega^+_{[#1]}\rangle}

\newcommand\Omnmp[2]{|\Omega^{#2,\mp}_{[#1]}\rangle}

\newcommand\Omnpm[2]{|\Omega^{#2,\pm}_{[#1]}\rangle}

\def\one           {\mbox{\small $1\!\!$}1}
\def\onedim        {one-dimensional}
\def\onehalf       {\mbox{$\frac12$}}
\def\onetol        {1,2,...\,,\el}
\def\onetolme      {1,2,...\,,\el-1}
\def\onetolmz      {1,2,...\,,\el-2}
\def\onetom        {1,2,...\,,M}
\def\onetomme      {1,2,...\,,M-1}

\def\onetoN        {1,2,...\,,N}
\newcommand\Oonmp[1] {|\Omega^{#1,\mp}\Null\rangle}
\newcommand\Oonpm[1] {|\Omega^{#1,\pm}\Null\rangle}
\newcommand\Osnpm[1] {|\Omega^{#1,\pm}_{\rm s}\rangle}
\newcommand\Osnmp[1] {|\Omega^{#1,\mp}_{\rm s}\rangle}

\def\otol          {0,1,...\,,\el}
\def\otolme        {0,1,...\,,\el-1}
\def\Ovvo           {|\Omega_{{\rm v}}\rangle}
\def\Ovo           {|\Omega^{0,\pm}_{{\rm v}}\rangle}

\newcommand\Ovnpm[1] {|\Omega^{#1,\pm}_{\rm v}\rangle}
\newcommand\Ovnmp[1] {|\Omega^{#1,\mp}_{\rm v}\rangle}
\def\Oz            {\mathit{O}(2)}

\newcommand\pfj[1] {\phi_{[#1]}}

\def\Phila         {|\Phi_\Lambda\rangle}

\def\PJ            {\mbox{$P^{}_{\!J}$}}
\def\Po            {\mbox{$P^{}_{\!0}$}}
\newcommand\Pm[1]  {\mbox{$P^{-}_{\![#1]}$}}
\newcommand\Pp[1]  {\mbox{$P^{+}_{\![#1]}$}}
\newcommand\Ppm[1] {\mbox{$P^{\pm}_{\![#1]}$}}
\def\psiemq        {\mbox{$\psi^{}_{1,0}(-q)$}}
\newcommand\psim[1]{\mbox{$\psi^{}_{M,#1}$}}

\newcommand\psinh[1]{\mbox{$\psi^{}_{\el,#1}$}}
\newcommand\psinz[1]{\mbox{$\psi^{}_{2N,#1}$}}

\def\qdi           {\mbox{$\cal D$}}

\def\qlo           {q^{L_0^{(\rm NS)_{}}}}
\def\qmh           {q^{m+1/2}}
\def\qzme          {q^{2m+1}}

\def\reals         {{\mathbb R}}
\newcommand\Reh[1] {W(1,#1)}
\def\rep           {representation}

\def\resp          {respectively}

\def\rmo           {\circ}
\def\Rn            {\mbox{${\cal R}\kz_{\rm NS}$}}
\def\Ro            {\mbox{${\cal R}^{}_{\Oz}$}}
\def\role          {r\^ole}
\newcommand\rp[1]  {\Phi_{[#1]}}
\def\rpj           {\Phi_J}
\def\rpo           {\Phi_0}
\def\Rw            {\mbox{${\cal R}\kz_{\rm WZW}$}}
\newcommand\sect[1] {\section{#1}\setcounter{equation}{0}}
\newcommand\Sect[2] {\sect{#1}\label{s.#2}
                   \ifnum\draftcontrol=1 \query{s.#2} \fi}
\def\sgam          {\gamma}

\def\sint          {\sin(t)\,}

\newcommand\sN[1]  {\Theta^{}_{N,#1}(q)}

\def\subseT        {\!\subset\!}
\newcommand\summN[1]{{\displaystyle\sum_{\scriptstyle m_1,m_2,...,m_N\in\zet
                   \atop \scriptstyle m_1+m_2+...+m_N=#1}\!\!\!}}
\newcommand\sumMN[1]{{\displaystyle\sum_{\scriptstyle \vecm\in\zet^N
                   \atop \scriptstyle \sum m_i=#1}\!}}
\def\summZ         {\sum_{m\in\zet}}
\newcommand\sumni[1]{\sum_{n=#1}^\infty}
\def\sumnZ         {\sum_{n\in\zet}}
\def\sumq          {{\displaystyle\sum_{q=1}^2}}

\def\sun           {\mathfrak{su}(N)}
\def\sunh          {\widehat{\mathfrak{su}}(N)}
\def\sunhe         {\widehat{\mathfrak{su}}(N)_1}

\newcommand\tte[1] {t^{#1}_{\eps}}
\newcommand\ttee[2]{t^{#1,#2}_{\eps,\eta}}

\newcommand\ttp[1] {t^{#1}_+}
\newcommand\ttpm[1]{t^{#1}_{\pm}}
\def\trNS          {{\rm tr}^{}_{\HNSh}\!}
\def\twodim        {two-di\-men\-si\-o\-nal}

\def\Uc            {\Hhzc}

\def\Uj            {\Hhzj}  
\def\Uk            {\mathit{U}(\kv)}
\def\unh           {\widehat{\mathfrak{u}}(N)}
\def\Uo            {\Hhzo}  
\def\Us            {\Hhzs}  
\def\Uv            {\Hhzv}  

\def\vecm          {{\bf m}}
\def\vi            {\varphi}

\def\wi            {\xi}
\def\wrt           {with respect to }
\def\wrtt          {with respect to the }

\def\wzwt          {WZW theory}
\def\wzwts         {WZW theories}
\newcommand\x[2]   {x^{#1,+}_{#2}}
\newcommand\xpm[2] {x^{#1,\pm}_{#2}}
\newcommand\xmp[2] {x^{#1,\mp}_{#2}}

\newcommand\Xpm[2] {X^{#1,\pm}_{#2}}
\newcommand\Xmp[2] {X^{#1,\mp}_{#2}}
\newcommand\xb[2]  {\bar x^{#1,+}_{#2}}
\newcommand\xbpm[2]{\bar x^{#1,\pm}_{#2}}
\newcommand\xbmp[2]{\bar x^{#1,\mp}_{#2}}

\newcommand\Xbpm[2]{\bar X^{#1,\pm}_{#2}}
\newcommand\Xbmp[2]{\bar X^{#1,\mp}_{#2}}
\newcommand\y[2]   {x^{#1,-}_{#2}}

\newcommand\yb[2]  {\bar x^{#1,-}_{#2}}

\def\zet           {{\mathbb Z}}

\documentclass[12pt]{report}
\usepackage{amssymb,amsfonts,latexsym} 
\begin{document}

%%%%%%%%%%%%%%%%%%%%%%%%%%%%%%%%%%%%%%%%%%%%%%%%%%%%%%%%%%%%%%%%%%%%%%%%%%%%%%

\begin{titlepage}
\title{Superselection Sectors\\of\\
$\son$ Wess-Zumino-Witten Models\\ \vspace{1.3cm}}
\author{Dissertation\\zur Erlangung des Doktorgrades\\
des Fachbereichs Physik\\der Universit\"at Hamburg\\
\vspace{2cm}}
\date{vorgelegt von\\
{\sc Jens B\"ockenhauer}\\aus Hamburg\\
\vspace{1.7cm}
Hamburg\\1996}
\cleardoublepage
\end{titlepage}
\maketitle
\cleardoublepage

\thispagestyle{empty}

\begin{center}
{\bf Abstract}
\end{center}
The superselection structure of $\son$ WZW models is investigated from
the point of view of algebraic quantum field theory. At level $1$ it
turns out that the observable algebras of the WZW theory can be
constructed in terms of even CAR algebras. This fact allows to
give a formulation of these models close to the DHR framework.
Localized endomorphisms are constructed explicitly in terms of
Bogoliubov transformations, and the WZW fusion rules are proven
using the DHR sector product.

At level $2$ it is shown that most of the sectors are realized in
$\HNSh=\HNS\otimes\HNS$ where $\HNS$ is the Neveu-Schwarz sector
of the level $1$ theory. The level $2$ characters are derived and
$\HNSh$ is decomposed completely into tensor products of the sectors
of the WZW chiral algebra and irreducible representation spaces
of the coset Virasoro algebra. Crucial for this analysis is the
DHR decomposition of $\HNSh$ into sectors of a gauge invariant
fermion algebra since the WZW chiral algebra as well as the coset
Virasoro algebra are invariant under the gauge group $\Oz$.
\vspace{1.5cm}

\begin{center}
{\bf Zusammenfassung}
\end{center}
Es wird die Superauswahlstruktur von $\son$ WZW Modellen unter dem
Gesichtspunkt der algebraischen Quantenfeldtheorie untersucht. Es
stellt sich heraus, da{\ss} sich f\"ur Level $1$ die Observablenalgebren
der WZW Theorie durch gerade CAR Algebren konstruieren lassen. Diese
Tatsache erlaubt eine Formulierung f\"ur diese Modelle dicht am
DHR Rahmen. Lokalisierte Endomorphismen werden explizit als
Bogoliubov Transformationen konstruiert, und die WZW Fusionsregeln
werden mithilfe des DHR Sektorproduktes bewiesen.

Es wird gezeigt, da{\ss} f\"ur Level $2$ die meisten Sektoren realisiert
sind in $\HNSh=\HNS\otimes\HNS$, wobei $\HNS$ der Neveu-Schwarz Sektor
der Level $1$ Theorie ist. Die Level $2$ Charaktere werden abgeleitet,
und $\HNSh$ wird vollst\"andig in Tensorprodukte von Sektoren der
WZW chiralen Algebra und irreduziblen Darstellungsr\"aumen der Coset
Virasoro Algebra zerlegt. Entscheidend ist bei dieser Analyse die DHR
Zerlegung von $\HNSh$ in Sektoren einer eichinvarianten
Fermionalgebra, da sowohl die WZW chirale Algebra als auch die
Coset Virasoro Algebra unter der Eichgruppe $\Oz$ invariant sind.

\pagenumbering{Roman}
\setcounter{page}{0}
\tableofcontents
\newpage
\pagenumbering{arabic}

%%%%%%%%%%%%%%%%%%%%%%%%%%%%%%%%%%%%%%%%%%%%%%%%%%%%%%%%%%%%%%%%%%%%%%%%%%%%%%

\newtheorem{definition}{Definition}[chapter]
\newtheorem{lemma}[definition]{Lemma}
\newtheorem{corollary}[definition]{Corollary}
\newtheorem{theorem}[definition]{Theorem}
\newtheorem{proposition}[definition]{Proposition}

%%%%%%%%%%%%%%%%%%%%%%%%%%%%%%%%%%%%%%%%%%%%%%%%%%%%%%%%%%%%%%%%%%%%%%%%%%%%%%

\chapter{Introduction}
This dissertation is concerned with the application of
methods of algebraic quantum field theory (AQFT) to concrete
models of conformal field theory (CFT). Since the basic principles
and the mathematical framework of AQFT and CFT are very
different it is to be feared that the number of readers being
familiar with both settings is rather small. Therefore we feel
obliged to give some introduction to both, AQFT and CFT.
However, it is not reasonable to present largely extended reviews
here; for more detailed discussions of these topics we will
refer the reader to the literature.

\section{\sloppy The Algebraic Approach to Quantum Field Theory}
In the early sixties Haag and Kastler \cite{HK} began to develop
the algebraic approach to quantum field theory. The aim
of this program was to understand quantum field theory
(QFT) in a mathematically rigorous way, and this can be seen
in contrast to the Lagrangian approach which is, although being
very successful in high energy physics, always
accompanied by serious mathematical problems. Employing only
the basic principles of special relativity and quantum theory,
they showed that it is natural to formulate QFT  
in terms of $C^*$-algebras which are associated to bounded regions
in Minkowski spacetime and represent physical quantities that
are observable by measurements within these regions.
Superselection sectors, conventionally defined as (minimal)
subspaces of the Hilbert space of physical states so that
observables do not make transitions between them \cite{WWW},
arise naturally in this framework as the inequivalent irreducible
representation spaces of the algebra of observables.

\subsection{DHR Theory: Starting from the Field Algebra}
In the first paper \cite{DHR1} of an important series,
Doplicher, Haag and Roberts 
(DHR for short) started their analysis from the assumption of a
given field algebra and a gauge group (of the first kind)
acting on it, and defined the observables to be the gauge 
invariant part of this field algebra. More precisely, they
considered a Hilbert space $\cH$ of physical states with
associated algebra $\fB(\cH)$ of bounded operators. They
assumed that to each bounded spacetime region $\cO$ there is
an associated von Neumann algebra $\fF(\cO)\subset\fB(\cH)$,
so that one has a net of {\it local field algebras} 
fulfilling {\it isotony}, i.e.\ $\cO_1\subset\cO_2$ implies 
$\fF(\cO_1)\subset\fF(\cO_2)$.
The total field algebra $\fFg$ is, by definition, the norm 
closure of the union of all local field algebras,
  \[ \fFg = \overline{\bigcup_\cO \fF(\cO)} \,, \]
and is assumed to act irreducibly on $\cH$,
  \[ \fFg'' = \fB(\cH)\,. \]
(A prime always denotes the commutant in the algebra of
bounded operators in the corresponding Hilbert space.) 
Further the following assumptions were made: There
is a strongly continuous unitary representation $U$ of the
covering $\Point$ of the
Poincar{\'e} group $\Poin$ such that the energy operator
$P_0$ has non-negative spectrum and the eigenvalue zero
belongs to a unique (up to a phase) vector $\Omego\in\cH$,
the {\it vacuum state}. The action $U$ of $\Point$ 
transforms the fields covariantly,\footnote{This
assumption of a Poincar{\'e} covariant field algebra is 
rather restrictive, it excludes for instance Quantum
Electrodynamics from the beginning. However, the
setting was tailored to describe strong interaction
physics with short range forces where the assumptions
are coherent.}
  \[ U(L) \, \fF(\cO) \, U(L)^{-1} = \fF(L \cO)\,,
  \qquad L \in \Point \,, \]
and leaves the vacuum invariant, $U(L)\Omego=\Omego$.
Further there is a strongly continuous representation $Q$
of a compact gauge group $\cG$, commuting with $U$, 
acting locally on $\fFg$ in the sense that
  \[ Q(g) \, \fF(\cO) \, Q(g)^{-1} = \fF(\cO)\,,
  \qquad g\in\cG \,, \]
and leaving the vacuum invariant,
$Q(g)\Omego=\Omego$.
The local observable algebras $\fA(\cO)$ are then defined
as the gauge invariant parts of $\fF(\cO)$,
  \[ \fA(\cO) = \fF(\cO) \cap Q(\cG)' \,,\]
and the total observable algebra is
  \[ \fAg = \overline{\bigcup_\cO \fA(\cO)}\,. \]
The fields are local relatively to the observables,
  \[ \fF(\cO) \subset \fA(\cO')' \,, \]
where $\fA(\cO')$ denotes the $C^*$-algebra generated
by all $\fA(\cO_1)$ with $\cO_1$ space-like separated
from $\cO$.
Note that this implies in particular locality of the
observables which is the manifestation of Einstein
causality of observable quantities in QFT,
 \[ \fA(\cO) \subset \fA(\cO')' \,.\]
Two further assumptions were made; a certain
``cluster property'' and the {\it Reeh-Schlieder property}
which states that every analytic vector for the energy
operator $P_0$ (i.e.\ a vector $\Psik\in\cH$ such that the
power series $\sum_n \| P_0^n \Psik \| t^n/n!$ has non-zero 
radius of convergence in $t$) is cyclic and separating
for each $\fF(\cO)$.
Under these assumptions it could be shown that the Hilbert
space decomposes as follows into superselection sectors
$\cH_\xi$,
  \[ \cH = \bigoplus_{\xi\in\hat{\cG}} \cH_\xi
  \otimes H_\xi \,. \]
Here the sum runs over the spectrum $\hat{\cG}$ of the
gauge group, i.e.\ the set of unitary equivalence classes 
$\xi$ of all (unitary, continuous)
irreducible representations of $\cG$,
and each multiplicity space $H_\xi$ carries a representation
$Q_\xi$ of class $\xi$ with (finite) dimension $d_\xi$,
so that $Q(g)$, $g\in\cG$, takes the form
  \[ Q(g) = \bigoplus_{\xi\in\hat{\cG}} \bfe
  \otimes Q_\xi(g) \,, \]
whereas $A\in\fAg$ acts as
  \[ A = \bigoplus_{\xi\in\hat{\cG}} \pi_\xi(A)
  \otimes \one_{d_\xi} \,. \]
Here $\pi_\xi$ are irreducible representations of
$\fAg$, and $\pi_\xi$ and $\pi_{\xi'}$ are inequivalent
for $\xi\neq\xi'$.
The representation corresponding to the trivial class in
$\hat{\cG}$ is called the {\it vacuum representation}
and is denoted by $\pi_0$.
Note that the correspondence between the irreducible
representations of the gauge group and those of the
observable algebra induces a product of the superselection
sectors of $\fAg$ describing the composition of charge
quantum numbers: It comes from the representation ring
$\cR_\cG=\zet\hat{\cG}$ of the gauge group and can be 
expressed in terms of fusion rules,
  \[ \xi_i \times \xi_j = \sum_k N_{ij} {}^k \, \xi_k \,,\]
where the non-negative integers $N_{ij} {}^k$, the 
{\it fusion coefficients}, describe the multiplicity of the
representation class $\xi_k$ appearing in the decomposition 
of the tensor product of $\xi_i$ and $\xi_j$.

With an additional maximality relation for the local 
observable algebras (Haag duality) and for the case that
$\cG$ is abelian it could also be shown in \cite{DHR1}
that for all $\xi\in\hat{\cG}$ there are certain 
{\it localized automorphisms} $\varrho_\xi$ of $\fAg$
which are implemented by unitaries
$\psi_\xi$ in some $\fF(\cO)$, 
$\varrho_\xi(A)=\psi_\xi A \psi_\xi^*$, $A\in\fAg$,
such that $\pi_\xi$ is unitarily equivalent to $\pi_0$
composed with $\varrho_\xi$, 
$\pi_\xi\simeq\pi_0\circ\varrho_\xi$. Later, 
Doplicher and Roberts \cite{DRu} were able to 
generalize this to non-abelian sectors
(i.e.\ sectors with $d_\xi>1$ when the gauge group is 
non-abelian): These sectors correspond to
non-surjective {\it localized endomorphisms} which are
implemented by multiplets of isometries $\psi_\xi^i$
in some $\fF(\cO)$, $i=1,2,...\,,d_\xi$, transforming
according to a representation of class $\xi$ of the
gauge group (and satisfying the relations of a Cuntz 
algebra). Since endomorphisms can be composed one 
gets a product of unitary equivalence
classes $ [ \pi_\xi ] $ of representations $\pi_\xi$ of
$\fAg$ by defining
  \[ [ \pi_\xi ] \times [ \pi_{\xi'} ]  = [ \pi_0 \circ
  \varrho_\xi \varrho_{\xi'} ] \,. \]
Indeed, this product precisely reproduces the above
sector product, we have
  \[ [ \pi_{\xi_i} ] \times [ \pi_{\xi_j} ]  = 
  \bigoplus_k N_{ij} {}^k \, [ \pi_{\xi_k} ] \,. \]
However, first the investigations were tackled 
from a different direction.

\subsection{DHR Theory: Starting from the Observables}
Already in their second paper \cite{DHR1b}, Doplicher,
Haag and Roberts began to investigate the theory when it
is given in terms of observable algebras 
corresponding to the representation of $\fAg$ in the
vacuum sector. The reason for this new direction was that
it seemed to be more natural to start only from observable
quantities whereas unobservable fields  and the gauge
group may be regarded as auxiliary objects. We will
not give a complete historical review here but sketch
some of the results of \cite{DHR1b, DHR2, DHR2b, DR};
for a detailed treatment of these topics we refer
to Haag's book \cite{Haag}. The theory is now
given by a net of local von Neumann algebras $\cR(\cO)$
acting on a Hilbert space $\cH_0$ and labelled by
$\cO\in\DK$, the set of open double 
cones\footnote{A double cone is a non-void intersection of
a forward and a backward light-cone; the restriction to
these special regions is just of a technical nature.}
in Minkowski space, and the algebra of 
{\it quasilocal observables} $\cA$ is defined 
as the $C^*$-algebra generated by all $\cR(\cO)$,
  \[ \cA = \overline{\bigcup_{\cO\in\DK} \cR(\cO)}\,, \]
where it is assumed that $\cA''=\fB(\cH_0)$. Since
one interprets $\cR(\cO)$ as $\pi_0(\fA(\cO))$ in the
previous framework, this requirement simply expresses
irreducibility of the vacuum representation. One also
assumes that there is a unitary, strongly continuous 
representation $U_0$ of $\Poin$ such that the observables
transform covariantly and that the spectrum condition
is fulfilled. The locality condition for observables
will now be strengthened to {\it Haag duality},
  \[ \cR(\cO) = \cA(\cO')' \]
where $\cA(\cO')$ again denotes the $C^*$-algebra 
generated by all $\cR(\cO_1)$ with $\cO_1$
spacelike to $\cO$.
While being a result in the previous setting, the
{\it DHR selection criterion}
  \be \pi |_{\cA(\cO')} \simeq \pi_0 |_{\cA(\cO')}
  \labl{DHRc} 
for some double cone $\cO$ and $\pi_0=\id$
will now serve as a selection criterion for 
physical representations $\pi$ that have to be 
considered.\footnote{It is now the
DHR criterion which is too stringent for Quantum
Electrodynamics as it excludes all states with
non-vanishing charge by virtue of Gauss' law.
Moreover, Buchholz and Fredenhagen pointed out
\cite{BF} that also in purely massive theories there
can exist charges which are measurable at arbitrarily
large distances. However, such a situation can be
treated \cite{BF} by replacing double cones by
spacelike cones as (unbounded) localization
regions, but this will not be considered here.}
If one denotes the unitary operator realizing the 
equivalence (\ref{DHRc}) by $V$ then one obtains by
  \[ \varrho (A) = V^* A \, V \,,
  \qquad A \in \cA\,, \]
a localized endomorphism (i.e.\ $\varrho(A)=A$
whenever $A\in\cA(\cO')$) satisfying
  \[ \pi \simeq \pi_0 \circ \varrho \,. \]
An important feature of these localized endomorphisms 
of the observable algebra $\cA$ is that they
allow to define a product of DHR sectors, i.e.\ of 
equivalence classes $ [ \pi ] $ of representations
$\pi$ satisfying (\ref{DHRc}), without the presence
of a gauge group a priori; namely again by 
$ [ \pi ] \times [ \pi' ]  = [ \pi_0 \circ
\varrho \varrho' ] $. So we still have a
product structure of charge quantum numbers $\xi$ which
label the DHR sectors $ [ \pi_\xi ] $. 
Assuming also the {\it Borchers property}\footnote{The
Borchers property states that
for each non-zero projection $E$ in some $\cR(\cO)$ there
is a double cone $\cO_1$ containing $\cO$ properly so that
$E$ is equivalent to the identity within $\cR(\cO_1)$.}
it could be shown in \cite{DHR2,DHR2b} 
that to every charge $\xi$ there is a unique
conjugate charge $\overline{\xi}$ which means 
that $ [ \pi_0 ] $
appears (precisely once) in the product
$ [ \pi_\xi ] \times [ \pi_{\overline{\xi}} ] $. 
Moreover, to each charge sector there is an associated 
representation of the permutation group which is
characterized by its statistical dimension $d_\xi$,
the order of parastatistics, and a sign which 
distinguishes para-Bose from para-Fermi statistics.
Thus simple sectors ($d_\xi=1$) obey ordinary
Bose or Fermi statistics. One observes that such
superselection structures are completely analogous
to the ring structure of (equivalence classes of) irreducible,
continuous, unitary representations of compact groups:
The sector product corresponds to the tensor product
of group representations, and the statistical dimension
to the ordinary dimension of group representations.
It took some years until Doplicher and Roberts \cite{DR}
succeeded in proving that the assumed structure is 
indeed enough to reconstruct a field net and a compact
gauge group so that the observables can be recovered
as the gauge invariant fields.

We have to mention that some deviations of the structure
described above arise if one considers localizable charges
in two-dimensional spacetime. Owing to the fact that the
spacelike complement of a double cone then has two
connected components one obtains, in general, braid group
statistics instead of permutation symmetry of the
sectors \cite{FRS1}. This leads to theories containing
particles like ``anyons'' and ``plektons''.

\section{Conformal Field Theory}
We will give some brief introductory remarks on a few
topics of CFT here, and we hope that it
will make the comprehension of the following chapters easier.
For a detailed treatment we refer the reader to the literature
on CFT and its mathematical background, e.g.\ \cite{FST,Mack}
or Fuchs' book \cite{Fuchsb}.

\subsection{CFT Basics}
A CFT is a quantum field theory
where the fields transform covariantly under the conformal
group. The conformal group is the group of transformations
$f:\bbM^D\rightarrow\bbM^D$ of $D$-dimensional spacetime
$\bbM^D$ so that the metric $g$ transforming to 
$g'=g\circ f^{-1}$ remains invariant up to a scalar factor,
$g_{\mu\nu}'(x')=\Omega(x)g_{\mu\nu}(x)$, $\Omega(x)\neq0$.
The conformal group contains the Poincar{\'e} group
($\Omega(x)=1$) as a subgroup. If $D\ge3$ the conformal group
is generated besides translations, ${x'}^\mu=x^\mu+a^\mu$,
and Lorentz transformations, ${x'}^\mu=\Lambda^\mu{}_\nu x^\nu$,
also by scale transformations ${x'}^\mu=\lambda x^\mu$,
$\lambda>0$, and special conformal transformations
  \[ {x'}^\mu = \frac{x^\mu + b^\mu x^2}{1+ 2 b\cdot x + b^2x^2},
  \qquad b^\mu \in \reals^D \,. \]
These transformations generate a group of dimension
$(D+1)(D+2)/2$ which, however, does not act properly on
$\bbM^D$ since special conformal transformations can map
finite points to infinity. A proper action can be 
arranged by a suitable compactification $\tilde{\bbM}^D$
of the spacetime. For $D=2$ the situation is rather different; the
conformal symmetry then is infinite dimensional. The corresponding
Lie algebra consists of two copies of the Witt algebra. The
Witt algebra is the Lie algebra with generators $\el_m$,
$m\in\zet$, subject to relations
  \[ [ \el_m, \el_n ] = (m-n) \el_{m+n} \,. \]
If the Minkowski space is compactified to
$\tilde{\bbM}^2=S^1\times S^1$, where each circle is the
image of light ray coordinates $x_\pm=x^0\pm x^1$ via
the Cayley transformation
  \[ x_\pm \longmapsto z,\zb\in S^1, \qquad 
  z=\frac{1+\I x_+}{1-\I x_+} \,, \qquad
  \zb=\frac{1+\I x_-}{1-\I x_-} \,,\]
then the Witt algebra action can be realized by
  \[ \el_m = -z^{m+1} \frac{\D}{\D z}\,,\qquad
  \bar{\el}_m = -\zb^{m+1} \frac{\D}{\D \zb}. \]
At the quantum level it is no longer the
Witt algebra which implements the conformal symmetry but
the Virasoro algebra $\Vir$ with generators $L_m$, $m\in\zet$,
and a central element $C$ satisfying relations
  \[ [ L_m,L_n ] = (m-n)L_{m+n} + \Frac1{12}\, \del m{-n} \,
  m(m^2-1) \, C \,. \]
The Virasoro algebra is the unique non-trivial central extension 
of the Witt algebra. The extension is necessary because one requires
the conformal energy operator $H=L_0+\bar{L}_0$ to be
bounded from below. Here $\bar{L}_0\in\Virb$, the second
copy of the Virasoro algebra. In general, the Hilbert space
of physical states of a two-dimensional CFT splits as
  \[ \Hph = \bigoplus_{i,j}
  \cH_i \otimes \bar{\cH}_j \]
where $\cH_i$ and $\bar{\cH}_j$ carry unitary irreducible
representations of $\Vir$ and $\Virb$. Let us consider one
copy of the Virasoro algebra alone for a while.
Unitarity means that $L_m^*=L_{-m}$. The
positive energy condition now means that the spaces
$\cH\equiv\cH_i$ have to be 
{\it highest weight modules},
i.e.\ there is a vector $\Omd\in\cH$ such that
  \[ L_0 \Omd = \Delta \Omd \]
with a scalar $\Delta$, and
  \[ L_m \Omd = 0 \,, \qquad m>0 \,. \]
By irreducibility, the central element $C$ acts as a
scalar $c$. Moreover, $\cH$ is spanned by vectors
  \be L_{-m_1} L_{-m_2} \cdots L_{-m_k} \Omd \,,
  \qquad m_1 \ge m_2 \ge \cdots \ge m_k >0 \,.\labl{virvec}
In general these vectors span the {\it Verma module}
$M(c,\Delta)$ freely. However, the unitarity condition
may imply linear dependencies of vectors (\ref{virvec}),
or, equivalently, the existence of {\it null states} with
vanishing norm. Unitarity imposes also restrictions
on the possible values of $c$ and $\Delta$.

Let us now turn to the picture of quantum fields
acting in the Hilbert space $\Hph$. We first would like
to emphasize that unitary modules over
Lie algebras possess a pre-Hilbert space structure and are,
for example, generated by the finite linear span of vectors
like (\ref{virvec}). Since in quantum physics one has to deal with
Hilbert spaces we will use the corresponding Hilbert
space completions so that the modules appear as their dense
subspaces. When there is no confusion with domains of
unbounded operators etc.\ we will sometimes sloppily use
the term ``modules'' also for the completed spaces.
The first field one may introduce is the conformal 
{\it stress energy tensor}, defined by the (formal) series
  \[ T(z)=\sum_{m\in\zet} z^{-m-2} L_m \]
and analogously $\bar{T}(\zb)$. Very important
objects in CFT are the {\it primary fields}. A 
primary field $\phi(z,\zb)$ of weight
$(\Delta,\bar{\Delta})$ transforms, by definition,
covariantly relative to the stress energy tensor, i.e.
  \be
  \bearl [ L_m , \phi (z,\zb) ] = z^m \left(
  z \Frac{\partial}{\partial z} + (m+1)\Delta \right)
  \phi (z,\zb) \,, \\{}\\[-.4em]
  [ \bar{L}_m , \phi (z,\zb) ] = \zb^m \left(
  \zb \Frac{\partial}{\partial \zb} + (m+1)\bar{\Delta}
  \right) \phi (z,\zb) \,. \eear \labl{prim}
Note that the stress energy tensor itself is not a primary
field. The fields are understood to be ``analytically
continued'' to $\bbC\setminus \{0\}$ in both variables
$z,\zb$. In a fixed CFT the value of $c$ is fixed and it is
also assumed that there is a vacuum vector $\Omego\in\Hph$
of highest weight zero, $L_0\Omego=\bar{L}_0\Omego=0$.
An important feature of a primary field $\phi$ with weight
$(\Delta,\bar{\Delta})$ is that it creates a highest
weight vector with weight $(\Delta,\bar{\Delta})$ out
of the vacuum in the sense that
  \[ \lim_{z,\zb\to 0} \phi (z,\zb)   \Omego 
  = |\Omega_{\Delta,\bar{\Delta}}\rangle \,. \]
Thus we have a state-field-correspondence between
highest weight states and primary fields. Similarly,
fields that correspond to non-highest weight vectors
are called {\it descendants}. For the primary fields
one introduces operator product expansions,
  \[ \phi_i (z,\zb) \phi_j (w,\bar{w}) = \sum_k
  c_{ij} {}^k (z,\zb,w,\bar{w})  \phi_k (w,\bar{w}) 
  + \ldots  \]
in terms of radially ordered products, i.e.\
$|z|>|w|$, $|\zb|>|\bar{w}|$. Here the dots stand for
contributions of descendants. Conformal covariance
implies that
  \[ c_{ij} {}^k (z,\zb,w,\bar{w}) = 
  (z-w)^{\Delta_k-\Delta_i-\Delta_j}
  (\zb-\bar{w})^{\bar{\Delta}_k-\bar{\Delta}_i
  -\bar{\Delta}_j} \, C_{ij} {}^k \,. \]
The constants $C_{ij} {}^k$ may be interpreted as an
analogy to Clebsch-Gordan coefficients. Somewhat coarser
information is encoded in the {\it fusion rules}
  \[ \phi_i \times \phi_j = \sum_k N_{ij} {}^k \phi_k \]
where the (non-negative integer) fusion coefficients indicate
whether the primary field $\phi_k$ appears in the operator
product expansion of $\phi_i$ and $\phi_j$
($N_{ij} {}^k \neq 0$) or not ($N_{ij} {}^k =0$).
The primary fields of a CFT model generate an
associated {\it fusion ring} $\mathcal{R}$ (over $\zet$),
see e.g.\ \cite{Fuchs2}. Since such fusion rings are
abelian there are only one-dimensional representations.
There is a distinguished positive representation
$\mathcal{D}$ which associates to each primary field
$\phi_i$ a {\it quantum dimension} 
$\mathcal{D}_i\equiv\mathcal{D}(\phi_i)>0$.

In generic CFTs the two copies $\Vir\times\Virb$
are extended to a larger {\it symmetry algebra} $\fW\times\fWb$
and each {\it chiral half} $\fW,\fWb$ contains a Virasoro
algebra as a subalgebra. In such theories there can appear
fusion coefficients $N_{ij} {}^k >1$. This corresponds to
the fact that there might be additional contributions of
descendants in the operator product expansions that would
be excluded in the case $N_{ij} {}^k =1$ by 
$\fW,\fWb$-symmetry, see e.g.\ \cite[Sect.\ 5.1]{Fuchsb}.
Examples of CFTs with enlarged symmetry algebras
are the {\it Wess-Zumino-Witten} (WZW) theories
that will be discussed now.  Moreover, one may treat each chiral
algebra for its own so that one has a chiral CFT on the circle.

\subsection{WZW Theories}
In a WZW theory the chiral algebra $\fW$ is given by the
semi-direct sum of an (untwisted) affine Kac-Moody
algebra $\gh$ and an associated Virasoro algebra. This has 
to be made a little bit more precise now. Let $\g$ be a simple
finite-dimensional Lie algebra. Recall that $\g$ has a 
canonical {\it triangular decomposition},
  \[ \g = \g_+ \oplus \fh \oplus \g_- \]
where the commutative {\it Cartan subalgebra} $\fh$ has a
basis $\{H^j,\,j=\onetol\}$ and the subalgebras
$\g_\pm$ are generated by Chevalley generators $E_\pm^j$, 
$j=\onetol$, subject to commutation relations
  \[ [ H^j,H^k ] = 0\,,\quad [ E^j_+, E^k_- ] = \del jk H^j\,,
  \quad [ H^j, E_\pm^k ] = \pm (\alpha^{(k)})^j E_\pm^k \,, \]
where the $\alpha^{(k)}$ are the simple roots of the
Lie algebra $\g$. The associated (infinite-dimensional)
affine Lie algebra $\gh$ is generated by a central element
$K$ and the range of linear mappings 
$J_m:\g\rightarrow\gh$, $T\mapsto J_m(T)$, $m\in\zet$,
so that commutation relations $ [ J_m(T),K ] =0$ and
  \[ [ J_m(T),J_n(T') ] = J_{m+n}( [ T,T' ] ) +
  m \, \del m{-n} (T|T') K \]
hold. Here $(\cdot|\cdot)$ denotes the invariant bilinear form
of $\g$. Adding one further element $D$, the ``derivation'',
satisfying
  \[ [ D, J_m(T) ] = m \, J_m(T) \,, \qquad [ D,K ] = 0 \]
one obtains the full affine Kac-Moody algebra according to
Cartan's classification, for simplicity, we will also refer to
$\gh$ as the affine Lie algebra when the derivation is
omitted, however. The affine Lie algebra inherits also a 
triangular decomposition,
  \[ \gh = \gh_+ \oplus \fhh \oplus \gh_- \]
where 
$\gh_\pm = J_0(\g_\pm) \oplus \bigoplus_{m=1}^\infty J_{\pm m} (\g)$
and
$\fhh = J_0(\fh) \oplus \bbC K \oplus \bbC D$.
Similarly, one has Chevalley generators $\EE j\pm\in\gh_\pm$,
$j=\otol$,
  \[ \EE j\pm = J_0 (E^j_\pm), \quad j=\onetol \,, \qquad
  \EE 0\pm = \pm J_{\pm 1} (E_{\pm \theta}) \,. \]
Here $E_\theta\in\g$ is the element corresponding to the
highest root $\theta$ of $\g$. We also introduce 
$\cH^j=J_0(H^j)$, $j=\onetol$. It is the triangular
decomposition which allows to define highest weight modules
over an affine Kac-Moody algebra quite parallel to those
over simple Lie algebras. A highest weight module over an
affine Lie algebra $\gh$ is a vector space $\Hl$ with a
highest weight vector $\Oml\in\Hl$ that is annihilated by 
the Chevalley generators of positive grade,
  \[ \EE j+ \Oml = 0 \,, \qquad j=\otol\,, \]
the Cartan subalgebra acts by scalars on $\Oml$,
  \[ \cH^j \Oml = \Lambda^j \Oml \,, \qquad
  K \Oml = k \Oml \,, \]
and the action of $\gh_-$ on $\Oml$ spans the whole
vector space $\Hl$. For {\it integrable} 
highest weight modules (i.e.\ those modules which
admit a certain definition of an exponential map)
the normalized eigenvalue of $K$, called the {\it level},
  \[ \kv = \frac{2k}{\theta^2} \,, \]
($\theta^2$ is the square length
of the highest root of $\g$) can only be a non-negative
integer. The interesting highest weight modules for CFT
are the unitary ones, i.e.\ those which possess a pre-Hilbert
space structure such that $(\EE j+)^*=\EE j-$ and 
$(\cH^j)^*=\cH^j$. From the unitarity requirement arise
severe restrictions on the possible weights 
$\Lambda=(\Lambda^j)$: At a fixed level $\kv$, there is
only a finite number of admissible weights $\Lambda$.

It is the {\it Sugawara construction} which associates to
an affine Lie algebra a realization of the Virasoro algebra.
This works as follows: Fix a basis 
$\{ T^a, \,a=1,2,...\,,\dimg \}$ of $\g$ such that
$(T^a|T^b)=\del ab$ and set $J^a_m=J_m(T^a)$. In particular in
a highest weight module the following expression is well
defined,
  \[ L_m = \frac{1}{\theta^2(\kv+\gv)}
  \sum_{a=1}^\dimg \sum_{n\in\zet} 
  \normord{J^a_n J^a_{m-n}}\,, \qquad m\in \zet\,. \]
Here $\gv$ is the dual Coxeter number of $\g$, and we used the
normal ordering prescription
  \[ \normord{J^a_m J^a_n} = \left\{ \bearll
  J^a_m J^a_n \qquad & m<0 \\
  J^a_n J^a_m & m\ge 0 \eear \right. .\]
One checks that
  \be [ L_m , J^a_n ] = -n \, J^a_{m+n} \,, \labl{semi}
in particular, the generator $-L_0$ can be identified with
the derivation $D$. Moreover, at a fixed level $\kv$, the
$L_m$ satisfy the relations of the Virasoro algebra with
fixed value of the central charge
  \[ c= \frac{\kv \, \dimg}{\kv+\gv} \,. \]
As mentioned, for a WZW theory the chiral algebra is given by
  \[ \fW = \gh \rtimes \Vir \]
acting on a (pre-) Hilbert space which is the 
direct sum over the unitary
highest weight modules of $\gh$ at a fixed level, and $\Vir$
is given by the Sugawara construction. (If we consider an affine
Lie algebra $\gh$ concretely realized at a fixed level $\kv$ it 
will often be denoted by $\gh_{\kv}$.)
Note that the {\it currents}
  \[ J^a(z) = \sum_{n\in\zet} z^{-n-1} J^a_n \]
from (\ref{semi}) in comparison with (\ref{prim}) are
(chiral) primary fields of unit weight. Physicists often
use the term ``current algebras'' instead of affine
Lie algebras.

Also very important objects in CFT are the
{\it Virasoro specialized characters}. By positivity of $L_0$
one can define in a module $\cH_i$ of the chiral algebra
  \be \chi_i (\tau) = \tr_{\cH_i}
   q^{L_0 - \frac{c}{24}} \labl{defchar}
where $q=\exp(2\pi\I\tau)$ and $\tau$ is in the upper complex
half plane. Such characters possess simple transformation
properties with respect to the {\it modular group}
$\PSLZ$ which can be generated by elements $S,T$ with
relations $S^2=(ST)^3=1$ and realized by transformations
$T:\tau\mapsto\tau+1$ and $S:\tau\mapsto -1/\tau$.
The characters form a (projective) representation
of $\PSLZ$ in the sense that $S$ acts as a matrix $(S_{ij})$,
  \[ \chi_i \left( - \Frac{1}{\tau} \right) = 
  \sum_j S_{ij} \, \chi_j (\tau) \]
whereas $T$ acts diagonally,
  \[ \chi_i(\tau+1) = \exp \left( 2\pi \I \left(
  \Delta_i - \Frac{c}{24} \right) \right) \chi_i(\tau)\,.\]
Such characters are very helpful tools in the analysis of
CFTs. There is an amazing connection between the modular 
transformation matrix $S$ and the fusion rules; namely
$S$ ``diagonalizes the fusion rules'' in the sense that
  \[ N_{ij} {}^k = \sum_l 
  \frac{S_{il}S_{jl}(S^{-1})_{lk}}{S_{0l}} \,. \]
This is called the Verlinde formula (since it goes back
to a conjecture of E. Verlinde), and sometimes it is
convenient to use this formula for the explicit
computation of the fusion rules of WZW models.

\subsection{Coset Theories}
An important tool to generate new CFTs from WZW theories is
the {\it coset construction} of Goddard, Kent and Olive
\cite{goko1,goko2}. We will briefly sketch some of the ideas
of \cite{goko1,goko2} and also \cite{kawa} here.
First we have to note that the Sugawara construction can be
generalized to affine Lie algebras $\gh$ based on semisimple
Lie algebras $\g$, i.e.\ Lie algebras of the form
$\g=\g_1\oplus\g_2\oplus\cdots\oplus\g_n$ with
$\g_i$ simple. Modules of such $\gh$ are typically of the
form $\cH = \bigotimes_{i=1}^n \cH_i$
where $\cH_i$ are modules of the affine Lie algebras
$\gh_i$ associated to $\g_i$. Suppose that the $\cH_i$
are (unitary) highest weight modules of $\gh_i$ at
fixed levels $k_i$. Then we can define
  \[ L_m^\g = \sum_{i=1}^n L_m^{\g_i} \]
where $L_m^{\g_i}$ is the Virasoro operator associated to
$\gh_i$ (acting non-trivially on the $i$-th tensor factor
$\cH_i$). Such operators satisfy the relations of a Virasoro
algebra with central charge $c^\g=\sum_{i=1}^n c^{\g_i}$.
Now suppose that there is a subalgebra $\fh\subset\g$ with
associated affine Lie algebra $\fhh$ and Virasoro operators
$L_m^\fh$. The operators
  \[ L_m^{\g / \fh}=L_m^\g-L_m^\fh \]
generate the {\it coset Virasoro algebra} $\Vir^{\g/\fh}$
which is a Virasoro algebra of central charge
$c^{\g/\fh}=c^\g-c^\fh$. Further one checks
  \be  [ L_m^{\g / \fh}, J_n^a ] = 0 \labl{nix} 
where $J_n^a$, $a=1,2,...\,,\dim\fh$, are the generators
of $\fhh$. Let $\cH_\Lambda$ be a highest weight module over
$\gh$ of weight $\Lambda$. As a module over $\fhh$, one can
decompose $\cH_\Lambda$ as
  \be \cH_\Lambda = \bigoplus_\lambda \cH_{\Lambda;\lambda}
  \otimes \cH_\lambda^{\fhh} \labl{coset}
into tensor products of branching spaces 
$\cH_{\Lambda;\lambda}$ and highest weight modules
$\cH_\lambda^{\fhh}$ over $\fhh$. The branching spaces
$\cH_{\Lambda;\lambda}$ constitute modules of  $\Vir^{\g/\fh}$
due to (\ref{nix}); they are in general not irreducible
but can be interpreted as the representation spaces
of an enlarged algebra $\Cos$
of a certain coset CFT. On the level of characters, the 
decomposition (\ref{coset}) reads
  \[ \chi_\Lambda (\tau) = \sum_\lambda
  b_{\Lambda;\lambda} (\tau) \, \chi_\lambda^{\fhh} (\tau) \,,\]
and the {\it branching functions} $b_{\Lambda;\lambda}$
are to be interpreted as (sums of) the 
characters of the coset CFT.

Let us illustrate this at a special example. Let
$\g=\fn\oplus\fn$ with $\fn$ a simple Lie algebra.
Consider a (highest weight) module over $\gh$ of the form
  \[ \cH_{\Lambda,\Lambda'} = \cH_\Lambda^{(1)}
  \otimes \cH_{\Lambda'}^{(k)} \]
where $\cH_\Lambda^{(1)}$ is a level $1$, and
$\cH_{\Lambda'}^{(k)}$ is
a level $k$ module over $\fnh$. Further, let $\fh$ be the
diagonal embedding of $\fn$ in $\g$. Then $\fhh$ acts on
$\cH_{\Lambda,\Lambda'}$ at level $k+1$.
According to (\ref{coset}) we may decompose
$\cH_{\Lambda,\Lambda'}$ into highest weight modules
$\cH_\lambda^{(k+1)}$ over $\fhh_{k+1}$,
  \[ \cH_{\Lambda,\Lambda'} = \bigoplus_\lambda
  \cH_{\Lambda,\Lambda';\lambda} \otimes
  \cH_\lambda^{(k+1)} \]
where the branching spaces $\cH_{\Lambda,\Lambda';\lambda}$
constitute modules of $\Cos$ of the coset theory which is
in this special case denoted by
$(\fnh_1\oplus\fnh_k)/ \fnh_{k+1}$. For the characters the
branching reads
  \[ \chi_{\Lambda,\Lambda'} (\tau) \equiv \chi_\Lambda^{(1)}
  (\tau) \, \chi_{\Lambda'}^{(k)} (\tau) = \sum_\lambda
  b_{\Lambda,\Lambda';\lambda} (\tau) \,
  \chi_\lambda^{(k+1)} (\tau) \,. \]
Here $\chi_\Lambda^{(1)}$, $\chi_{\Lambda'}^{(k)}$,
$\chi_\lambda^{(k+1)}$ denote the characters of
highest weight modules over $\fnh$ at levels $1$,
$k$, $k+1$, respectively, and the branching functions
$b_{\Lambda,\Lambda';\lambda}$ are characters of the
coset CFT $(\fnh_1\oplus\fnh_k)/ \fnh_{k+1}$. In \cite{goko2}
this procedure was successfully used to realize the full
discrete series of highest weight modules over $\Vir$
with central charge $0<c<1$ with $\fn=\mathfrak{su}(2)$.
In Chapter 4 we will concentrate to the special coset
CFT $(\sonh_1\oplus\sonh_1)/ \sonh_2$.

\section{CFT within the Algebraic Approach}
It is perhaps not too surprising that there are always 
difficulties to apply the abstract and conceptually clear
mathematical setting of AQFT to concrete 
quantum field theoretical models.
However, models of CFT seem to be a fruitful area of application
of the algebraic methods. Although the technical tools that are
used in the algebraic approach and in CFT are rather different, a lot
of structural similarities has been observed for a long time.
The appearance of braid group statistics in two-dimensional
spacetime is only one aspect.
Indeed, the symmetry (or chiral) algebras seem to play a
\role\ parallel to that of the observable algebras in the
algebraic approach, and their highest weight modules 
appear as the perfect analogue to the superselection
sectors. Therefore a suitable formulation of
CFT models in the algebraic framework is expected to
reproduce the conformal fusion rule structure by
the DHR sector product. Unfortunately, these natural
translation prescriptions do not have the status of a
proven mathematical theorem. Moreover, certain deviations
of the canonical DHR framework will necessarily arise because
the fusion rings of CFT models are in general not associated
to DHR gauge groups but to more general ``quantum symmetry''
objects like quantum groups. For instance, the quantum
dimensions $\mathcal{D}$ which would correspond to the
representation dimension (or statistical dimension) in
the DHR framework are in general non-integral.
But there is a small number
of CFT models which have been successfully investigated
in the algebraic framework and where these correspondences
could be established. These models are chiral theories,
so we have to discuss how the algebraic framework looks
on the circle.

\subsection{DHR Theory on the Circle}
We have seen that the circle $S^1$ as a ``spacetime''
arises as the compactification of a light cone axis.
Since double cones in $\bbM^2$ project to intervals on
the light cone, the natural localization regions on the
circle are non-void, open proper subintervals $I\subset S^1$;
the set of those will be denoted by $\mathcal{J}$. We give a
formulation starting from the observables in the vacuum
sector parallel to the discussion in Subsection 1.1.2.
Acting in some Hilbert space $\cH_0$ we have for each
$I\in\mathcal{J}$ a local von Neumann algebra $\cR(I)$
so that isotony holds, $I_1\subset I_0$ implies 
$\cR(I_1)\subset\cR(I_0)$. Instead of the
Poincar{\'e} group here the group $\PSU$ acts as the
spacetime symmetry,\footnote{The explicit action of
$\PSU$ on the circle will be discussed in Chapter 3,
Subsection 3.1.4.} i.e. there is a strongly continuous
representation $U$ of $\PSU$ in $\cH_0$. The observables
transform covariantly,
  \[ U(g) \cR(I) U(g)^{-1} = \cR(gI)\,,
  \qquad g\in\PSU\,, \]
and the generator $L_0$ of rotations has non-negative
spectrum. The eigenvalue zero belongs to a unique
vacuum vector $\Omego\in\cH_0$. Since spacelike
separated double cones project to disjoint intervals,
the locality requirement becomes now
  \[ \cR(I_1)\subset\cR(I_2)' \,,\qquad
  I_1\cap I_2 = \emptyset \,. \]
A total algebra cannot be defined as the norm closure
of the union of all local algebras since the set
$\mathcal{J}$ is not directed by inclusion. However,
for $\zeta\in S^1$, the set $\Jz$ of intervals
$I\in\mathcal{J}$ that do not contain $\zeta$ in
their closure is directed; we obtain a bundle of 
$C^*$-algebras
  \[ \cA_\zeta = \overline{\bigcup_{I\in\Jz}
  \cR(I)} \,, \]
each invariant with respect to the subgroup of
$\PSU$ leaving the point $\zeta$ invariant.
Following \cite{FRS2}, one could also introduce the 
universal algebra; however, for our applications it
turned out to be more convenient to restrict to a
fixed algebra $\cA_\zeta$. We call such an algebra
$\cA_\zeta$ the observable algebra of the punctured
circle, the point $\zeta$ will then be called 
``point at infinity''.

\subsection{Models}
The first example of a CFT treated in the algebraic framework
was the $\ue$ current algebra on the circle \cite{BMT},
generated by a current $J(z)$, $z\in S^1$, with
commutation relations
  \[ [ J(z),J(z') ] = - \delta' (z-z') \,, \]
or, in the language of Fourier-Laurent components
  \[ [ J_m,J_n ] = m \, \del m{-n} \,, \qquad
  J(z)=\sum_{n\in\zet} z^{-n-1} J_n \,. \]
The sectors of the theory are the highest weight modules
of the $\ue$ current algebra and are specified by a highest
weight $\lambda\in\reals$ which we call the charge in this
context. Introducing {\it Weyl operators}
  \[ W(f) = \E^{\I J(f)} \,, \qquad
  J(f) = \oint \frac{\D z}{2\pi\I}\, J(z) f(z) \,, \]
for real test functions $f$ on the circle, one
arrives at a $C^*$-algebra description and has
relations
  \[ W(f)W(g)=\E^{-A(f,g)/2} W(f+g) \,,   \qquad 
  A(f,g)=\oint \frac{\D z}{2\pi\I}\, f'(z)g(z)\,.\]
Local observable algebras are then defined to be generated by
Weyl operators $W(f)$ with locally supported test functions
$f$. The relevant endomorphisms $\varrho_q$ (which are 
indeed automorphisms here) are induced by certain ``charge
functions'' $q$ on $S^1$ with $zq(z)\in\reals$; they are
defined by their action on Weyl operators
  \[ \varrho_q (W(f)) = \E^{\I q(f)} W(f)\,,\qquad
  q(f)= \oint \frac{\D z}{2\pi\I}\, q(z)f(z)\,.\]
Such endomorphisms indeed generate a sector of charge
  \[ \lambda_q = \oint \frac{\D z}{2\pi\I} \, q(z)\,.\]

A somewhat more interesting model, namely the chiral
Ising model, was first investigated
in \cite{MS1} from the algebraic point of view. This
model is, roughly speaking, given by a $c=1/2$
Virasoro algebra, and its three
inequivalent, unitary, irreducible highest weight modules
are realized in fermionic representation spaces. More
precisely, there appear two different representations
of the canonical anticommutation relations (CAR), the 
{\it Neveu-Schwarz} and the {\it Ramond sector}. The
$\Vir$-modules arise as their subspaces related to the fact
that the Virasoro generators act as unbounded expressions
in fermionic creation and annihilation operators. This
explicit realization allowed to employ the structure
of the underlying fermions to construct the relevant
endomorphisms which indeed reproduced the Ising fusion
rules. However, the use of observable algebras containing
unbounded elements and also the use of non-localized 
endomorphisms did not really fit into the DHR framework so
that some open questions were left. In particular,
non-localized endomorphisms do not generate a unique
fusion ring. A formulation more close to the DHR
framework was given in \cite{leci}. The foundation
of the analysis carried out there is the fact that
the representation theory of the even CAR algebra 
(i.e.\ the subalgebra of the CAR algebra generated
by bilinear expressions in creation and annihilation
operators) reproduces precisely the modules of the
$c=1/2$ Virasoro algebra. Therefore local even CAR
algebras could be identified as local observable
algebras and the localized endomorphisms that 
generate the sectors could be defined in terms of 
Bogoliubov transformations of the CAR algebra, moreover,
since local normality could be proven,
their action could be lifted to a net of local
von Neumann algebras. Having the well-developed
mathematics of Bogoliubov transformations and
quasi-free states of CAR at hand, various equivalences
could be proven; as a result, the Ising fusion rule
algebra was rigorously verified. We provide a
generalization of this analysis to the level $1$
$\son$ WZW models in this dissertation in Chapter 3.

There is another important operator algebraic approach
to CFT, namely Wassermann's analysis \cite{wass?} of
fusion of positive energy representations using bounded
operators. Although making essential use of the fermionic
construction in WZW models, the algebras of bounded operators
do not arise as certain fermion algebras in this analysis
but from representations of the Lie group which corresponds
to the chiral (Lie) algebra. The Lie group associated
to a WZW model with chiral algebra $\gh\rtimes\Vir$ is
given by a semidirect product $\LGer\rtimes\DiffS$ where
$\LGer$ is a central extension of the loop group $\LG$
and $\DiffS$ is the diffeomorphism group of the circle.
Let $\Gr$ be a compact Lie group with associated Lie
algebra $\g$. The elements of the loop group $\LG$ are
the smooth mappings from the circle $S^1$ into the group
$\Gr$ (``loops''), \linebreak[2]
and the group product is given by point-wise
multiplication \cite{PrSe}. The interesting representations of 
$\LG$ are those of {\it positive energy} i.e.\ unitary, projective
representations that extend to $\LG\rtimes\DiffS$ so that
the generator of rotations is bounded from below. Such
projective representations come from genuine representations
of $\LGer$, a central extension of $\LG$ by the circle group
$\bbT$. Indeed, the positive energy representations of
$\LG$ are just the ``integrated versions'' of the unitary
integrable highest weight modules of the Lie algebra
$\gh\rtimes\Vir$. Wasser\-mann's analysis essentially
concentrates on the special case $\Gr=\SUN$ where the
positive energy representations of the loop group
$\LSUN$ are realized in fermionic Fock spaces. Local
algebras arise in the framework of loop groups naturally
as the von Neumann algebras which are generated by
{\it local loop groups} $\LIG$ in a certain representation;
for some interval $I\subset S^1$ the local loop group
$\LIG$ consists of those loops equal to the identity
off $I$. However, Wassermann does not use endomorphisms
to fuse the positive energy representations of
$\LSUN$ but the so-called ``Connes fusion'' coming
from the theory of bimodules over von Neumann algebras.
With this machinery he can indeed reproduce the fusion
rules of $\LSUN$ at arbitrary level given by the
Verlinde formula. Moreover, he can prove that the
Connes fusion is equivalent to the DHR sector product
of localized endomorphisms. An approach to the discrete
series representations of $\DiffS$ following Wassermann's
ideas is given in \cite{Loke}.

\section{Summary of Results and Overview}
We have already discussed that the unitary highest
weight modules of the chiral algebra play the \role\
of the superselection sectors of the WZW theory.
In this dissertation we investigate the superselection
structure of $\son$ WZW models at level $1$ and $2$.
At level $1$ we are able to present an operator algebraic
formulation of these models close to the DHR framework.
We did not succeed in developing an analogous program at
level $2$, however, we will show that the mathematical
methods of the DHR analysis can be successfully applied 
to figure out a lot of information on the superselection
structure of the WZW models at level $2$ and, possibly,
also at higher level.

\subsection{Results (Level 1)}
At level $1$ the $\son$ WZW models have only a small
number of sectors, the basic ($\circ$), the vector
($\rmv$) and two spinor modules ($\rms,\rmc$) if $N$
is even, $N=2\el$, respectively the basic, the vector
and one spinor module ($\sigma$) if $N$ is odd,
$N=2\el+1$. It is known that these modules are
realized in two representation spaces of CAR, namely in
the Neveu-Schwarz sector $\HNS$ ($\circ,\rmv$) and in the 
Ramond sector $\HR$ ($\rms,\rmc$ resp.\ $\sigma$).
For the explicit realization, we use
Araki's selfdual formalism of CAR. The selfdual CAR
algebra $\CKG$ is the $C^*$-algebra generated by
elements $B(f)$, $B$ linear in vectors $f$ of some
Hilbert space $\KK$ which is endowed with a complex
conjugation $\GG$, so that
$\{B(f)^*,B(g)\}=\langle f,g \rangle \bfe$ and
$B(f)^*=B(\GG f)$, $f,g\in\KK$, holds.
The notion of creation and
annihilation operators appears in this framework
only in a chosen (Fock) representation of $\CKG$.
In our case the underlying
``pre-quantized'' Hilbert space is $\KK=\LSN$.
In $\HNS$ and $\HR$ act representations 
$\PiNS$ respectively $\PiR$ of $\CKG$, and the
actions of $\sonh$ in $\HNS$ and $\HR$ arise via
second quantization from a natural action in the
pre-quantized Hilbert space. The second quantized
currents and also their Sugawara operators possess
expressions as sums over bilinears in creation
and annihilation operators. Using some well-known facts on 
the CAR algebra and applying also some of our own results
on the decomposition of certain CAR representations
into irreducibles, we can show that the representations
of the even CAR algebra $\CKG^+$
(the subalgebra generated by
bilinear expressions $B(f)B(g)$, $f,g\in\KK$) 
reproduce precisely the sectors of the chiral algebra
in $\HNS$ and $\HR$. Thus we can identify the even
CAR algebra as observables, and a local net structure
comes from the natural embedding of the underlying
pre-quantized spaces $\LIN$ with $I\subset S^1$ open
intervals. Thereby we distinguish a point at infinity
$\zeta\in S^1$ and restrict to ``finite intervals''
i.e.\ those intervals such that $\zeta$ is not contained
in their closures. The WZW sectors then appear as
irreducible summands of the restrictions of $\PiNS$
and $\PiR$ to $\CKG^+$; the vacuum sector of the theory
corresponds to the basic module. Therefore the localized
endomorphisms that generate the relevant sectors can be
defined as endomorphisms of the (even) CAR algebra; we
can realize them explicitly in terms of Bogoliubov
transformations. Using again CAR mathematics, we can
prove that their composition realizes the WZW fusion rules.
A crucial point is that {\it local normality} holds,
i.e.\ we prove that the representations which represent
the sectors are quasi-equivalent in restriction to any
local algebra. This fact allows to extend the 
representations and endomorphisms to a net of local
von Neumann algebras which is obtained by weak
closure of local $C^*$-algebras in the vacuum sector.
It is shown that the net obeys Haag duality on the
punctured circle. Thus we are in the position that
the composition of our special localized endomorphisms
generalizes indeed to their unitary equivalence
classes, i.e.\ we have a well-defined DHR sector
product, and we prove rigorously that it
reproduces the WZW fusion rules.

We believe that the proof of the WZW fusion rules
in terms of the DHR sector product is remarkable
because it is completely independent of the methods
that are used in CFT to derive fusion rules (and
which do not all possess a mathematically satisfactory
status). Therefore we believe the goal of this analysis
to be two-fold: Firstly, a non-trivial model of CFT could
be incorporated into the DHR framework and expected
correspondences between CFT and AQFT could be 
established explicitly. Secondly, the conformal fusion
rules could be proven in a rather independent way.

\subsection{Results (Level 2)}
It is well-known that $\kv$-fold tensor products of
level $1$ modules contain level $\kv$ modules (with
an infinite multiplicity) and that all level $\kv$ modules
can be realized in this way. More precisely, the tensor
products of $\g_1$-modules decompose into tensor products 
of $\g_{\kv}$-modules and modules of the coset CFT
$ (\g_1)^{\oplus \kv} / \g_{\kv}$. We employ this idea to 
realize sectors of $\sonh$ WZW models at level $2$: In the 
tensor product space $\HNSh=\HNS\otimes\HNS$ (the ``big Fock 
space'') we have a canonical action of $\sonh$ at level $2$.
Hence all the level $2$ modules that appear in tensor products
of the level $1$  basic and vector modules are realized
in the big Fock space $\HNSh$. It is somewhat surprising
that, although we do not consider tensor products of the
level $1$ spinor modules, ``nearly all'' level $2$ WZW
sectors are realized in $\HNSh$; $\el+3$ of $\el+7$ if 
$N=2\el$ respectively $\el+2$ of $\el+4$ if $N=2\el+1$.
However, since they appear with infinite multiplicities it
would in general be a rather hopeless task to identify all
irreducible components within $\HNSh$. It is in fact the DHR
theory that yields a very useful order structure in the big
Fock space: On $\HNSh$ their acts a canonical fermion
field algebra $\fFg$ which can be built out of a net of local
field algebras. The fermionic expressions of the level
$2$ currents are invariant under the gauge group $\Oz$.
Hence the full chiral algebra of the WZW models is
gauge invariant. Moreover, the Virasoro generators of
the relevant coset CFT
  \[ \Cos = (\sonh_1\oplus\sonh_1)/\sonh_2  \]
living in $\HNSh$ are gauge invariant as well.
We introduce an algebra $\fAg$ as the gauge invariant
part of $\fFg$ and we can decompose the big Fock space
into sectors of the gauge invariant fermion algebra
{\`a} la DHR. By gauge invariance, neither the currents of 
$\son_2$ nor the coset Virasoro operators
can make transitions between these sectors. However,
the $\fAg$ sectors are not irreducible with respect to
the action of $\sonh_2$. Indeed we have proper 
inclusions\footnote{That the inclusion
$\AW \subset \fAg$ is proper is clear since at least
the coset Virasoro algebra which does not belong to the
WZW chiral algebra is gauge invariant as well. This is
in contrast to the situation
at level $1$ where the gauge ($\zet_2$) invariant
fermion algebra is identified to be the observable
algebra of the WZW model.}
  \[ \AW \subset \fAg \subset \fFg \,. \]
Here $\AW$ denotes the algebra of bounded operators which
corresponds to the chiral algebra of the WZW model: It can be
constructed from the action of local loop groups $\LIG$ with
$\Gr=\SON$ in $\HNSh$. The algebra $\AW$ contains the
bounded functions of locally smeared currents and the
irreducible spaces of the positive energy representations
are precisely the highest weight modules of the chiral algebra.

Fortunately, the central charge of the Virasoro algebra of the
coset CFT $(\sonh_1\oplus\sonh_1)/\sonh_2$ equals one, and the
$c=1$ CFTs are all classified. In our context, it easily turns 
out to be a so-called $c=1$ $\zet_2$ orbifold
which is well understood.
In particular, the characters of its irreducible modules
are given in the literature. However, the branching rules
corresponding to the embedding of $\sonh_2$ modules in
the tensor products of $\sonh_1$ basic and vector
modules were not completely known; they turn out
as a result of our analysis.
We present a large set of simultaneous highest 
weight vectors of $\sonh_2$ and the coset Virasoro 
algebra. Moreover, a detailed analysis of the
Virasoro specialized characters corresponding to the
sectors of $\fAg$ allows to puzzle out the complete
decomposition of $\HNSh$ into tensor products of
$\sonh_2$ highest weight modules and irreducible
modules of the coset Virasoro algebra. The solution of
this puzzle includes also an independent derivation of
the level $2$ characters which are, however, already
known in the literature. We believe that besides these
results the explicit construction of level $2$ modules
and modules of the coset CFT within tensor products of
level $1$ modules is of some particular interest because
all these structures become relatively transparent due
to the fermionic realization.

\subsection{Overview}
This dissertation is organized as follows. In Chapter 2
we present two rather different mathematical tools which
will be intensively employed in our analysis of the WZW
models. Firstly, we introduce Araki's selfdual CAR algebra
and present some well-known results on quasi-free states,
Fock representations and Bogoliubov transformations. We
have included some of our own results on the decomposition
of representations induced by Bogoliubov endomorphisms into
irreducibles, because these theorems will be applied in
Chapter 3. Secondly, we present several technical details of
the simple Lie algebra $\son$ and the associated affine
Lie algebra $\sonh$ since this will become relevant in the
following chapters, especially in Chapter 4. We also list
the unitary highest weight modules and fusion rules of the
$\son$ WZW models at level $1$ and $2$.

In Chapter 3 we analyze the $\son$ WZW models at level $1$.
We describe how the realizations of $\sonh_1$ arise via
second quantization from a natural action of $\sonh$ in the
pre-quantized space $\LSN$. Employing the representation
theory of even CAR we discuss in detail the treatment of
the WZW models in the algebraic framework: We introduce
a net of local $C^*$--algebras, construct localized
endomorphisms and extend them to a net of local von Neumann
algebras. Finally we prove the WZW fusion rules in terms
of the DHR sector product.

In Chapter 4 we analyze the $\son$ WZW models at level $2$.
We introduce the big Fock space and, acting on it, the fermion
algebra and its gauge invariant subalgebra. 
Employing the representation theory of the gauge
group $\Oz$ we decompose the big Fock space into the DHR
sectors of the gauge invariant fermion algebra. We present
a large set of simultaneous highest weight vectors of
$\sonh_2$ and the coset Virasoro algebra $\Vir\coset$ in
terms of fermionic creation operators acting on the vacuum.
A detailed analysis of the Virasoro specialized characters
ends up with the decomposition of the big Fock space into
tensor products of level $2$ highest weight modules and 
irreducible modules of the coset Virasoro algebra.

\chapter{Mathematical Preliminaries}
In this chapter we prepare some mathematical
background because it will be needed in the following
chapters. In order to keep these mathematical technicalities
out of the later discussion of the WZW models, we treat
two rather different mathematical fields here. Firstly,
we discuss the CAR algebra, quasi-free states and
Bogoliubov transformations in Araki's selfdual formulation
of CAR. Secondly, we treat the Lie algebras $\son$ and $\sonh$.

\section{The Selfdual CAR Algebra}
We follow essentially Araki's articles \cite{Ara1,Ara2}.
For a detailed discussion of the CAR algebra and the
connections between its different formulations
see also \cite{EK}.

\subsection{Basic Notations and Results}
Let $\cK$ be some infinite dimensional separable 
Hilbert space endowed with
an antiunitary involution $\GG$ (complex conjugation),
$\GG^2={\bf 1}$, which obeys
\[ \langle \GG f, \GG g \rangle = \langle g ,
f \rangle \,, \qquad f,g\in\cK \,. \]
The selfdual CAR algebra $\CKG$
is defined to be the $C^*$-norm closure of the algebra
which is generated by the range 
of a linear mapping $B:f\mapsto B(f)$,
such that
\[ \{ B(f)^*,B(g) \} = \langle f,g \rangle {\bf 1} \,,
\qquad B(f)^*=B(\GG f) \,, \qquad f,g\in\cK \,, \]
holds. The $C^*$-norm satisfies
\be
\|B(f)\| \le \|f\|, \qquad f \in \cK.
\labl{Cnorm}
The states of $\CKG$ we are
interested in are called quasi-free states. By definition,
a quasi-free state $\omega$ fulfills for $n\in \bbN$
\begin{eqnarray*}
\omega (B(f_1) \cdots B(f_{2n+1})) &=& 0, \\
\omega (B(f_1) \cdots B(f_{2n})) &=&
(-1)^{n(n-1)/2} \sum_\sigma
\mbox{\rm sign} \sigma\prod_{j=1}^n \omega
(B(f_{\sigma(j)})B(f_{\sigma(n+j)}))\,\,\,
\end{eqnarray*}
where the sum runs over all permutations
$\sigma \in \mathcal{S}_{2n}$ with the property
\[ \sigma (1) < \sigma (2) < \cdots < \sigma (n),
\qquad \sigma (j) < \sigma (j+n), \qquad j=1,2,...\,,n.\]
Clearly, quasi-free states are completely characterized by
their two point functions. Moreover, there is a
correspondence between the set of quasi-free
states and the set
\[ \QKG=\{S\in\mathfrak{B}(\cK)
\,|\, S=S^*,\, 0 \le S \le {\bf 1},\,
S + \overline{S} = {\bf 1} \}, \]
(we have used the notation $\overline{A}=\GG A\GG$ 
for operators $A$ on $\cK$) 
given by the formula
\be
\omega (B(f)^* B(g)) = \langle f,Sg \rangle .
\labl{phiS}
So it is convenient to denote the quasi-free state 
characterized by Eq.~(\ref{phiS}) by $\omega_S$.
The projections in $\QKG$
are called basis projections or polarizations. 
For a basis projection $P$, the state $\omega_P$ 
is pure and is called a Fock state. The corresponding 
GNS representation $(\cH_P,\pi_P,|\Omega_P\rangle)$
is irreducible, it is called a Fock representation.
The space $\cH_P$ can be canonically identified
with the antisymmetric Fock space 
$\mathcal{F}_-(P\cK)$. 

A projection $E\in\mathfrak{B}(\cK)$
with the property that $E\overline{E}=0$ 
is called a partial basis projection with
$\GG$-codimension 
${\rm dim}\,{\rm ker}(E+\overline{E})$.
Note that $E$ defines a Fock representation 
$(\cH_E,\pi_E,|\Omega_E\rangle)$ of
$\mathcal{C}((E+\overline{E})\cK,\GG)$.
Now let $E$ be a partial basis projection with
$\GG$-codimension 1, and choose a
$\GG$-invariant unit vector 
$e_0\in{\rm ker}(E+\overline{E})$.
Following Araki, pseudo Fock representations 
$\pi_{E,+}$ and $\pi_{E,-}$ of
$\CKG$ are defined in
$\cH_E$ by
\be
\pi_{E,\pm}(B(f)) = \pm \fsqz
\langle e_0,f \rangle Q_E(-{\bf 1}) +
\pi_E(B((E+\overline{E})f), \qquad f\in\cK,
\labl{piE}
where $Q_E(-{\bf 1})\in\mathfrak{B}(\cK)$ is the
unitary, self-adjoint implementer of the automorphism 
$\alpha_{-1}$ of $\CKG$ 
defined by $\alpha_{-1}(B(f))=-B(f)$ (which
restricts also to an automorphism of
$\mathcal{C}((E+\overline{E})\cK,\GG)$).
Pseudo Fock representations $\pi_{E,+}$ and 
$\pi_{E,-}$ are inequivalent and irreducible. 
Araki proved
\cite{Ara1}
\begin{lemma} 
Let $E$ be a partial basis projection with
$\GG$-codimension 1, and let $e_0\in\cK$
be a $\GG$-invariant unit vector of 
${\rm ker}(E+\overline{E})$. Define
$S\in\QKG$ by
\be
S = \half |e_0\rangle\langle e_0| + E.
\labl{mittel}
Then a GNS representation 
$(\cH_S,\pi_S,|\Omega_S\rangle)$ of the
quasi-free state $\omega_S$ is given by the 
direct sum of two inequivalent, irreducible 
pseudo Fock representations,
\be
(\cH_S,\pi_S,|\Omega_S\rangle) =
\left( \cH_E \oplus \cH_E,\pi_{E,+} 
\oplus \pi_{E,-}, \fsqz (|\Omega_E
\rangle \oplus |\Omega_E \rangle) \right) .
\ee
\lablth{pFock}
\end{lemma}

There is an important quasi-equivalence criterion 
for GNS representations of quasi-free states.  
Quasi-equivalence will be denoted by "$\approx$"
and unitary equivalence by "$\simeq$". Let us denote 
by $\JHS$ the ideal of Hilbert-Schmidt
operators in $\mathfrak{B}(\cK)$ and for 
$A\in\mathfrak{B}(\cK)$ by $[A]_2$ its 
Hilbert-Schmidt equivalence class
$[A]_2=A+\JHS$. 
Araki proved \cite{Ara1,Ara2}
\begin{theorem}
For quasi-free states $\omega_{S_1}$ and
$\omega_{S_2}$ of $\CKG$
we have quasi\-equivalence $\pi_{S_1}\approx\pi_{S_2}$
if and only if 
$[S_1^\h]_2=[S_2^\h]_2$.
\lablth{Araki}
\end{theorem}

Next we define the set
\[ \IKG = \{ V\in \mathfrak{B}(\cK)
\,|\, V^*V={\bf 1},\,\, V=\overline{V} \} \]
of Bogoliubov operators. Bogoliubov operators 
$V\in \IKG$ induce unital
$\ast$-endomorphisms $\varrho_V$ of 
$\CKG$, defined by their action 
on the canonical generators,
\[ \varrho_V(B(f))=B(Vf). \]
Moreover, if $V\in \IKG$ is surjective
(i.e.~unitary), then $\varrho_V$ is an automorphism.
The group of unitary Bogoliubov operators is denoted by $\IoKG$.
A Bogoliubov automorphism $\varrho_U$ is called implementable
in a Fock representation $\pi_P$ if there exists a unitary
$Q_P(U)\in\mathfrak{B}(\mathcal{H}_P)$ such that
  \be \pi_P\circ\varrho_U(x)=Q_P(U)\pi_P(x)Q_P(U)^*,
  \qquad x\in\CKG . \ee
For a given basis projection $P$, let $\IPoKG$ be the set of unitary
Bogoliubov operators which induce automorphisms being implementable
in $\pi_P$,
  \be \IPoKG = \{ U \in \IoKG \,\, | \,\, PU\bP \in \JHS \}. \ee
$\IPoKG$ is a topological group with respect to the $P$-norm topology,
given by the distance $\|U-U'\|+\|P(U-U')\bP\|_2$. The $P$-strong
topology is defined by the seminorms $\|(U-U')f\|+\|P(U-U')\bP\|_2$,
$f\in\cK$. Consider the (real) Lie algebra
  \be \iPoKG = \{ H \in \BK \,\,|\,\, H^*=-H,\,\,H=\overline{H},\,\,
  PH\bP \in \JHS \}. \ee
The following is well known (see e.g.~\cite{Ara2,CR})
\begin{theorem}
(1) $U_t$ is a continuous one-parameter subgroup of $\IPoKG$ relative
to the $P$-norm topology if and only if 
$U_t=\E^{tH}$ for some $H\in\iPoKG$.

(2) For $H\in\iPoKG$ there is a unique skew self-adjoint $\dQP (H)$ on
$\cH_P$ such that the unitary $\E^{t\dQP (H)}$ implements
$\E^{tH}$, $\OmP$ is in its domain and
$\langle \Omega_P | \dQP (H) \OmP = 0$. The map $H\mapsto\dQP (H)$ is
real linear.

(3) Any vector in the (finite) linear span of finite particle vectors
is analytic for $-\I\,\dQP (H)$ for any $H\in\iPoKG$ and we have
  \be [ \dQP (H_1), \dQP (H_2) ] = \dQP ([H_1,H_2]) +
  c_P(H_1,H_2)\,\bfe , \ee
where
  \be c_P(H_1,H_2) = \half \tr 
  (PH_2\bP H_1P - PH_1\bP H_2P). \labl{cP}

(4) For $H\in\iPoKG$, $\E^{t\dQP(H)}$ is strongly continuous
with respect to the $P$-strong topology.
\lablth{bounded}
\end{theorem}
The scalar term (\ref{cP}) which appears here is called the
``Schwinger term''. There is also a generalization to the case 
that $H$ is unbounded \cite{Ara2}.
\begin{theorem}
(1) $U_t$ is a continuous one-parameter subgroup of $\IPoKG$ relative
to the $P$-strong topology if $U_t=\E^{tH}$ for a skew self-adjoint
operator $H$ such that $H=\overline{H}$, $PHP$ skew self-adjoint and
the closure of $\bP HP$ is Hilbert-Schmidt class.

(2) For such an $H$, there is a unique skew self-adjoint $\dQP (H)$ on
$\cH_P$ such that the unitary $\E^{t\dQP (H)}$ implements
$\E^{tH}$, $\OmP$ is in its domain and
$\langle \Omega_P | \dQP (H) \OmP = 0$.
\lablth{unbounded}
\end{theorem}
We conclude that $\iPoKG$ is just the ``bounded part''
of the Lie algebra corresponding to the group $\IPoKG$.

\subsection{Decomposition of Representations}
We now investigate the decomposition of representations
$\pi_P\circ\varrho_V$ into irreducible subrepresentations;
here $\pi_P$ is a Fock representation and $\varrho_V$ a
Bogoliubov endomorphism. The discussion is based on
\cite{Decom,awzw}, compare also \cite{Binnenneu}.

A quasi-free state, composed with a
Bogoliubov endomorphism is again a quasi-free state,
namely we have $\omega_S\circ\varrho_V=\omega_{V^*SV}$. 
In the following we are interested in representations
of the form $\pi_P\circ\varrho_V$ instead of GNS
representations $\pi_{V^*PV}$ of states
$\omega_{V^*PV}=\omega_P\circ\varrho_V$. 
Indeed, the former are multiples of the latter,
in particular, we have \cite{Binnenneu,Rideau}
\be
\pi_P\circ\varrho_V\simeq 2^{N_V}\pi_{V^*PV},
\qquad N_V={\rm dim}({\rm ker}V^*\cap P\cK).
\labl{cyclics}
Thus the identification of the Hilbert-Schmidt
equivalence class $[(V^*PV)^\h]_2$ is
the identification of the quasi-equivalence class
of $\pi_P\circ\varrho_V$. For the identification
of the unitary equivalence class, we need a
decomposition of $\pi_P\circ\varrho_V$ into 
irreducible subrepresentations which will now be
elaborated.
We will see that only Fock and pseudo Fock 
representations appear in the decomposition of
representations $\pi_P\circ\varrho_V$ if the
Bogoliubov operator has finite corank.
\begin{theorem}
Let $P$ be a basis projection and
let $V$ be a Bogoliubov operator with
$M_V=\mbox{\rm dim ker}V^* < \infty$. If $M_V$ is
an even integer we have (with notations as above)
\be
\pi_P\circ\varrho_V \simeq 2^{M_V/2} \pi_{P'}
\labl{decomeven}
where $\pi_{P'}$ is an (irreducible) Fock representation.
If $M_V$ is odd then we have
\be
\pi_P\circ\varrho_V \simeq 2^{(M_V-1)/2}
(\pi_{E,+} \oplus \pi_{E,-})
\labl{decomodd}
where $\pi_{E,+}$ and $\pi_{E,-}$ are inequivalent
(irreducible) pseudo Fock representations.
\lablth{evenodd}
\end{theorem}
\bproof
We present a brief version of the proofs in
\cite{Decom} here. First recall the following
well-known fact.
If the test function space possesses an orthogonal
decomposition $\cK=\cK_1\oplus\cK_2$
into $\GG$-invariant subspaces then we can regard
$\mathcal{C}(\cK_j,\GG_j)$ with
$\GG_j=\GG|_{\cK_j}$, $j=1,2$, as
subalgebras of $\CKG$,
and $\CKG$ is canonically
isomorphic to the $\mathbb{Z}_2$-graded tensor
product of $\mathcal{C}(\cK_1,\GG_1)$ with
$\mathcal{C}(\cK_2,\GG_2)$ by the identification
$x_1\otimes x_2$ with $x_1x_2$,
$x_j\in\mathcal{C}(\cK_j,\GG_j)$, $j=1,2$.
Let $S_j\in\mathcal{Q}(\cK_j,\GG_j)$, $j=1,2$.
If $S\in\QKG$ is of the form
$S=S_1\oplus S_2$ then $\omega_S$ is a product state,
$\omega_S=\omega_{S_1}\otimes\omega_{S_2}$ which means that
$\omega_S(x_1x_2)=\omega_{S_1}(x_1)\omega_{S_2}(x_2)$
whenever $x_j\in\mathcal{C}(\cK_j,\GG_j)$, 
$j=1,2$, \cite{Manu,Pow}. 
In our application, $S_1=P_1$ is a basis
projection, and we assume $\cK_2$ to be
two-dimensional here. Then choose an
orthonormal base (ONB)
$\{f_+,f_-\}$ of $\cK_2$ with the property 
$\GG_2 f_+=f_-$. One checks easily that 
$S_2\in\mathcal{Q}(\cK_2,\GG_2)$ implies that
it is of the form
\[ S_2=\left( \begin{array}{cc} \lambda_+ & 0 \\
0 & \lambda_- \end{array} \right), \qquad
0 \le \lambda_\pm \le 1, \qquad \lambda_+ +
\lambda_- =1, \]
with respect to the decomposition
$\cK_2=\mathbb{C}f_+\oplus\mathbb{C}f_-$.
Using the only non-trivial evaluation of
$\omega_{S_2}$ on $B(f_+)B(f_-)$ it is obvious
that
\[ \omega_{S_2} = \lambda_+ \omega_{F_+} + 
\lambda_- \omega_{F_-} \]
with Fock states $\omega_{F_\pm}$ corresponding
to basis projections 
$F_\pm=|f_\pm\rangle\langle f_\pm|$. Hence, if
both $\lambda_+$ and $\lambda_-$ are non-zero,
then $\omega_S$ is a mixture of two Fock states
of $\CKG$,
\[ \omega_S = \lambda_+ \omega_{P_+} + 
\lambda_-  \omega_{P_-},
\qquad P_\pm = P_1 \oplus F_\pm. \]
Since $\omega_{P_+}\neq\omega_{P_-}$ these states
are orthogonal, and hence $\pi_S$ is the direct sum
of two equivalent Fock representations.
Now let $P$ be a basis projection of $\cK$ and 
let $V\in\IKG$ be a Bogoliubov
operator with $M_V={\rm dim ker}V^*=2$. Choose 
an ONB $\{q_+,q_-\}$ of ${\rm ker}V^*$ such that
$\GG q_+=q_-$. Then $\langle q_+,Pq_- \rangle=0$.
Let $\lambda_\pm=\langle q_\pm,Pq_\pm\rangle$, hence
$\lambda_+ + \lambda_-=1$, and define  
$Q_\pm=|q_\pm\rangle\langle q_\pm|$. 
Note that $VV^*={\bf 1}-Q_+-Q_-$.
Suppose $\lambda_\pm\neq 0$. Then
\[ P_1=V^*P({\bf 1}-\lambda_+^{-1}Q_+ -
\lambda_-^{-1}Q_-)PV \]
is a partial basis projection with 
$\GG$-codimension 2. Define one-dimensional
projections
\[ F_\pm = \lambda_+^{-1} \lambda_-^{-1}
V^*PQ_\mp PV, \]
$F_+=\overline{F_-}$, $P_1F_\pm=F_+F_-=0$, and set
\[ S_2 = \lambda_+ F_+ + \lambda_- F_-. \]
We identify $\cK_1={\rm ran}(P_1+\overline{P_1})$
and $\cK_2={\rm ran}(F_++F_-)$ and check that
$\cK=\cK_1\oplus\cK_2$. Hence
\[ S=V^*PV=P_1+S_2 \]
is as discussed above. Because $\lambda_\pm\neq 0$
we have $q_\pm\notin P\cK$, hence $N_V=0$
and $\pi_P\circ\varrho_V\simeq\pi_{V^*PV}$ by
Eq.~(\ref{cyclics}). So it follows 
$\pi_P\circ\varrho_V\simeq 2\pi_{P'}$ with
some Fock representation $\pi_{P'}$. 
On the other hand, if 
$\lambda_-=0$ ($\lambda_+=0$) then $q_+\in P\cK$ 
($q_-\in P\cK$) and hence
$N_V=1$. It follows $\pi_P\circ\varrho_V
\simeq 2\pi_{V^*PV}$ by Eq.~(\ref{cyclics}).
But $q_+\in P\cK$ ($q_-\in P\cK$) 
implies that $V^*PV$ is a
projection and hence $\pi_{V^*PV}$ is a Fock 
representation. We conclude that $\pi_P\circ\varrho_V
\simeq 2\pi_{P'}$ holds generally if $M_V=2$. Let now
$M_V=2N$, $N\in\mathbb{N}$. Then choose a $\GG$-invariant 
ONB $\{v_n,n\in\mathbb{N}\}$ of $\cK$, 
i.e.~$v_n=\GG v_n$, $n\in\mathbb{N}$. 
Further we choose a $\GG$-invariant ONB 
$\{w_n,n=1,2,...\,,2N\}$ of ker$V^*$, and
we define $w_{2N+n}= V v_n$ for $n\in \mathbb{N}$.
Since $\cK={\rm ran}V\oplus{\rm ker}V^*$
the set $\{w_n,n\in\mathbb{N}\}$
forms another $\GG$-invariant ONB of $\cK$ and
we can write
\[ V=\sum_{n=1}^\infty |w_{2N+n}\rangle\langle v_n|. \]
We introduce Bogoliubov operators 
$V_0,V_2\in\IKG$,
\[ V_0 = \sum_{n=1}^\infty |w_n\rangle\langle v_n|,\qquad
V_2 = \sum_{n=1}^\infty |v_{n+2}\rangle\langle v_n|, \]
such that $M_{V_j}=j$, $j=0,2$, and $V=V_0 V_2^N$.
Since $V_0$ is unitary, $P_0=V_0^*PV_0$ is again
a basis projection, and now we can conclude iteratively
\[ \pi_P\circ\varrho_V=\pi_{P_0}\circ\varrho_{V_2}^N
\simeq 2^N \pi_{P'} \]
with some Fock representation $\pi_{P'}$,
and this is Eq.~(\ref{decomeven}). Now assume 
$V\in\IKG$ has $M_V=1$. Since
${\rm ker}V^*$ is $\GG$-invariant we have $N_V=0$
and hence $\pi_P\circ\varrho_V\simeq\pi_{V^*PV}$ by
Eq.~(\ref{cyclics}). Moreover, 
$VV^*={\bf 1}-Q_0$ where $Q_0$ is a one-dimensional
$\GG$-invariant projection. Note that
$\langle q_0,Pq_0\rangle=\frac{1}{2}$ for each
$\GG$-invariant unit vector $q_0$. Using
$Q_0PQ_0=\frac{1}{2}Q_0$ one finds easily
\[ S=V^*PV= \half E_0+E \]
where $E=2S^2-S$ is a partial basis projection with
$\GG$-codimension 1, and $E_0=4(S-S^2)=4V^*PQ_0PV$ 
is the one-dimensional $\GG$-invariant projection
on the vector $V^*Pq_0$, $EE_0=\overline{E}E_0=0$,
thus $S=V^*PV$ is of the form (\ref{mittel}) and hence
\[ \pi_P\circ\varrho_V \simeq \pi_{E,+}
\oplus \pi_{E,-} \]
with pseudo Fock representations $\pi_{E,\pm}$.
Now let $V\in\IKG$ be a
Bogoliubov operator with $M_V=2N+1$,
$N\in\mathbb{N}$. Recall our $\GG$-invariant ONB
$\{v_n,n\in\mathbb{N}\}$ of $\cK$ and choose a
$\GG$-invariant ONB $\{w_n,n=0,1,...\,,2N\}$
of ker$V^*$. We define $w_{2N+n}=V v_n$ for 
$n\in\mathbb{N}$. Then 
$\{w_n,n\in\mathbb{N}_0\}$ is an ONB of $\cK$,
too, and we can write
\[ V=\sum_{n=1}^\infty |w_{2N+n}\rangle\langle v_n|. \]
We introduce Bogoliubov operators
$V_1,V_2\in\IKG$, 
\[ V_1 = \sum_{n=1}^\infty  |w_n\rangle\langle v_n|,
\qquad V_2 = \sum_{n=0}^\infty |w_{n+2}\rangle
\langle w_n|, \]
such that $M_{V_j}=j$, $j=1,2$, and $V=V_2^N V_1$.
It follows that
\[ \pi_P\circ\varrho_V 
\simeq 2^N \pi_{P'}\circ\varrho_{V_1} \]
with $\pi_{P'}$ some Fock representation. Since
$M_{V_1}=1$ we conclude
\[ \pi_P\circ\varrho_V \simeq 2^N (\pi_{E,+}\oplus
\pi_{E,-}) \]
with pseudo Fock representations $\pi_{E,\pm}$, and
this is Eq.~(\ref{decomodd}). \eproof

We define the even algebra 
$\CKG^+$
to be the subalgebra of $\alpha_{-1}$-fixpoints,
\[ \CKG^+ = \{ x \in \CKG \,\,|\,\, \alpha_{-1}(x)=x \}. \]
We now are interested in what happens when our
representations of $\CKG$ 
are restricted to the even algebra. For basis
projections $P_1,P_2$, with $[P_1]_2=[P_2]_2$,
Araki and D.E.~Evans \cite{AE} defined an index,
taking values $\pm 1$,
\[ \mbox{ind}(P_1,P_2)= (-1)^{{\rm dim}(
P_1\cK\cap({\bf 1}-P_2)\cK)}. \]
The automorphism $\alpha_{-1}$ leaves any
quasi-free state $\omega_S$ invariant. Hence 
$\alpha_{-1}$ is implemented in $\pi_S$. In particular,
in a Fock representation $\pi_P$, $\alpha_{-1}$ 
extends to an automorphism $\bar{\alpha}_{-1}$ of
$\pi_P(\CKG)''=
\mathfrak{B}(\cH_P)$. The following proposition
is taken from \cite{Ara2}.
\begin{proposition}
Let $P$ be a basis projection and let $U\in\IPoKG$. 
Denote by $Q_P(U)\in\mathfrak{B}(\cH_P)$ the
unitary which implements $\varrho_U$ in $\pi_P$. Then
\be
\bar{\alpha}_{-1}(Q_P(U)) = \sigma(U) Q_P(U),\qquad
\sigma(U)=\pm 1.
\ee
In particular, $\sigma(U)=\mbox{\rm ind}(P,U^*PU)$.
Moreover, given two unitary operators
$U_1,U_2\in\IPoKG$ of this type,
$\sigma$ is multiplicative, 
$\sigma(U_1U_2)=\sigma(U_1)\sigma(U_2)$.
\lablth{ind}
\end{proposition}
Furthermore, one has \cite{AE,Ara2}
\begin{theorem}
Restricted to the even algebra 
$\CKG^+$, a Fock representation
$\pi_P$ splits into two mutually inequivalent, irreducible
subrepresentations,
\be
\pi_P|_{\CKG^+}=\pi_P^+\oplus
\pi_P^-,
\ee
and the commutant is generated by
$Q_P(-{\bf 1})$.
Given two basis projections $P_1,P_2$, then 
$\pi_{P_1}^\pm \simeq \pi_{P_2}^\pm$
if and only if $[P_1]_2=[P_2]_2$ and
{\rm ind}$(P_1,P_2)=+1$, and
$\pi_{P_1}^\pm \simeq \pi_{P_2}^\mp$
if and only if $[P_1]_2=[P_2]_2$ and
{\rm ind}$(P_1,P_2)=-1$.
\lablth{resteven}
\end{theorem}
For some real $v\in\cK$, i.e.~$\GG v=v$, and
$\|v\|=1$ define $U\in\IoKG$ by
\be
U = 2|v\rangle\langle v|-{\bf 1}. 
\labl{Uvec}
Then $\varrho_U$ is implemented in each Fock
representation $\pi_P$ by the unitary self-adjoint
$Q_P(U)=\sqrt{2}\pi_P(B(v))$, since $\varrho_U$ is
implemented in $\CKG$ by
$q(U)=\sqrt{2}B(v)$,
\begin{eqnarray*}
q(U)B(f)q(U)
&=& 2B(v)B(f)B(v) \\
&=& 2\{B(v),B(f)\}B(v) - 2B(f)B(v)B(v) \\
&=& 2\langle v,f \rangle B(v)  - B(f) \\
&=& B( 2\langle v, f \rangle v - f)  \\
&=& B(Uf).
\end{eqnarray*}
Hence $\sigma(U)=-1$ and we immediately have the
following
\begin{corollary}
Let $U\in\IKG$ be as in 
Eq.~(\ref{Uvec}). Then, in restriction to
$\CKG^+$, we have for each Fock
representation $\pi_P$ equivalence
$\pi_P^\pm \circ \varrho_U \simeq \pi_P^\mp$.
\lablth{Bogvec}
\end{corollary}
Now let us consider the restrictions of pseudo
Fock representations.
\begin{lemma}
The pseudo Fock representations $\pi_{E,+}$ and 
$\pi_{E,-}$ of Eq.~(\ref{piE}), when restricted to the 
even algebra $\CKG^+$, remain
irreducible and become equivalent.
\lablth{restpseudo}
\end{lemma} 
\bproof 
Without loss of generality, we prove that
$\pi_{E,+}$, when restricted to
$\CKG^+$, remains irreducible.
Let $T\in\pi_{E,+}(\CKG^+)'$.
Then, in particular, 
$T\in\pi_E(\mathcal{C}((E+\overline{E})\cK,\GG)^+)'$,
hence $T=\lambda {\bf 1} + \mu Q_E(-{\bf 1})$,
$\lambda,\mu\in\mathbb{C}$. 
Now choose a non-zero
$f\in(E+\overline{E})\cK$. Then
\[ \pi_{E,+}(B(e_0)B(f)) = \fsqz 
Q_E(-{\bf 1}) \pi_E(B(f)), \]
so we compute
\[ [ T,\pi_{E,+}(B(e_0)B(f)) ] =
\sqrt{2} \mu \pi_E(B(f)).\]
This implies $\mu=0$, $T=\lambda{\bf 1}$,
proving irreducibility. It
remains to be shown that $\pi_{E,+}$ and
$\pi_{E,-}$, when restricted to
$\CKG^+$,
become equivalent. Now choose arbitrary
$f,g\in\cK$. It is not hard to check that
\[ \pi_{E,+}(B(f)B(g))=Q_E(-{\bf 1})
\pi_{E,-}(B(f)B(g))Q_E(-{\bf 1}). \]
Since $\CKG^+$ is generated
by such elements $B(f)B(g)$ the unitary
$Q_E(-{\bf 1})$ realizes the equivalence of the
restrictions of $\pi_{E,\pm}$. \eproof

Summarizing we obtain
\begin{theorem}
With notations of Theorem \ref{evenodd}, a representation
$\pi_P\circ\varrho_V$ restricts as follows to the even
algebra $\CKG^+$: If $M_V$ is even
we have
\be
\pi_P\circ\varrho_V|_{\CKG^+}
\simeq 2^{M_V/2}(\pi_{P'}^+\oplus\pi_{P'}^-)
\ee
with $\pi_{P'}^+$ and $\pi_{P'}^-$ mutually
inequivalent and irreducible. If $M_V$ is odd, then
\be
\pi_P\circ\varrho_V|_{\CKG^+}
\simeq 2^{(M_V+1)/2} \, \pi
\ee
with $\pi$ irreducible.
\lablth{restevenodd}
\end{theorem}

\newpage
%%%%%%%%%%%%%%%%%%%%%%%%%%%%%%%%%%%%%%%%%%%%%%%%%%%%%%%%%%%%%%%%%%%%%%%%%

\section{On $\son$ and $\sonh$}
We now turn to the mathematics of Lie algebras. In view
of applications in the following chapters we introduce some
notation and present some technical details of
the simple \lie\ $\son$ and the associated affine \lie\ 
$\sonh$ because we believe that it might be somewhat
laborious for the reader to collect these facts from
the literature (e.g.\ \cite{Kac,Mick}).

\subsection{The Simple Lie Algebra $\son$}
Let $E^{i,j}$ be the $(N\times N)$-matrix with entries 
$(E^{i,j})_{k,l}=\del ik\del jl$. 
Define
  \[  \T ij =\I\,(E^{i,j}-E^{j,i})  \]
for $i,j=\onetoN$. Elements $\T ij$, $1\le i<j\le N$, provide a basis
of $\son$, and we have
  \[  [\T ij,\T kl]=\I\,( \del jk \T il 
  + \del il \T jk - \del jl \T ik -   \del ik \T jl ) . \] 
Let $\ell=3,4,\ldots$ denote the rank of $\son$, i.e.~$N=2\ell$ 
and $N=2\ell+1$ for
even and odd $N$, respectively. Define
  \[  \te ij\eps\eta = \half(\eps\T{2i}{2j-1}
  +\eta\T{2i-1}{2j})   + \halfi
  (\T{2i-1}{2j-1}-\eps\eta\T{2i}{2j}) \]
for $i,j=\onetol$ and $\eps,\eta=\pm1$ and
  \[  \tee j\eps = -\fsqz (\eps\T{2j-1}{2\ell+1} 
  - \I\T{2j}{2\ell+1}) \] 
for $j=\onetol$ and $\eps=\pm1$. Further define
  \[ \bearll
  H^j = \T{2j-1}{2j} & {\rm for}\ j=\onetol \,, \\
  \EO j\pm = \pm \te j{j+1}\pm\mp & {\rm for}\ j=\onetolme \,, 
  \eear \]
and
  \[  \EO\ell\pm = \left\{ \bearll
  \pm \te {\ell-1}\ell\pm\pm  & {\rm for}\ N=2\ell \,, \\
  \pm \tee \ell\pm & {\rm for}\ N=2\ell+1
  \,. \eear\right. \]
These matrices obey the commutation relations
  \[ \bearll
  [H^j,H^k]=0 \,, \\[.3em]
  [H^j,\EO k\pm]=\pm\,(\alpha^{(k)})^j_{}\,\EO k\pm \,, \\[.3em]
  [\EO j+,\EO k-]= \del jk\, H^j \,,\eear \]
for $j,k=\onetol$, with
  \[  (\alpha^{(k)})^j_{}= \del jk-\del j{k+1} 
  \hsp{.7}  \] 
for $k=\onetolme$ and
  \[  \hsp{1.2}  (\alpha^{(\ell)})^j_{}= \left\{ \bearll
  \del j{\ell-1}+\del j\ell & {\rm for}\ N=2\ell \,,  \\
  \del j\ell & {\rm for}\ N=2\ell+1 \,. \eear\right. \] 
Moreover, they also obey the Serre relations of $\son$. 
Hence they constitute a Cartan-Weyl basis of $\son$, and the 
$\alpha^{(k)}$ are the simple roots of $\son$.
The elements corresponding to positive roots are $\te ij+-$ 
and $\te ij++$ with $1\le i<j\le\ell$, additionally $\tee j+$, 
$j=\onetol$, if $N=2\ell+1$; the one corresponding to the 
highest root $\theta$ is $E_\theta=\te 12++$.

Also note that the invariant bilinear form on $\son$\ is 
  \be  ( \T ij | \T kl ) = \half {\rm tr} \, ( \T ij \T kl )
  = \del ik \del jl -   \del il \del jk \,. \labl{ibf}
In particular, we have
  \[  (H^i|H^j) = \del ij = (\EO i+|\EO j-)  \,, \qquad
  (\EO i\pm|\EO j\pm) = 0  \,. \]
The fundamental weights $\Lamf j$, $j=\onetol$,
of $\son$ are defined by
  \[ (\alpha^{(j)},\Lamf k ) = \left\{ \bearll
  \frac12 \del k\ell \quad & \mbox{for} \,\,\, j=\ell,\,\,
  N=2\ell+1,\\ \del kj & \mbox{else,} \eear\right. \]
and its components (in the orthogonal base) are
  \[ \Lamf j = (\underbrace{1,1,\ldots,1}_{j\;\mathrm{times}}
  ,0,0,\ldots,0) \]
for $j=\onetolmz$ and also for $j=\ell-1$ if $N=2\ell+1$,
as well as
  \[ \Lamf {\ell-1} = \half (1,1,\ldots,1,1,-1) \]
for $N=2\ell$, and
  \[ \Lamf \ell = \half (1,1,\ldots,1,1,1) \,. \]

\subsection{The Affine Lie Algebra $\sonh$}
The infinite-dimensional Lie algebra $\sonh$ is, by definition,
the algebra generated by elements $\J ijm$, $i,j=\onetoN$ and 
a central element $K$ satisfying relations
  \be [ \J ijm , \J kln ] = \I  (\del jk \, \J il{m+n}
  + \del il \, \J jk{m+n} - \del jl \, \J ik{m+n} -
  \del ik \, \J jl{m+n}) + 
  m \, \del m{-n} (\T ij | \T ji) K, \labl{JmJnK}
and $[\J ijm , K ] = 0$. The elements $\J ijm$,
$1\le i<j\le N$, and $K$ provide a basis of $\sonh$ as a
vector space. Introducing one further element $D$ (``derivation'') obeying
  \[ [D, \J ijm ] = m \, \J ijm , \qquad
  [D,K]=0, \]
one obtains the full affine Lie algebra $D^{(1)}_\ell$ if 
$N=2\ell$ respectively $B^{(1)}_\ell$ if $N=2\ell+1$.
Note that via identification $J_m(\T ij)\equiv\J ijm$
and defining $J_m(T)$ linear in $T\in\son$, $m\in\zet$,
the relations (\ref{JmJnK}) can also be written as follows,
  \[ [J_m(T),J_n(T')] = J_{m+n}([T,T']) + m \,
  \del m{-n} (T|T') \, K, \]
and $[ J_m(T),K]=0$. Also the affine \lie\ $\sonh$ 
possesses a  Chevalley basis; the \csa\ generators are
  \[  \HH j= \Jo{2j-1}{2j}  \]
for $j=\onetol$, and the Chevalley generators are given by
  \[  \EE j\pm=\pm\Jot j{j+1}\pm\mp \qquad {\rm for}\ j=\onetolme\,,\]
and
  \[  \EE\el\pm = \left\{ \bearll \pm\Jot{\el-1}\el\pm\pm & \forzl \,,
  \nline7 \pm J_0(\ttpm \el) & \forzle\,, \eear\right.  \]
further
  \[ \EE0\pm= \pm\Jt{\pm1}12\mp\mp \,. \]
for $i,j=\onetol$ and $\eps,\eta=\pm1$.

The unitary integrable highest weight modules of $\sonh$ at
level $1$ are listed in table \ref{T1}. There $\Lambda$ 
denotes the \hw\ \wrtt horizontal subalgebra $\son$, 
$\Delta$ the conformal weight, and \qdi\ the 
quantum dimension.
In the first column we provide a ``name'' for the 
associated primary field of the
relevant \wzwt; in the following chapters we will use these names 
as labels for the \ihwm s, i.e.\
write $\cH_\Lambda=\cH_\circ$ for $\Lambda=0$ etc.,
and analogously for other quantities such as
characters. (We find it convenient to use identical names for some of the 
fields at level one and at level two; when required to avoid ambiguities in 
the notation, we will always also specify the level.)

\begin{table}[bpth]\caption{Unitary \hwm s of $\sonh$  at level 1 
for $N=2\el$ (left) and for $N=2\el+1$ (right).} \label{T1}
\begin{center}
  \begin{tabular}{|c|c|c|c|} \hline &&&\\[-.9em]
  field & $\Lambda$   & $\Delta$      & \qdi \\ \hline\hline &&&\\[-.9em]
  $\rmo$&   0         & 0             & 1    \\ &&&\\[-.9em]
  $\rmv$& $\Lj1$      & $\Frac12$     & 1    
  \\ &&&\\[-.8em] \hline &&&\\[-.9em]
  $\rms$& $\Lj{\el-1}$& $\Frac N{16}$ & 1    \\ &&&\\[-.8em]
  $\rmc$& $\Lj\el$    & $\Frac N{16}$ & 1    
  \\[-.8em]&&&\\ \hline \end{tabular}
  \hsp5
  \begin{tabular}{|c|c|c|c|} \hline &&&\\[-.9em]
  field & $\Lambda$   & $\Delta$      & \qdi \\ \hline\hline &&&\\[-.9em]
  $\rmo$&   0         & 0             & 1    \\ &&&\\[-.9em]
  $\rmv$& $\Lj1$      & $\Frac12$     & 1    
  \\ &&&\\[-.8em] \hline &&&\\[-.9em]
  $\sigma$ & $\Lj\el$    & $\Frac N{16}$ & $\sqrt2$
  \\[-.8em]&&&\\ \hline \multicolumn4c {} \\[.5em] \end{tabular}
\end{center} \end{table}
  
In the tables we have separated the modules by a horizontal line into two 
classes. In the fermionic description, the modules in the first part are in 
the \NS sector, while those in the second part are in the Ramond sector.

Let us now turn to the fusion rules of the WZW model
based on $\sonhe$. Since the basic module ($\circ$) always 
represents the unit of the fusion ring it is denoted by $1$
in this context.

If $N=2\ell$ the fusion rules\footnote{The fusion rules which are
not listed explicitly all follow from the commutativity 
and the associativity of the fusion product and from the 
fact that $1$ is the unit of the fusion ring.} read
  \be  \bearl 
  \rmv \times \rmv = 1 \,, \qquad\quad \rmv \times \rms = \rmc \,, \nline7
  \rms \times \rms = \rmc \times \rmc = \left\{ \bearll
       1  & {\rm for}\ \el\iN2\zet\,, \\[.2em] 
       \rmv & {\rm for}\ \el\iN2\zet+1\,, \eear\right. \nline7
  \rms \times \rmc = \left\{ \bearll
       \rmv & {\rm for}\ \el\iN2\zet\,, \\[.2em] 
       1  & {\rm for}\ \el\iN2\zet+1\,. \eear\right. 
  \eear \labl{toteven}
All sectors are simple i.e.\ have unit quantum dimension.

If $N=2\ell+1$ the fusion rules read
  \be \rmv\times\rmv=1, \qquad \sigma\times
  \rmv=\sigma, \qquad \sigma\times\sigma
  =1+\rmv. \labl{odd}
Only the $\sigma$ sector is not simple; indeed we have
$\mathcal{D}_\sigma=\sqrt{2}$. 

We are going to describe the situation at level $2$ now.
If $N=2\ell$ ($N=2\el+1$) we have $\el+7$ ($\el+4$) integrable
highest weight modules which are listed in the following 
tables \ref{t1}, \ref{t2}. Again we have separated the
modules by a horizontal line into two classes. In the
fermionic description, the modules in the first part
appear in the ``doubled'' Neveu-Schwarz sector
$\HNS\otimes\HNS$ while those in the second part
involve the Ramond sector.
\begin{table}[ptbh]\caption{Unitary \hwm s of $\sonh$  at level 2 
for $N=2\el$.} \label{t1}
\begin{center}
  \begin{tabular}{|c|c|c|c|} \hline &&&\\[-.9em]
  field & $\Lambda$    & $\Delta$     & \qdi \\ \hline\hline &&&\\[-.9em]
  $\rmo$&   0          & 0            & 1    \\ &&&\\[-.9em]
  $\rmv$& $2\Lj1$      & 1            & 1    \\ &&&\\[-.8em]
  $\rms$& $2\Lj{\el-1}$& $\Frac N8$   & 1    \\ &&&\\[-.8em] 
  $\rmc$& $2\Lj\el$    & $\Frac N8$   & 1    
  \\ &&&\\[-.8em]%\hline &&&\\[-.9em]
  $\pfj j$   & $\left\{ \bearl \Lj j\ \;{\rm for}\ j=\onetolmz, \\[.1em]
          \Lj{\el-1}+\Lj\el\ {\rm for}\ j=\el-1 \eear\right.$
             & $\Frac{j(N-j)}{2N}$ & 2  \\ &&&\\[-.8em] \hline &&&\\[-.9em]
  $\sigma$ & $\Lj{\el-1}$ & $\Frac{N-1}{16}$ & $\sqrt\el$   \\ &&&\\[-.8em]
  $\tau$   & $\Lj\el$     & $\Frac{N-1}{16}$ & $\sqrt\el$   \\ &&&\\[-.8em]
  $\sigma'$& $\Lj1+\Lj{\el-1}$ & $\Frac{N+7}{16}$ & $\sqrt\el$ 
  \\ &&&\\[-.8em]
  $\tau'$  & $\Lj1+\Lj\el$     & $\Frac{N+7}{16}$ & $\sqrt\el$ 
  \\[-.8em]&&&\\ \hline \end{tabular}
\end{center} \end{table}

\begin{table}[ptbh]\caption{Unitary \hwm s of $\sonh$  at level 2 
for $N=2\el+1$.} \label{t2}
\begin{center}
  \begin{tabular}{|c|c|c|c|} \hline &&&\\[-.9em]
  field & $\Lambda$    & $\Delta$     & \qdi \\ \hline\hline &&&\\[-.9em]
  $\rmo$&   0          & 0            & 1    \\ &&&\\[-.9em]
  $\rmv$& $2\Lj1$      & 1            & 1    
  \\ &&&\\[-.8em]%\hline &&&\\[-.9em]
  $\pfj j$   & $\left\{ \bearl \Lj j\ \;{\rm for}\ j=\onetolme, \\[.1em]
           2\Lj\el\ {\rm for}\ j=\el \eear\right.$
           & $\Frac{j(N-j)}{2N}$ & 2  \\ &&&\\[-.8em] \hline &&&\\[-.9em]
  $\sigma$ & $\Lj{\el-1}$ & $\Frac{N-1}{16}$ & $\sqrt\el$ \\ &&&\\[-.8em]
  $\sigma'$& $\Lj1+\Lj{\el-1}$ & $\Frac{N+7}{16}$ & $\sqrt\el$ 
  \\[-.8em]&&&\\ \hline \end{tabular}
\end{center} \end{table}
Omitting the twisted sectors $\sigma,\sigma',(\tau,\tau')$ all other
sectors generate a fusion subring $\Rn$ of the full fusion ring
$\Rw$ of the WZW model based on $\sonhz$. The fusion rules of this 
fusion subring are the following:
For $N=2\el$ we have
  \be  \bearl 
  \rmv \times \rmv = 1 \,, \qquad\quad \rmv \times \rms = \rmc \,, \nline7
  \rms \times \rms = \rmc \times \rmc = \left\{ \bearll
       1  & {\rm for}\ \el\iN2\zet\,, \\[.2em] 
       \rmv & {\rm for}\ \el\iN2\zet+1\,, \eear\right. \nline7
  \rms \times \rmc = \left\{ \bearll
       \rmv & {\rm for}\ \el\iN2\zet\,, \\[.2em] 
       1  & {\rm for}\ \el\iN2\zet+1\,, \eear\right. \nline6
  \rmv \times \pfj j = \pfj j\,,  \qquad\quad
  \rms \times \pfj j = \rmc \times \pfj j = \pfj{\el-j}\,, \nline4
  \pfj i \times \pfj j = \pfj{\,|i-j|\,} + \pfj{i+j} \,.
  \eear \ee
Here it is to be understood that whenever on the right hand side a label
$j$ appears which is larger than $\el$, it must be interpreted as the number 
  \[  j' = N-j \,, \] 
and when the label equals zero or $\el$, one has to identify $\pfj j$ as
the sum
  \[  \pfj0 \equiv 1  + \rmv\,,  \qquad  
   \pfj\el \equiv \rms + \rmc \,. \]
For $N=2\el+1$ the fusion rules read 
  \be  \bearl
  \rmv \times \rmv = 1 \,,  \qquad\quad
  \rmv \times \pfj j = \pfj j \,, \nline6
  \pfj i \times \pfj j = \pfj{\,|i-j|\,} + \pfj{i+j} \,.
  \hsp4  \eear \ee
This time it is understood that when $j$ is larger than $\el$,
it stands for the number $j' = N-j$, and again 
that $\pfj0 \equiv 1  + \rmv$.

\chapter{$\son$ Wess-Zumino-Witten Models at Level 1}
In this chapter we give a formulation of the
level $1$ $\son$ WZW models. We discuss the realization
of $\sonh\rtimes\Vir$ in the Neveu-Schwarz and in the
Ramond sector. Employing the fact that the representation
theory of the even fermion algebra reproduces the sectors
of the chiral algebra, we can introduce a net of local
$C^*$-algebras in terms of even CAR algebras and define 
endomorphisms that generate the WZW sectors. We extend their
action to a net of local von Neumann algebras, and then
we can prove the WZW fusion rules in terms of the DHR
sector product.

This analysis is based on
\cite{awzw} and is a generalization of the program
carried out for the Ising model \cite{leci}. As
\cite{leci} was motivated from the earlier work \cite{MS1}
of Mack and Schomerus, \cite{awzw} is motivated from the
ideas of Fuchs, Ganchev and Vecserny{\'e}s \cite{FGV}.

\section{Realization of $\sonh\rtimes\Vir$}
We begin our analysis with the fermionic realization
of the level $1$ modules coming from second quantization
in the Neveu-Schwarz and the Ramond sector.

\subsection{Representation in $\cK$}
From now on, let $\cK=\LSN\equiv\LSE\otimes\bbC^N$. 
We define a (Fourier) orthonormal base
\[ \left\{ \ef ir, \quad r\in\zet+\half ,\quad
i=\onetoN \right\} \]
by the definition
\[ \ef ir = e_r \otimes u^i \]
where $e_r\in\LSE$ are defined by
$e_r(z)=z^r$, $z=\E^{\I\phi}$, $-\pi<\phi\le\pi$,
and $u^i$ denote the canonical unit
vectors of $\bbC^N$. Further we denote by $\GG$
the canonical complex conjugation in $\LSN$ so
that $\GG\ef ir=\ef i{-r}$. We then define  the
Neveu-Schwarz operator
$\PNS\in\QKG$ to be the basis projection
\[ \PNS = \sum_{i=1}^N \sumrNp \kete i{-r} \brae i{-r}. \]
For $i,j=\onetoN$ and $m\in\zet$, we define the 
following operators in $\BK$,
  \[ \bet ijm = \sumrZp \kete i{r+m} \brae jr. \]
One checks by direct computation
  \[ [ \bet ijm , \bet kln ] = \del jk \, \bet il{m+n}
  - \del il \, \bet kj{m+n}. \]
Defining
  \[ \ttau ijm = \I \, (\bet ijm - \bet jim ), \]
(the $\ttau ijm$ act as multiplication operators
$z^m\otimes T^{i,j}$ on $\LSE\otimes\bbC^N$)
we obtain a realization of $\sonh$ at level zero,
  \[ [ \ttau ijm , \ttau kln ] = \I \, ( \del jk \, 
  \ttau il{m+n}   + \del il \, \ttau jk{m+n} - \del jl 
  \, \ttau ik{m+n} -   \del ik \, \ttau jk{m+n} ). \]
Note that skew self-adjoint combinations
  \[ \I \, \ttau ij0, \qquad
  \tau^{i,j}_{m,+} = \I \, (\ttau ijm + \ttau ij{-m}),
  \qquad \tau^{i,j}_{m,-} = \ttau ijm - \ttau ij{-m},
  \qquad m=1,2,\ldots \]
are elements of $\iPNSoKG$. 
Similarly, we define on $\cK$ operators $\lambda_m$,
$m\in\zet$, which act as 
$-z^m \left(z\frac{\D}{\D z}+\frac{m}{2}\right)$ 
in each component,
  \[ \lambda_m = - \sum_{i=1}^N \sumrZp 
  (r+\sfrac m2) \kete i{r+m} \brae ir. \]
Hence
  \[ [ \lambda_m , \lambda_n ] = (m-n) \lambda_{m+n}, \]
i.e.~we obtain a realization of $\Vir$ with zero central
charge (Witt algebra). Note that skew self-adjoint combinations
  \[ \I \, \lambda_0, \qquad
  \lambda_{m,+} = \I \, (\lambda_m + \lambda_{-m}),
  \qquad \lambda_{m,-} = \lambda_m - \lambda_{-m},
  \qquad m=1,2,\ldots \]
are as in Theorem \ref{unbounded}. Since also
  \[ [\lambda_m , \ttau ijn ] = -n \, \ttau ij{m+n} \]
holds we have together a realization of
$\sonh_0 \rtimes \Vir_0$.

\subsection{Realization of $\sonh\rtimes\Vir$ in the Neveu-Schwarz Sector}
We now go on in defining
a realization of $\sonh_1$ by the procedure of second quantization.
Let us denote by $(\HNS,\PiNS,\ONS)$ the GNS 
representation of the quasi-free state
$\omega_{\PNS }$, and then we
define Fourier modes acting on $\HNS$,
\[ \bbe ir =\PiNS(B(\ef ir)),\qquad r\in\zet + \half,
\quad i=\onetoN. \]
Hence we have $(\bbe ir)^*=\bbe i{-r}$ and 
anticommutation relations 
\[ \{\bbe ir, \bbe js \} = \del ij \, \del {r+s}0 \, \bfe, \]
and Fourier modes with positive grade                   
act as annihilation operators in $\HNS$,
\[ \bbe ir \, \ONS = 0, \quad r>0. \]
{\it Finite energy vectors}
  \be \bbe {i_m}{-r_m} \cdots \bbe {i_2}{-r_2}
  \bbe {i_1}{-r_1} \ONS, \qquad r_l\in\bbNo +\half ,
  \qquad i_l=\onetoN \labl{VNS}
are total in $\HNS$ i.e.\ finite linear combinations 
produce a dense subspace $\HNSf$.
Denoting normal ordering by colons,
\[ \normord{\bbe ir \bbe js}\,\,\, = 
\left\{ \begin{array}{rl}
\bbe ir \bbe js & r<0 \\ -\bbe js \bbe ir & r>0
\end{array} \right. , \qquad r,s\in\zet + \half, \]    
we introduce unbounded operators on $\HNS$
  \[ \BBe ijm = \half \sumrZp \normord{\bbe ir \bbe j{m-r}}\,. \]
Note that these infinite series terminate on finite
energy vectors (\ref{VNS}), i.e.\ $\HNSf$ is an invariant
dense domain of these expressions. 
For $T\in\son$ define current operators $J_m(T)$ by
  \[ J_m(T) = \sum_{i,j=1}^N (T)_{i,j} \BBe ijm . \]
In particular, $\J ijm \equiv J_m(\T ij)$,
  \[ \J ijm = \I \, (\BBe ijm - \BBe jim). \]
Then one checks by direct computation
  \[ [ \J ijm , \bbe kr ] = \I \, (\del jk \, \bbe i{r+m}
  - \del ik \, \bbe j{r+m}), \]
or equivalently
  \[ [ \J ijm , \PiNS (B(f)) ] = \PiNS (B(\ttau ijm f)). \]
Moreover,
  \be [ \J ijm , \J kln ] = \I \, (\del jk \, \J il{m+n}
  + \del il \, \J jk{m+n} - \del jl \, \J ik{m+n} -
  \del ik \, \J jl{m+n}) + 
  m \, \del m{-n} (\T ij | \T ji), \labl{JmJn}
i.e.~we have a realization of $\sonh$ at level 1. It is 
also straightforward to check that the scalar term on 
the r.h.s.\ of Eq.\ (\ref{JmJn}) is indeed the
Schwinger term (\ref{cP}),
  \[ c_{\PNS}(\ttau ijm , \ttau kln) =
  m \, \del m{-n} (\T ij | \T kl), \]
and hence we identify
  \[ \I \, \J ij0 = \D Q_{\PNS} (\I \, \ttau ij0),\]
and
  \[ \I\, ( \J ijm + \J ij{-m}) =   
  \D Q_{\PNS} (\ttau ij{m,+}),\quad
  \J ijm - \J ij{-m} = \D Q_{\PNS} (\ttau ij{m,-}), \quad 
  m=1,2,\ldots \]
Further, we define unbounded operators $L_m$, $m\in\zet$,
on $\HNS$
  \[ L_m = - \half \sum_{i=1}^N \sumrZp 
  (r- \sfrac m2) \normord{\bbe ir \bbe i{m-r}}\,. \]
Note that also these series terminate on finite energy vectors.
One checks by direct computation
  \[ [ L_m , \bbe ir ] = -(r + \sfrac m2) 
  \bbe i{r+m} \]
or, for $f$ in the domain of $\lambda_m$,
  \[ [ L_m , \PiNS (B(f)) ] = \PiNS (B(\lambda_m f)). \]
Moreover,
  \be [ L_m , L_n ] = (m-n) L_{m+n} + m(m^2-1) \, \del m{-n}
  \, \sfrac N{24}, \labl{LmLn}
and
  \be [ L_m , \J ijn ] = -n\,  \J ij{m+n}, \labl{LmJn} 
i.e.\ together we have a realization of 
$\sonh_1\rtimes\Vir_{N/2}$.
Indeed, we identify
  \[ \I \, L_0 = \D Q_{\PNS} (\I \, \lambda_0),\]
and
  \[ \I\, ( L_m + L_{-m}) =   
  \D Q_{\PNS} (\lambda_{m,+}),\quad
  L_m - L_{-m} = \D Q_{\PNS} (\lambda_{m,-}), \quad 
  m=1,2,\ldots \]

\subsection{Characters}
It is known that $\HNS$ decomposes as a $\sonh$ module
into the basic module $\cH_\circ$ and the vector module $\cH_\rmv$
with highest weight vectors $\Omo=\ONS$ and 
$\Ovvo=2^{-\h} (\bbe 1{-\h} +\I\bbe 2{-\h})\ONS$, respectively.
By $\zet_2$-invariance of the current operators this is
precisely the decomposition into the even and the odd
Fock space, respectively. The corresponding projections
$P_\circ\equiv P_+$ and $P_\rmv\equiv P_-$ can be written as
  \[ P_\pm = \half ( \bfe \pm Q_{\PNS}(-\bfe)). \]
It is instructive to compute the Virasoro specialized
characters of the irreducible modules.
We introduce Euler's product function,
  \be \varphi (q) = \prodni 1 (1-q^n). \labl{Euler}
Thus for the character\footnote{For simplicity, we use
the argument $q=\exp (2\pi\I\tau)$, $|q|<1$, directly instead
of the upper complex half plane variable $\tau$. We define the
characters simply as the trace of $q^{L_0}$ here, i.e.\ we
also neglect the additional term $-c/24$ in (\ref{defchar}).}
$\chi_\circ^{(1)}(q)=\tr_{\HNS}(P_\circ q^{L_0})$
of the basic module we obtain
  \beaa \chi_\circ^{(1)} (q) &=& \frac12 \left[ 
  \prodni 0 (1+q^{n+\h})^N + \prodni 0 (1-q^{n+\h})^N 
  \right] \\
  &=&  \frac{(\varphi(-q^\h))^N}{2\,(\varphi(q))^N}
  + \frac{(\varphi(q^\h))^N}{2\,(\varphi(q))^N} , \eeaa
while for the character 
$\chi_\rmv^{(1)}(q)=\tr_{\HNS}(P_\rmv q^{L_0})$
of the vector module we get
  \beaa \chi_\rmv^{(1)} (q) &=& \frac12 \left[ 
  \prodni 0 (1+q^{n+\h})^N - \prodni 0 (1-q^{n+\h})^N 
  \right] \\
  &=& \frac{(\varphi(-q^\h))^N}{2\,(\varphi(q))^N}
  - \frac{(\varphi(q^\h))^N}{2\,(\varphi(q))^N} . \eeaa

\subsection{M\"obius Covariance} 
Let us give some brief remarks
on M\"obius covariance of the 
vacuum sector. Although this is always treated as
standard knowledge we have not found a complete
proof of M\"obius covariance in the literature;
therefore we present our own calculations here.
The M\"obius symmetry on the circle 
$S^1$ is given by the group 
$\PSU=\SU/\zet_2$ where
\[ \SU = \left\{ \left. g= \left( 
\begin{array}{cc}
\alpha & \beta \\ \overline{\beta} & \overline{\alpha}
\end{array} \right) \in \GLZ \,\,
\right| \,\, |\alpha|^2-|\beta|^2=1 \right\}. \]
Its action on the circle is
\[ gz= \frac{\overline{\alpha}z-
\overline{\beta}}{-\beta z + \alpha}, \qquad
z \in S^1. \]
Consider the one-parameter-group of rotations
$a_0(t)$,
\[ a_0(t) = \left( \begin{array}{cc}
\E^{-\I t/2} & 0 \\ 0 &
\E^{\I t/2} \end{array} \right),
\qquad t \in {\mathbb R}. \]
Any element $g\in\SU$ can be decomposed
into a rotation $a_0(t)$ and a transformation
$g'=a_0(-t)g$ leaving the point $z=-1$ invariant,
\[ g=a_0(t)g', \qquad  g'= \left( 
\begin{array}{cc} \alpha' & \beta' \\ 
\overline{\beta'} & \overline{\alpha'}
\end{array} \right), \qquad
\frac{\overline{\alpha'}+
\overline{\beta'}}{\alpha' + \beta'}=1. \]
Since $a_0(t+2\pi)=-a_0(t)$ we can determine
$t$, $-2\pi<t\le 2\pi$ uniquely by the
additional requirement ${\rm Re}(\alpha')>0$. Then
a representation $U$ of $\SU$ in our 
Hilbert space $\cK$ of test functions 
$f=(f^i)_{i=\onetoN}$ is defined component-wise by
  \be \left( U (g) f \right)^i (z) = \eps (g;z)
  (\alpha + \overline{\beta}
  \overline{z})^{-\h }(\overline{\alpha}
  +\beta z)^{-\h }
  f^i \left( \frac{\alpha z + \overline{\beta}}{\beta z
  + \overline{\alpha}} \right), \labl{repr}
where for $z=\E^{\I\phi}$, $-\pi<\phi\le\pi$,
\[ \eps (g;z)= -\,{\rm sign}(t-\pi-\phi)
\,\,{\rm sign}(t+\pi-\phi), \]
and ${\rm sign}(x)=1$ if $x\ge 0$,
${\rm sign}(x)=-1$ if $x<0$. 
We observe that
$\eps(g;z)$ is discontinuous at $z=-1$
and $z=g(-1)=-(\overline{\alpha}+\overline{\beta})
(\alpha+\beta)^{-1}$.
Up to this $\eps$-factor, Eq.~(\ref{repr}) 
is a well-known definition of a representation
of $\SU$. So it remains to be checked 
that
\[ \eps(g_1;z) \eps(g_2;g_1^{-1}z)
=\eps(g_1g_2;z). \]
Since both sides have their discontinuities at
$z=-1$ and $z=g_1g_2(-1)$ they can differ only
by a global sign. But this possibility is easily
excluded by an argument of $L^2$-continuity in 
$g$. We want to show that $U$ is also unitary.
Since the action of $U(g)$ is the same in each 
component we need only consider the case $N=1$.
Hence we have to establish 
$\langle U(g) e_r, U(g) e_s \rangle = \delta_{r,s}$
for $r,s\in\zet+\half$,
\begin{eqnarray*}
\langle U(g) e_r , U(g) e_s \rangle &=&
\oint_{S^1} \frac{\D z}{2\pi\I z}
(\alpha + \overline{\beta}\overline{z})^{-1}
(\overline{\alpha} + \beta z)^{-1}
\left( \frac{\alpha z + \overline{\beta}}
{\beta z + \overline{\alpha}} \right)^{s-r} \\
&=& \frac{1}{2\pi\I} \oint_{S^1} \D z \,\,
(\alpha z + \overline{\beta})^{s-r-1}
(\beta z + \overline{\alpha})^{r-s-1} \\
&=& \left\{ \begin{array}{cc}
0 & \quad s>r \\ 
\frac{\alpha^{s-r-1}}{(r-s)!}
\frac{\D^{r-s}}{\D z^{r-s}}(\beta z+
\overline{\alpha})^{r-s-1}\big|_{z=-
\frac{\overline{\beta}}{\alpha}} &
\quad s \le r \end{array} \right. \\
&=& \delta_{r,s}
\end{eqnarray*}
by Cauchy's integral formula, respecting that
$|\alpha|^2>|\beta|^2$ since $|\alpha|^2-
|\beta|^2=1$. Since the prefactor
on the right-hand side in Eq.~(\ref{repr})
is real we observe $[U(g),\GG]=0$ and hence
each $U(g)$, $g\in\SU$ induces a
Bogoliubov automorphism $\alpha_g=\varrho_{U(g)}$
of $\CKG$. Hence
$\SU$ is represented by automorphisms
of $\CKG$, and this restricts
to a representation of $\PSU$ by automorphisms
of $\CKG^+$.
In order to establish
M\"obius invariance of the vacuum state and hence
covariance of the vacuum sector we show that
\[ [\PNS,U(g)]=0,\qquad g\in\SU,\]
i.e.~that $U(g)$ respects the polarization of
$\cK$ induced by $\PNS$. Again we need
only consider the case $N=1$. It is 
sufficient to show that
\[ \langle e_{-r},U(g) e_s \rangle =0,
\quad r,s\in\bbNo + \half, \qquad
g \in \SU. \]
The functions $e_r(z)$, $r\in\zet +\half$
are smooth on $S^1$ except at their cut at
$z=-1$. The prefactor $\eps(g;z)$ in
Eq.~(\ref{repr}) achieves that 
$(U(g)e_r)(z)$ remains a smooth function except
at $z=-1$, i.e. that the cut is not transported
to $g(-1)$. Hence we have
\[ (U(g)e_r)(z)= \pm(\alpha 
+\overline{\beta}\overline{z})^{-\h}
(\overline{\alpha}+\beta z)^{-\h} \left( \frac{\alpha z + 
\overline{\beta}}{\beta z + \overline{\alpha}}
\right)^r, \]
where all the half-odd integer powers are to be
taken in the same branch with cut at
$z=-1$. So we can compute as follows:
\begin{eqnarray*}
\langle e_{-r}, U(g) e_s \rangle &=& \pm
\oint_{S^1} \frac{\D z}{2\pi\I z} z^r 
(\alpha z+\overline{\beta})^{-\h}
z^\h(\overline{\alpha}+\beta z
)^{-\h} \left( \frac{\alpha z + 
\overline{\beta}}{\beta z + \overline{\alpha}}
\right)^s \\
&=& \pm\frac{1}{2\pi\I} \oint_{S^1} \D z \,\, 
z^{r-\h} (\alpha z +\overline{\beta}
)^{s-\h} (\overline{\alpha} +
\beta z)^{-s-\h} = 0,
\end{eqnarray*}
again by Cauchy's formula, respecting
$|\alpha|^2>|\beta|^2$ and that $r,s$ are
positive half-odd integers here.

Consider further one-parameter-subgroups 
\[ a_+(t) = \left( \begin{array}{cc}
\cosh t & \I \, \sinh t \\ - \I \, \sinh t &
\cosh t \end{array} \right), \qquad
a_-(t) = \left( \begin{array}{cc}
\cosh t &  \sinh t \\ - \sinh t &
\cosh t \end{array} \right), \]
$t\in{\mathbb R}$.
It is not hard to check that $a_0$ and $a_\pm$
correspond to infinitesimal generators 
$\I\,\lambda_0$ and $\lambda_{1,\pm}$,
respectively. More precisely,
\[ U(a_0(t))=\exp (\I t\lambda_0), \qquad 
U(a_\pm(t))=\exp (t\lambda_{1,\pm}) \]
by Stone's Theorem.

\subsection{Realization of $\sonh\rtimes\Vir$ in the Ramond Sector}
We define another (Fourier) orthonormal base
\[ \left\{ \ef in, \quad n\in\zet,\quad
i=\onetoN \right\} \]
by the definition
\[ \ef in = e_n \otimes u^i \]
where also $e_n\in\LSE$ are defined by
$e_n(z)=z^n$. Now we define the Ramond operator
$\SR\in\QKG$ by
\[ \SR = \sum_{i=1}^N \left( \half \kete i0 \brae i0
+ \sumnN \kete i{-n} \brae i{-n} \right) \]
and denote by $(\HR,\PiR,\OR)$ the GNS 
representation of the associated quasi-free state
$\omega_{\SR}$. Further introduce Fourier modes
acting on $\HR$,
\[ \bbe in =\PiR (B(\ef in)),\qquad n\in\zet,
\quad i=\onetoN. \]
Hence we have $(\bbe in)^*=\bbe i{-n}$ and 
anticommutation relations 
\[ \{\bbe im, \bbe jn \} = \del ij \, \del {m+n}0 \, \bfe, \]
also
\[ \bbe in \, \OR = 0, \quad n>0. \]
Finite energy vectors
  \be \bbe {i_m}{-n_m} \cdots \bbe {i_2}{-n_2}
  \bbe {i_1}{-n_1} \OR, \qquad n_l\in\bbNo,
  \qquad i_l=\onetoN \labl{VR}
are total in $\HR$ i.e.~finite linear combinations 
produce a dense subspace $\HRf$. Similar to the
situation in $\HNS$ we denote normal ordering by colons,
\[ \normord{\bbe im \bbe jn}\,\,\, = 
\left\{ \begin{array}{rl}
\bbe im \bbe jn & m<0 \\ -\bbe jn \bbe im & m\ge 0
\end{array} \right. , \qquad m,n\in\zet, \]     
and introduce unbounded operators on $\HR$ 
(by some abuse of notation we employ the same
symbols as in the Neveu-Schwarz sector),
  \[ \BBe ijm = \half \sumnZ \normord{\bbe in \bbe j{m-n}}\,. \]
Analogous to the situation in the Neveu-Schwarz sector,
$\HRf$ is an invariant dense domain of these expressions.
For $T\in\son$ define current operators $J_m(T)$ by
  \[ J_m(T) = \sum_{i,j=1}^N (T)_{i,j} \BBe ijm . \]
In particular, $\J ijm \equiv J_m(\T ij)$,
  \[ \J ijm = \I \, (\BBe ijm - \BBe jim). \]
We also find by direct computation
  \[ [ \J ijm , \bbe kn ] = \I \, (\del jk \, \bbe i{n+m}
  - \del ik \, \bbe j{n+m}) \,. \]
Note that in $\HR$ the action of $\J ijm$ comes also from
an action $\ttau ijm$ in $\KK$ via
$ [ \J ijm , \PiR (B(f)) ] = \PiR (B(\ttau ijm f))$.
We have $\ttau ijm = \I \, (\bet ijm - \bet jim )$, 
and one easily verifies that the $\J ijm$ implement
indeed the same action as those in the Neveu-Schwarz
sector, i.e.
  \[ \bet ijm = \sumnZ \kete i{n+m} \brae jn
  =\sumrZp \kete i{r+m} \brae jr \,. \]
Also (\ref{JmJn}) holds in the Ramond sector, but it is not
a direct consequence of Theorem \ref{bounded} since the
Ramond state $\omega_{{\SR}}$ is not pure i.e.\ not a Fock
state. However, there is also a generalization to
second quantization in non-Fock states; the Schwinger term
(\ref{cP}) is just slightly modified in this case,
see e.g.\ \cite{EK}.
We also define unbounded operators $L_m$, $m\in\zet$,
on $\HR$ (with invariant dense domain $\HRf$)
  \[ L_m = - \half \sum_{i=1}^N \sumnZ 
  (n- \sfrac m2) \normord{\bbe in \bbe i{m-n}} + \, \del m0 \,
  \sfrac{N}{16} \, \bfe , \]
Then one checks by direct computation
  \[ [ L_m , \bbe in ] = -(n + \sfrac m2) 
  \bbe i{n+m},\]
and also (\ref{LmLn}) and (\ref{LmJn}) hold 
in the Ramond sector. Note that the $L_m$ in $\HR$
come also from an action in $\KK$ which is
$-z^m \left( z \frac{\D}{\D z} + \frac{m}{2}\right)$
in each component, however, these differential
operators respect periodic boundary conditions
here in contrast to antiperiodic boundary
conditions in the Neveu-Schwarz case.

\subsection{Comparison to the Sectors of the
Even CAR Algebra}
As a $\sonh$ module $\HNS$ decomposes into
the basic ($\circ$) and the vector ($\rmv$) module. It is
also known that $\HR$ decomposes into
the direct sum of $2^\ell$ spinor ($\rms$) and 
$2^\ell$ conjugate spinor ($\rmc$) modules 
if $N=2\ell$ and into $2^{\ell+1}$ spinor
modules ($\sigma$) if $N=2\ell+1$. Using our previous
results of CAR theory, we can easily verify that
exactly the same happens if we restrict the
representations $\PiNS$ ($\pi_{\rm R}$) of
$\CKG$ in $\HNS$
(${\cal H}_{\rm R}$) to the even subalgebra
$\CKG^+$: Since $\PNS $
is a basis projection we have by Theorem
\ref{resteven}
\be
\PiNS|_{\CKG^+} =
\PiNS^+ \oplus \PiNS^-.
\labl{NSp}
Now $\PiNS^+$ acts in the even Fock space
\cite{Ara2} which corresponds to the basic 
module. Thus we may use the same symbols which 
label the sectors,
$\pi_0\equiv\PiNS^+$ ($\pi_0$ being the
basic, i.e.~vacuum representation) and
$\pi_\rmv\equiv\PiNS^-$. Consider
the Bogoliubov operator 
$V_{1/2}\in{\cal I}(\cK,\GG)$,
\[ V_{1/2} = \sum_{i=1}^N \left( 
|\tilde{e}_+^i\rangle\langle
e_0^i| + \sum_{n=1}^\infty
\big( |e_{n+\h }^i\rangle\langle e_n^i|
+ |e_{-n-\h }^i\rangle\langle e_{-n}^i|
\big) \right), \]
where $\tilde{e}_+^i=2^{-\h }(e_\h ^i+
e_{-\h }^i)$.
It is not hard to see that 
$\SR =V_{1/2}^*\PNS V_{1/2}$, that
$M_{V_{1/2}}=N$ and that $N_{V_{1/2}}=0$. We find  
$\pi_{\rm R}\simeq\PiNS\circ\varrho_{V_{1/2}}$
by Eq.~(\ref{cyclics}), and hence by Theorem 
\ref{restevenodd},
\be
\pi_{\rm R}|_{\CKG^+}
\simeq \left\{ \begin{array}{cl}
2^\ell\, (\pi_{P'}^+\oplus\pi_{P'}^-) &
\qquad N=2\ell \\
2^{\ell+1} \, \pi & \qquad N=2\ell+1
\end{array} \right.
\labl{Rp}
for a basis projection $P'$, 
$[P']_2=[\SR ^\h ]_2$. Thus we use
notations $\pi_\rms\equiv\pi_{P'}^+$,
$\pi_\rmc\equiv\pi_{P'}^-$ and
$\pi_\sigma\equiv\pi$. (Recall that $\pi$ is one of
the equivalent restrictions of the pseudo Fock
representations $\pi_{E,\pm}$.) We have seen that the
CAR representations $\PiNS$ and $\pi_{\rm R}$,
when restricted to the even algebra, reproduce 
precisely the sectors of the chiral algebra. This
is not quite a surprise because the Kac-Moody and 
Virasoro generators are made of fermion bilinears.
Here we see that they indeed act irreducibly in the
(dense subspaces of the) sectors of even CAR. This
is the reason why we are allowed to identify the
elements of the even CAR algebras as the bounded
operators representing the observables of the WZW
model, and also that we identify the sectors of
the even CAR algebra to be the WZW sectors. Note that
the Bogoliubov endomorphism $\varrho_{V_{1/2}}$
induces a transition from the vacuum sector to
spinor sectors.

\section{Treatment in the Algebraic Framework}
Our incorporation of the level 1 $\son$ WZW models
in the framework of AQFT is based on the fact that
(local) even CAR algebras can be identified as (local)
observable algebras. We proceed as follows: We
introduce a system of local even CAR algebras on
the circle. Then we can define localized endomorphisms
in terms of Bogoliubov transformations. Later we
extend representations and endomorphisms to a net
of von Neumann algebras on the punctured circle,
and this will be the foundation for the proof of the
fusion rules using the DHR sector product.

\subsection{Localized Endomorphisms}
We introduce at first a local structure on $S^1$,
i.e.\ we define local algebras of observables.
Let us denote by ${\cal J}$ the set of open,
non-void proper subintervals of $S^1$. For
$I\in{\cal J}$ set 
$\cK(I)=L^2(I;\bbC^N)$ 
and define local $C^*$-algebras
\[ \cCA(I)=\CKIG^+ \]
so that we have inclusions
\[ \cCA(I_1) \subset \cCA(I_0), \qquad
I_1 \subset I_0, \]
inherited by the natural embedding of the
$L^2$-spaces; and also we have locality,
\[ [\cCA(I_1),\cCA(I_2)]=\{0\},
\qquad I_1\cap I_2 = \emptyset. \]
Our construction of localized endomorphisms
happens on the punctured circle. Consider
the interval $I_\zeta\in{\cal J}$ which is $S^1$
by removing one ``point at infinity''
$\zeta\in S^1$, $I_\zeta=S^1\setminus\{\zeta\}$.
Clearly, 
$\cCA(I_\zeta)=\CKG^+$.
Further denote by $\Jz$ the set of 
``finite'' intervals $I\in{\cal J}$ such that
their closure is contained in $I_\zeta$,
\[ \Jz = \{ I\in{\cal J} \,\,|\,\,
\bar{I}\subset I_\zeta \}. \]
An endomorphism $\varrho$ of $\cCA(I_\zeta)$
is called localized in some interval 
$I\in\Jz$ if it satisfies
\[ \varrho(A)=A, \qquad A\in\cCA(I_1), \qquad
I_1\in\Jz, \qquad I_1 \cap I = \emptyset. \]
The construction of localized endomorphisms by 
means of Bogoliubov transformations leads to
the concept of pseudo-localized isometries
\cite{MS1}. For $I\in\Jz$ denote
by $I_+$ and $I_-$ the two connected components
of $I'\cap I_\zeta$ ($I'$ always denotes the
interior of the complement of $I$ in $S^1$,
$I'=I^\rmc\setminus\partial I^\rmc$).
A Bogoliubov operator 
$V\in{\cal I}(\cK,\GG)$ is called even
(resp.~odd) pseudo-localized in $I\in\Jz$
if 
\[ Vf = \epsilon_\pm f,\qquad  f\in \cK(I_\pm),
\qquad \epsilon_\pm \in \{-1,1\}, \]
and $\epsilon_+=\epsilon_-$ 
(resp.~$\epsilon_+=-\epsilon_-$).
Then, as obvious, $\varrho_V$ is localized in $I$
in restriction to $\cCA(I_\zeta)$.
Now we are ready to define our localized vector
endomorphism.
\begin{definition}
For some $I\in\Jz$ choose a real
$v\in\cK(I)$, $\GG v=v$ and $\|v\|=1$.
Define the unitary self-adjoint Bogoliubov
operator $U\in{\cal I}(\cK,\GG)$ by
\be
U= 2|v\rangle\langle v|-{\bf 1},
\ee
and the localized vector endomorphism
(automorphism) $\varrho_\rmv$ by
$\varrho_\rmv=\varrho_U$.
\lablth{rhov}
\end{definition}
Since $U$ is even pseudo-localized, and by
Corollary \ref{Bogvec}, $\varrho_\rmv$ is
indeed a localized vector endomorphism, 
i.e.~$\pi_0\circ\varrho_\rmv\simeq
\pi_\rmv$.
Further, by $U^2={\bf 1}$ we have
$\pi_0\circ\varrho_\rmv^2\simeq\pi_0$.
It follows also from Corollary \ref{Bogvec}
that $\pi_\rms\circ\varrho_\rmv
\simeq \pi_\rmc$.
The construction of a localized spinor
endomorphism is a little bit more costly.
Without loss of generality, we choose
$\zeta=-1$ and the localization region
to be $I_2$,
\[ I_2 = \left\{ z=\E^{\I \phi} \in S^1 \,\,\left| 
\,\, -\sfrac{\pi}{2} < 
\phi < \sfrac{\pi}{2} \right. \right\} \]
such that the connected components $I_\pm$ of
$I_2'\cap I_\zeta$ are given by
\beaa
I_- &=& \left\{ z=\E^{\I \phi} \in S^1 \,\,\left| \,\,
-\pi < \phi < -\sfrac{\pi}{2} \right. \right\}, \\
I_+ &=& \left\{ z=\E^{\I \phi} \in S^1 \,\, \left| \,\,
\sfrac{\pi}{2} < \phi < \pi \right. \right\}.
\eeaa
Our Hilbert space $\cK=\cK(I_\zeta)$
decomposes into a direct sum,
\[ \cK = \cK(I_-) \oplus \cK(I_2) 
\oplus \cK(I_+).\]
By $P_{I_+}$, $P_{I_-}$ we denote the projections onto the
subspaces $\cK(I_+)$, $\cK(I_-)$, respectively.
Define functions on $S^1$ by
\[ f_p(z)= \left\{ \begin{array}{cl}
\sqrt{2}\, z^{2p} & \qquad z\in I_2 \\
0 & \qquad z\notin I_2 \end{array} \right.,
\qquad p\in \half\zet, \]
and
\[ f_p^i = f_p \otimes u^i, \qquad
p\in \half\zet, \qquad i=\onetoN,\]
such that we obtain two ONB of the subspace
$\cK(I_2)\subset\cK$,
\[ \{ f_r^i, r\in\zet+\half ,
i=\onetoN \},\qquad \{ f_n^i, n\in\zet,
i=\onetoN \}. \]
Now define the odd pseudo-localized Bogoliubov operator
$V\in{\cal I}(\cK,\GG)$,
\beaa
V &=& P_{I_-}-P_{I_+}+ V^{(2)}, \\
V^{(2)} &=& \sum_{j\le N \atop j \,\, {\rm odd}}
(\I r^j + \I R^j) -
\sum_{j\le N \atop j \,\, {\rm even}}
(t^j + \I T^j),\\
r^j &=& \fsqz |f_\h ^j
\rangle\langle f_0^j| - \fsqz
|f_{-\h }^j \rangle\langle f_0^j|, \\
R^j &=& \sum_{n=1}^\infty \Big( |f_{n+\h }^j
\rangle\langle f_n^j| - |f_{-n-\h }^j
\rangle\langle f_{-n}^j|\Big),\\
t^j &=& \fsqz |f_\h ^{j-1}
\rangle\langle f_0^j| + \fsqz
|f_{-\h }^{j-1} \rangle\langle f_0^j|, \\
T^j &=& \sum_{n=1}^\infty\Big( |f_{n-\h }^j
\rangle\langle f_n^j| - |f_{-n+\h }^j
\rangle\langle f_{-n}^j|\Big).
\eeaa
Note that $V$ is unitary if $N=2\ell$. More
precisely, we have
\[ M_V= \left\{ \begin{array}{cl} 0 & \qquad
N=2\ell \\
1 & \qquad N=2\ell+1 \end{array} \right. .\]
Furthermore, we claim
\begin{lemma}
With notations as above,
\bea
[(V^*\PNS V)^\h ]_2 &=&
[\SR ^\h ]_2,\\
{[}(V^*V^*\PNS VV)^\h {]}_2 &=&
[\PNS ]_2.
\eea
\lablth{HS}
\end{lemma}
\bproof Let us first point out that that we
do not have to take care about the positive square
roots because for any basis projection $P$ and any
Bogoliubov operator $W\in{\cal I}(\cK,\GG)$
with $M_W<\infty$ we have
\[ [ (W^*PW)^\h  ]_2 = [ W^*PW ]_2 \]
since
\beaa
\| (W^*PW)^\h  - W^*PW \|_2^2 &\le&
\| W^*PW - (W^*PW)^2 \|_1 \\
&=& \| W^*P({\bf 1}-WW^*)PW\|_1 \\
&\le& \|W\|^2 \|P\|^2 \|{\bf 1}-WW^* \|_1 = M_W.
\eeaa
We used the trace norm and Hilbert Schmidt norm
$\|A\|_n=(\mbox{tr}(A^*A)^{n/2})^{1/n}$, 
$n=1,2$, respectively, and also an estimate \cite{PS}
\be
\|A^\h -B^\h \|_2^2\le\|A-B\|_1,
\qquad A,B \in \mathfrak{B}(\cK), \qquad A,B \ge 0.
\labl{HST} 
It was proven in \cite{leci}, Lemma 3.10, that
\[ V^*\PNS V - \SR , \quad
V\PNS V^* - \SR , \quad {V'}^*\PNS V'
-\SR , \quad V'\PNS {V'}^* - \SR  \]
are Hilbert Schmidt operators for the case
$N=1$, where in our notation
\[ V=P_{I_-} - P_{I_+} + \I r^1 + \I R^1,
\qquad V'=P_{I_-} - P_{I_+} + \I (T^1)^*. \]
For arbitrary $N$, operators 
$V^*\PNS V-\SR $ and 
$V\PNS V^*-\SR $ are just direct sums of 
the above Hilbert Schmidt operators (up to finite
dimensional operators), hence we conclude for
arbitrary $N$
\[ V^*\PNS V - \SR  \in 
\mathfrak{J}_2(\cK), \qquad
V\PNS V^* - \SR  \in
\mathfrak{J}_2(\cK). \]
But both relations together imply that
$\PNS -V^*V^*\PNS VV$ is also Hilbert
Schmidt, and this proves the lemma. \eproof

Hence we conclude 
$\PiNS\circ\varrho_V\approx\pi_{\rm R}$.
For $N=2\ell$ the basis projection
$P'=V^*\PNS V$ is as in Eq.\ (\ref{Rp}). For
$N=2\ell+1$ the representation
$\PiNS\circ\varrho_V$, when restricted
to $\CKG^+$, decomposes into
two equivalent irreducibles. With our above
definitions and using Corollary \ref{Bogvec},
this suggests the following
\begin{definition}
Choose $U\in{\cal I}(\cK,\GG)$ for
$v\in\cK(I_2)$ as in Definition
\ref{rhov}. For $N=2\ell$ define the
localized spinor endomorphism $\varrho_\rms$
by $\varrho_\rms=\varrho_V$ and the
localized conjugate spinor endomorphism 
$\varrho_\rmc$ by
$\varrho_\rmc=\varrho_U\varrho_V$. For
$N=2\ell+1$ define the localized
spinor endomorphism $\varrho_\sigma$ by
$\varrho_\sigma=\varrho_V$.
\lablth{rhos}
\end{definition}
Note that this definition fixes the choice,
if $N$ is even, which of the two inequivalent
spinor sectors is called s and which c. This might 
seem to be somewhat inconsistent because for the highest
weight modules there is no ambiguity within $\HR$, for
instance, 
$2^{-\ell}\prod_{j=1}^\ell (\bfe - 2\,\I\, b_0^{2j} b_0^{2j-1}) \OR$
is a highest weight vector of weight
$\ls$. However, for the sectors of even CAR we take
the freedom to rename the sectors i.e.\ which of the
spinor sectors is called $\rms$ and which $\rmc$. There
is no problem with the proof of the fusion rules later on 
since they are invariant under simultaneous exchange
of $\rms$ and $\rmc$. Indeed, our considerations have shown
\begin{theorem}
The localized endomorphisms of Definitions
\ref{rhov} and \ref{rhos} satisfy
$\pi_0\circ\varrho_\xi\simeq\pi_\xi$,
$\xi={\rm v,s,c},\sigma$.
\end{theorem}

\subsection{Extension to Local von Neumann Algebras}
We have obtained the relevant localized endomorphisms
which generate the sectors ${\rm v,s,c},\sigma$. It is
our next aim to derive fusion rules in terms of DHR 
sectors i.e.~of unitary equivalence classes 
$[\pi_0\circ\varrho]$ for localized endomorphisms 
$\varrho$. For such a formulation one needs local 
intertwiners in the observable algebra.
So we have to keep close to the DHR framework, in 
particular, we should use local von Neumann algebras
instead of local $C^*$-algebras $\cCA(I)$.
We define
\[ \cR(I) = \pi_0(\cCA(I))'',\qquad
I\in{\cal J}. \]
By M\"obius covariance of the vacuum state, this
defines a so-called covariant precosheaf on the
circle \cite{BGL}. In particular, we have Haag duality,
\be
\cR(I)'=\cR(I').
\labl{HD}
Since the set ${\cal J}$ is not directed by inclusion we
cannot define a global algebra as the $C^*$-norm
closure of the union of all local algebras. However,
the set $\Jz$ is directed so that we can define 
the following algebra $\cAloc$ of
quasilocal observables in the usual manner,
\[ \cAloc = \overline{\bigcup_{I\in\Jz} \cR(I)}. \]
We want to prove that Haag duality holds also on the
punctured circle and need some technical preparation.
Recall that a function $k\in\LSE$ is in the Hardy
space $H^2$ if $\langle e_{-n},k \rangle =0$ for all
$n\in\bbN$ where $e_{-n}(z)=z^{-n}$. There is
a Theorem of Riesz \cite[Th.\,6.13]{Doug} which states
that $k(z)\neq 0$ almost everywhere if $k\in H^2$ is
non-zero. Now suppose $f\in\PNS\cK$. Then
$g^i\in H^2$ where $g^i(z)=z^\h 
\overline{f^i(z)}$ component-wise, $i=\onetoN$. 
We conclude
\begin{lemma}
If $f\in \PNS \cK$ then $f\in\cK(I)$
implies $f=0$ for any $I\in{\cal J}$.
\lablth{Hardy}
\end{lemma}
For some interval $I\in\Jz$, let us denote by
$\cA_\zeta(I')$ the norm closure of the algebra 
generated by all
$\cR(I_1)$, $I_1\in\Jz$, 
$I_1\cap I=\emptyset$. Obviously 
$\cA_\zeta (I')''\subset \cR(I')$; a key 
point of the analysis is the following
\begin{lemma}
Haag duality remains valid on the punctured circle, i.e.
\be
\cR(I)'=\cA_\zeta(I')''.
\labl{HDpunc}
\end{lemma}
\bproof We have to prove 
$\cA_\zeta(I')''=\cR(I')$. It is 
sufficient to show that each generator
$\pi_0(B(f)B(g))$, $f,g\in\cK(I')$ of
$\cR(I')$ is a weak limit point of a net
in $\cA_\zeta(I')$. Note that the subspace
$\cK^{(\zeta)}(I')\subset\cK(I')$ of
functions which vanish in a neighborhood of
$\zeta$ is dense. So by Eq.\ (\ref{Cnorm}) we
conclude that it is sufficient to establish
this fact only for such generators with
$f,g\in\cK^{(\zeta)}(I')$, because these
generators approximate the arbitrary ones already 
in the norm topology. Let us again denote the
two connected components of $I'\setminus\{\zeta\}$
by $I_+$ and $I_-$, and the projections onto
corresponding subspaces $\cK(I_\pm)$ by
$P_\pm$. We also write $f_\pm=P_\pm f$ and
$g_\pm = P_\pm g$ for our functions 
$f,g\in\cK^{(\zeta)}(I')$.
Then we have
\beaa
\pi_0(B(f)B(g)) &=& \pi_0(B(f_+)B(g_+)) +
\pi_0(B(f_-)B(g_-)) \\
&&  \qquad + \pi_0(B(f_+)B(g_-)) + 
\pi_0(B(f_-)B(g_+)).
\eeaa
Clearly, the first two terms on the r.h.s. are
elements of $\cA_\zeta(I')$. We show that
the third term $Y=\pi_0(B(f_+)B(g_-))$ (then, by
symmetry, also the fourth one) is in
$\cA_\zeta(I')''$. In the same way as in
the proof of Lemma 4.1 in \cite{leci} one
constructs a sequence $\{X_n,n\in\bbN\}$,
\[ X_n = \pi_0 (B(h_n^+)B(h_n^-)) \]
where unit vectors $h_n^\pm\in\cK(I_n^\pm)$ 
are related by M{\"o}bius transformations such that 
intervals $I_n^\pm\subset I_\pm$ shrink to the point 
$\zeta$. Since $\|X_n\|\le 1$ by Eq.~(\ref{Cnorm}) it
follows that there is a weakly convergent subnet
$\{Z_\alpha,\alpha\in\iota\}$ ($\iota$ a directed
set), w-$\lim_\alpha Z_\alpha=Z$. For each
$I_0\in\Jz$ elements $X_n$ commute with
each $A\in\cR(I_0)$ for sufficiently large
$n$. Hence $Z$ is in the commutant of
$\cAloc$ and this implies
$Z=\lambda {\bf 1}$. We have chosen the vectors
$h_n^\pm$ related by M{\"o}bius transformations.
By M{\"o}bius invariance of the vacuum state
we have
\[ \lambda = \langle \Omega_0 | X_1 | \Omega_0
\rangle = \langle \GG h_1^+, \PNS 
h_1^- \rangle . \]
We claim that we can choose $h_1^\pm$ such
that $\lambda\neq 0$. For given $h_1^-$ set
$k=\PNS h_1^-$. We have $k\neq 0$, otherwise
$\GG h_1^-\in \PNS \cK$ in contradiction 
to $h_1^-\in\cK(I_1^-)$ by Lemma \ref{Hardy}.
Again by Lemma \ref{Hardy} we conclude that $k$
cannot vanish almost everywhere. So we clearly
can choose a function $h_1^+\in\cK(I_1^+)$ such that
$\lambda=\langle\GG h_1^+,k\rangle\neq 0$.
Now we find 
$Y=\lambda^{-1}$w-$\lim_\alpha YZ_\alpha$
and also $YZ_\alpha\in\cA_\zeta(I')$
because
\beaa
YX_n &=& \pi_0(B(f_+)B(g_-)B(h_n^+)B(h_n^-)) \\
&=& -\pi_0(B(f_+)B(h_n^+)) \pi_0(B(g_-)B(h_n^-))
\eeaa
is in $\cA_\zeta(I')$ for all 
$n\in\bbN$. \eproof

Since the vacuum representation is faithful 
on $\cCA(I_\zeta)$ we can
identify observables $A$ in the usual manner with
their vacuum representers $\pi_0(A)$. Thus we
consider the vacuum representation as acting
as the identity on $\cAloc$, and, in
the same fashion, we treat local $C^*$-algebras
as subalgebras $\cCA(I)\subset\cR(I)$.
Now we have to check whether we can extend our
representations $\pi_\xi$ and endomorphisms
$\varrho_\xi$ from $\cCA(I)$ to
$\cR(I)=\cCA(I)''$, $I\in\Jz$,
$\xi={\rm v,s,c},\sigma$. That is that we have to
check local quasi-equivalence of the 
representations $\pi_\xi$ and this will now be elaborated.
Define $E_{\rm R}\in\mathfrak{B}(\cK)$ by
\[ E_{\rm R} = \sum_{i=1}^N \sum_{n\in\bbN}
|e_{-n}^i \rangle\langle e_{-n}^i| +
\sum_{j \le N \atop j \,\, {\rm even}}
|e_+^j\rangle\langle e_+^j| \]
where $e_+^j=2^{-\h }(e_0^j+\I e_0^{j-1})$.
\begin{lemma}
For $I\in{\cal J}$ the subspaces 
$\PNS \cK(I)\subset \PNS \cK$
and $E_{\rm R}\cK(I)\subset E_{\rm R}\cK$
are dense.
\lablth{dense}
\end{lemma}
\bproof Suppose that $\PNS \cK(I)$
is not dense in $\PNS \cK$. Then there
is a non-zero $f\in \PNS \cK$ such that
\[ \langle f, \PNS  g \rangle =
\langle f, g \rangle = 0 \]
for all $g\in \cK(I)$. Hence 
$f\in\cK(I)^\perp =\cK(I')$ in 
contradiction to Lemma \ref{Hardy}. As quite
obvious, Lemma \ref{Hardy} holds for 
$f\in E_{\rm R}\cK$ as well. So also
$E_{\rm R}\cK(I)$ is dense in
$E_{\rm R}\cK$. \eproof

Note that $E_{\rm R}$ is a basis projection if
$N$ is even. For $N$ odd, $E_{\rm R}$ is a partial
basis projection with $\GG$-codimension 1 and
corresponding $\GG$-invariant unit vector $e_0^N$.
In this case
\[ \SR ' = \half  |e_0^N \rangle
\langle e_0^N| + E_{\rm R} \]
is of the form (\ref{mittel}). Let us denote by
$({\cal H}_{\rm R'},\pi_{\rm R'},
|\Omega_{\rm R'}\rangle)$ the GNS representation of
the quasi-free state $\omega_{E_{\rm R}}$ if $N$ is
even and $\omega_{\SR '}$ if $N$ is odd.
We conclude
\[ \pi_{\rm R'}|_{\CKG^+}
\simeq \left\{ \begin{array}{cl}
\pi_\rms \oplus \pi_\rmc & N=2\ell\\
2 \pi_\sigma & N=2\ell+1 \end{array}
\right. \]
by Theorem \ref{resteven} and Lemma \ref{pFock} and
the fact that $[ E_{\rm R} ]_2 = [ \SR ^\h  
]_2 = [ \SR  ]_2$ ($N$ even) and
$[ {\SR '}^\h  ]_2 = [ \SR ' ]_2
= [ \SR  ]_2$ ($N$ odd).
\begin{lemma}
For $I\in\Jz$ we have local quasi-equivalence
\be
\PiNS|_{\CKIG} \approx
\pi_{\rm R'}|_{\CKIG}.
\ee
\lablth{locquasi}
\end{lemma}
\bproof We first claim that 
$\ONS$ and $|\Omega_{\rm R'}\rangle$
remain cyclic for
$\PiNS(\CKIG)$ and
$\pi_{\rm R'}(\CKIG)$,
respectively. By Lemma \ref{dense},
$\PNS \cK(I)\subset \PNS \cK$
is dense. It follows that vectors of the form
$\PiNS(B(f_1)\cdots B(f_n))
\ONS$, with
$f_1,f_2,...\,,f_n\in \PNS \cK(I)$, 
$n=0,1,2,\ldots,$
are total in $\HNS$. This proves
the required cyclicity of $\ONS$.
For $N$ even, cyclicity of $|\Omega_{\rm R'}\rangle$
for $\CKIG$ is proven in
the same way. For $N$ odd, we have
${\cal H}_{\rm R'}={\cal H}_{E_{\rm R}} \oplus 
{\cal H}_{E_{\rm R}}$, $\pi_{\rm R'} =\pi_{E,+} 
\oplus \pi_{E,-}$ and $|\Omega_{\rm R'} \rangle =
2^{-\h } (|\Omega_{E_{\rm R}} \rangle 
\oplus |\Omega_{E_{\rm R}} \rangle)$ as in Lemma
\ref{pFock}, and the corresponding 
$\GG$-invariant unit vector is given by 
$e_0^N$. In order to prove cyclicity of
$|\Omega_{\rm R'} \rangle$ we show that
$\langle \Psi | \pi_{\rm R'}(x) | 
\Omega_{\rm R'} \rangle =0$ for all
$x\in\CKIG$,
$|\Psi\rangle=|\Psi_+\rangle\oplus|\Psi_-\rangle
\in{\cal H}_{\rm R'}$, implies $|\Psi\rangle=0$.
We have
\[ \langle \Psi | \pi_{\rm R'}(x) | 
\Omega_{\rm R'} \rangle = \fsqz
\langle \Psi_+ | \pi_{E_{\rm R},+}(x) | 
\Omega_{E_{\rm R}} \rangle + \fsqz
\langle \Psi_- | \pi_{E_{\rm R},-}(x) | 
\Omega_{E_{\rm R}} \rangle = 0 \]
Again by Lemma \ref{dense},
$E_{\rm R}\cK(I)\subset E_{\rm R}\cK$
is dense, hence vectors of the form
$\pi_{E_{\rm R},\pm}(x)|\Omega_{E_{\rm R}}\rangle =
\pi_{E_{\rm R}}(x)|\Omega_{E_{\rm R}}\rangle$,
$x=B(f_1)\cdots B(f_n)$, 
$f_1,f_2,...\,,f_n\in E_{\rm R}\cK(I)$, 
$n=0,1,2,\ldots,$ are total in ${\cal H}_{E_{\rm R}}$.
It follows $|\Psi_-\rangle=-|\Psi_+\rangle$.
Hence 
\[ \langle \Psi_+ | (\pi_{E_{\rm R},+}(y)-
\pi_{E_{\rm R},-}(y))| \Omega_{E_{\rm R}} \rangle
= 0, \qquad y\in\CKIG. \]
Keep all $x=B(f_1)\cdots B(f_n)$ as above and 
choose an $f\in\cK(I)$ such that 
$\langle e_0^N,f\rangle=2^{-\h }$. 
Set $y=(-1)^n B(f)x$. Then, by Eq.~(\ref{piE}),
we compute
\[ \pi_{E_{\rm R},\pm}(y)= (-1)^n \left( 
\pm \half  Q_{E_{\rm R}}(-\bfe) + 
\pi_{E_{\rm R}}(B((E_{\rm R}+
{\overline{E}}_{\rm R})f)) \right)
\pi_{E_{\rm R}}(x) \]
and hence
\[ (\pi_{E_{\rm R},+}(y)- \pi_{E_{\rm R},-}(y))
| \Omega_{E_{\rm R}} \rangle =
\pi_{E_{\rm R}}(B(f_1)\cdots B(f_n))| \Omega_{E_{\rm R}}
\rangle . \]
Because such vectors are total in
${\cal H}_{E_{\rm R}}$ we find $|\Psi_+\rangle=0$
and hence $|\Psi\rangle=0$. We
have seen that vectors $\ONS$
and $|\Omega_{\rm R'}\rangle$ remain cyclic.
Thus we can prove the lemma by showing
that the restricted states
$\omega_{P_I\PNS P_I}$ and
$\omega_{P_IE_{\rm R}P_I}$ ($N=2\ell$)
respectively $\omega_{P_I\SR 'P_I}$
($N=2\ell+1$) give rise to
quasi-equivalent representations. Because 
they are quasi-free on $\CKIG$
we have to show that
\[ [ (P_I\PNS P_I)^\h  ]_2 = \left\{
\begin{array}{cl} {[} (P_IE_{\rm R}P_I)^\h 
{]_2} & \qquad N=2\ell \\ 
{[} (P_I\SR 'P_I)^\h  {]_2} & \qquad
N=2\ell+1 \end{array} \right..\]
By use of Eq.~(\ref{HST}) it is sufficient to
show that the difference of $P_I\PNS P_I$
and $P_IE_{\rm R}P_I$ respectively
$P_I\SR 'P_I$ is trace class for $I\in\Jz$.
It is obviously sufficient to prove that
$P_I\PNS P_I-P_I\SR P_I$ is trace
class for the case $N=1$ (since all 
the operators above are, up to
finite dimensional operators, direct sums
of those for $N=1$).
We use the parameterization $z=\E^{\I\phi}$, 
$-\pi<\phi\le\pi$ of $S^1$. Recall that Hilbert
Schmidt operators $A\in{\cal J}_2(L^2(S^1))$ can be
written as square integrable kernels
$A(\phi,\phi')$. For instance, a rank-one-projection
$|e_r\rangle\langle e_r|$ has kernel $\E^{\I r(\phi-\phi')}$.
For (small) $\epsilon>0$ define operators in 
$\PNS^{(\epsilon)},\SR^{(\epsilon)}\in
{\cal J}_2(L^2(S^1))$ by kernels 
\[ \PNS^{(\epsilon)}(\phi,\phi') = 
\sum_{n=0}^\infty 
\E^{-(n+\h)(\I\phi-\I\phi'+\epsilon)}=
\frac{\E^{-(\I\phi-\I\phi'+\epsilon)/2}}
{1-\E^{-(\I\phi-\I\phi'+\epsilon)}}, \]
and
\[ \SR^{(\epsilon)}(\phi,\phi') = 
\frac{1}{2} + \sum_{n=1}^\infty 
\E^{-n(\I\phi-\I\phi'+\epsilon)}=
\frac{1}{1-\E^{-(\I\phi-\I\phi'+\epsilon)}}-
\frac{1}{2}. \]
Note that $\epsilon>0$ regularizes the singularities
for $\phi-\phi'=0,\pm 2\pi$. Using Cauchy's integral
formula, it is easy to check that for
$r,s \in \zet + \frac{1}{2}$,
\begin{eqnarray*} 
\lim_{\epsilon\searrow 0} \,\, \langle e_r,
\PNS^{(\epsilon)} e_s \rangle &=&
\lim_{\epsilon\searrow 0} \oint_{S^1}
\frac{\D z}{2\pi\I z} \oint_{S^1} \frac{\D z'}{2\pi\I z'}\,\,
\frac{z^{-r+\h} {z'}^{s+\h} \E^{-\epsilon/2}}
{z-z' \E^{-\epsilon}} \\
&=& \left\{ \begin{array}{cc} 
\lim_{\epsilon\searrow 0} \,\, \E^{s\epsilon}\,\delta_{r-s,0} 
\qquad & s<0 \\[.5em] 0 \qquad & \mbox{otherwise} 
\end{array} \right. \\
&=& \langle e_r, \PNS e_s \rangle .
\end{eqnarray*}
Because $\E^{s\epsilon}<1$ for $s<0$ this result
can be generalized to 
\[ \lim_{\epsilon\searrow 0} \,\,
\langle f, \PNS^{(\epsilon)} g \rangle =
\langle f, \PNS g \rangle \]
for arbitrary $f,g \in L^2(S^1)$ by an argument 
of bounded convergence. So we have weak convergence
$\mbox{w-}\lim_{\epsilon\searrow 0}
\PNS^{(\epsilon)}=\PNS^{(0)}\equiv\PNS$. In an 
analogous way one obtains
$\mbox{w-}\lim_{\epsilon\searrow 0}
\SR^{(\epsilon)}=\SR^{(0)}\equiv\SR$. Thus the
difference $\Delta^{(\epsilon)}=
\SR^{(\epsilon)}-\PNS^{(\epsilon)}$
with kernel
\[ \Delta^{(\epsilon)}(\phi,\phi')=
\frac{1}{1+\E^{-(\I\phi-\I\phi'+\epsilon)/2}}-
\frac{1}{2} \]
converges weakly to $\Delta=\SR-\PNS$.
We have to show that $X=P_I\Delta P_I$ is trace class.
The operator $P_I$ acts as multiplication with
the characteristic function $\chi_I(\phi)$
corresponding to $z=\E^{\I\phi}\in I$. Now
$X^{(\epsilon)}=P_I \Delta^{(\epsilon)} P_I$,
converging weakly to $X$, has kernel
\[ X^{(\epsilon)}(\phi,\phi')= \chi_I(\phi) \left(
\frac{1}{1+\E^{-(\I\phi-\I\phi'+\epsilon)/2}}-
\frac{1}{2} \right) \chi_I(\phi') \]
and is no longer singular for $\epsilon\searrow 0$. 
Thus the kernel $X^{(0)}(\phi,\phi')$ that is obtained
from $X^{(\epsilon)}$ by putting $\epsilon=0$ is
well-defined and hence
\[ \lim_{\epsilon\searrow 0}\,\,
\langle f, X^{(\epsilon)} g \rangle 
= \int_{-\pi}^\pi \frac{\D\phi}{2\pi}
\int_{-\pi}^\pi \frac{\D\phi'}{2\pi} \,\,
\overline{f(\E^{\I\phi})} X^{(0)}(\phi,\phi')
g(\E^{\I\phi'}), \qquad f,g\in L^2(S^1), \]
by the theorem of bounded convergence. It follows
$X=X^{(0)}\in{\cal J}_2(L^2(S^1))$.
Let $\tilde{\chi}_I$ be a smooth function on
$[-\pi,\pi]$ which satisfies $\tilde{\chi}_I(\phi)=1$ 
for $z=\E^{\I\phi}\in I$ and vanishes
in a neighborhood of $\phi=\pm\pi$. We define
\[ \tilde{X}(\phi,\phi')=\tilde{\chi}_I(\phi)
\left( \frac{1}{1+\E^{-(\phi-\phi')/2}}
-\frac{1}{2} \right) \tilde{\chi}_I(\phi') \]
such that $X=P_I\tilde{X}P_I$ and hence
\[ \|X\|_1=\|P_I\tilde{X}P_I\|_1 \le \|P_I \| 
\|\tilde{X}\|_1 \|P_I\| = \|\tilde{X}\|_1. \]
Since $\tilde{X}(\phi,\phi')$ is a smooth function
in $\phi$ and $\phi'$ it has fast decreasing
Fourier coefficients which coincide with matrix
elements $\langle e_n,\tilde{X} e_m \rangle$,
$n,m\in\zet$. This proves the statement
$\|X\|_1 < \infty$. \eproof

In restriction to the local even algebra
$\cCA(I)=\CKIG^+$,
$I\in\Jz$ we find by Lemma
\ref{locquasi}
\[ (\pi_0\oplus\pi_\rmv)|_{\cCA(I)}
\approx \left\{ \begin{array}{cl}
(\pi_\rms \oplus \pi_\rmc)
|_{\cCA(I)} & \qquad N=2\ell\\
2 \pi_\sigma |_{\cCA(I)} & \qquad
N=2\ell+1 \end{array}
\right. \]
Recall that $\pi_\rmv\simeq\pi_0\circ\varrho_U$
with $U=2|v\rangle\langle v|-{\bf 1}$ as in
Corollary \ref{Bogvec}. Choose $v\in\cK(I')$.
Then $\varrho_U(x)=x$ for $x\in\cCA(I)$, hence
$\pi_0$ and $\pi_\rmv$ are equivalent on
$\cCA(I)$. In the same way we obtain local
equivalence of $\pi_\rms$ and $\pi_\rmc$.
We conclude that indeed local normality holds
for all sectors.
\begin{theorem}
Restricted to local $C^*$-algebras $\cCA(I)$,
$I\in\Jz$, the representations $\pi_\xi$ are
quasi-equivalent to the vacuum representation
$\pi_0=\id$,
\be
\pi_\xi|_{\cCA(I)} \approx \pi_0|_{\cCA(I)},
\qquad I\in\Jz,\quad \xi={\rm v,s,c},\sigma.
\ee
\lablth{locnorm}
\end{theorem} 
We have seen that we have an extension of our 
representations $\pi_\xi$ to local von Neumann
algebras $\cR(I)$, $I\in\Jz$, and
thus to the quasilocal algebra
$\cAloc$ they generate. By unitary
equivalence $\varrho_\xi\simeq\pi_\xi$ on
$\cCA(I_\zeta)$ we have an extension of
$\varrho_\xi$ to $\cAloc$, too,
$\xi={\rm v,s,c},\sigma$. Being localized in
some $I\in\Jz$, they inherit
properties
\[ \varrho_\xi(A)=A, \qquad A\in\cA_\zeta(I'), \]
and
\[ \varrho_\xi(\cR(I_0)) \subset \cR(I_0),
\qquad I_0\in\Jz, \qquad I\subset I_0, \]
from the underlying $C^*$-algebras. So our
endomorphisms $\varrho_\xi$ are well-defined 
localized endomorphisms of $\cAloc$ in the common sense.
Moreover, they are transportable. This follows
because the precosheaf $\{\cR(I)\}$ is
M\"obius covariant. Hence $\cAloc$
is covariant with respect to the subgroup of
M\"obius transformations leaving $\zeta$
invariant.

\subsection{Fusion Rules}
In this subsection we prove the fusion rules of our
sectors $1,{\rm v,s,c},\sigma$ in terms of unitary
equivalence classes of localized endomorphisms
$[\varrho]\equiv [\pi_0\circ\varrho]$ (or,
equivalently, in terms of equivalence classes
$[\pi]$ of representations $\pi$ satisfying an
DHR criterion). Because we deal with an Haag dual
net of local von Neumann algebras, by standard 
arguments, it suffices to check a fusion rule
$[\varrho_\xi \varrho_{\xi'}]$ for special
representatives $\varrho_\xi\in [\varrho_\xi]$,
$\varrho_{\xi'}\in [\varrho_{\xi'}]$. This will be
done by our examples of Definitions \ref{rhov}
and \ref{rhos}. For instance,
we clearly have $\rmv\times\rmv=1$ for all
$N\in\bbN$. Let us first consider the even
case, $N=2\ell$. By Corollary \ref{Bogvec}
we easily find $\rmv\times\rms=\rmc$,
$\rmv\times\rmc=\rms$. Since $V$ then is
unitary and by Lemma \ref{HS} we
have $\PiNS\circ\varrho_V^2\simeq\PiNS$.
Now $\PiNS$, when restricted to 
$\cCA(I_\zeta)\equiv\CKG^+$, 
decomposes into the basic
and the vector representation. Hence only the
possibilities $\rms\times\rms=1$ or
$\rms\times\rms=\rmv$ are left, i.e.~we
have to check whether 
$\PiNS^+\circ\varrho_V^2$ is equivalent to
$\PiNS^+$ or $\PiNS^-$, 
i.e.~whether $\varrho_\rms$ is a
self-conjugate endomorphism or not. For $N=2\ell$
the action of $V$ in the $(2j-1)^\mathrm{th}$ and the
$2j^\mathrm{th}$ component, $j=\onetol$, is the 
same as in the $1^\mathrm{st}$ and the $2^\mathrm{nd}$
component, respectively. So we can write the
square $W=V^2$ as a product,
\[ W=W_{1,2}W_{3,4}\cdots W_{N-1,N} \]
where $W_{1,2}$ acts as $W$ in the first two
components and as the identity in the others, etc.
Since $\sigma$ of Prop.~\ref{ind} is multiplicative
and clearly all $W_{2j-1,2j}$ lead to implementable
automorphisms we have
\[ \sigma(W) = \sigma(W_{1,2})\sigma(W_{3,4})
\cdots \sigma(W_{N-1,N}). \]
All $W_{2j-1,2j}$ are built in the same way, hence
all the $\sigma(W_{2j-1,2j})$ are equal
i.e.~$\sigma(W)=\sigma(W_{1,2})^{\ell}$.
Since $\sigma$ takes only values $\pm 1$ this
is $\rms\times\rms=1$ if $\ell$ is even.
But for odd $\ell$ this reads
$\sigma(W)=\sigma(W_{1,2})$. Thus we first
check the case $N=2$. If $\sigma(W_{1,2})=+1$
then $\varrho_\rms$ is self-conjugate,
otherwise it is not self-conjugate,
i.e.~$\rms\times\rms=\rmv$.
It is a result of Guido and Longo \cite{GL} that a 
conjugate morphism $\overline{\varrho}$ is given by
\[ \overline{\varrho} = 
{\rm j} \circ \varrho \circ {\rm j} \]
where ${\rm j}$ is the antiautomorphism corresponding 
to the reflection $z\mapsto\overline{z}$ on the
circle (PCT transformation). In our model,
${\rm j}$ is the extension of the antilinear Bogoliubov
automorphism ${\rm j}_\Theta$,
\[ {\rm j}_\Theta (B(f))=B(\Theta f), \qquad
\Theta f \equiv \Theta \left( (f^i)_{i=1,2} \right) 
= \left( f_{\rm refl}^i \right)_{i=1,2}, \]
where $f\in L^2(S^1;{\mathbb C}^2)$
and $f_{\rm refl}^i(z)=
\overline{f^i(\overline{z})}$ for $z\in S^1$.
So we have a candidate 
$\overline{\varrho_\rms}\equiv\overline{\varrho_V}
=\varrho_{\Theta V\Theta}$. It is quite obvious
that $\Theta P_{I_\pm} \Theta = P_{I_\mp}$ and
that $\Theta f_p^i = f_p^i$, $p\in\half\zet$,
so it follows by antilinearity of $\Theta$ ($N=2$)
\[ \Theta V \Theta = -P_{I_-}+P_{I_+} +
(-\I r^1-\I R^1) - (t^2 - \I T^2). \]
It is not hard to see that this is
\[ \Theta V \Theta = U_{1,2} V, \qquad
U_{1,2} = 2 |v_\h ^1\rangle
\langle v_\h ^1|-{\bf 1}, \qquad
v_\h ^1=\fsqz
(f_\h ^1+f_{-\h }^1). \]
Now $U_{1,2}$ is as in Corollary \ref{Bogvec}
so that we find 
$\rms\times\rmv\times\rms=\rms\times\rmc=1$ 
for $N=2$. Hence $\sigma(W_{1,2})=-1$, so it follows
$\rms\times\rmc=1$ for all $N=2\ell$ with $\ell$ odd.
For the case $N=2\ell+1$ the situation is
different because $\varrho_V$ then is not an
automorphism. As discussed at the end of Section 5,
the representation $\PiNS\circ\varrho_V$
(and, of course, also 
$\PiNS\circ\varrho_U\varrho_V$) decomposes,
in restriction to $\CKG^+$,
into two equivalent irreducibles corresponding to
the spinor sector $\sigma$. So we find at first
$\rmv\times\sigma=\sigma$. Let us consider
$\PiNS\circ\varrho_V^2$. We have
$M_{V^2}=2M_V=2$, hence by Theorem \ref{evenodd}
and Lemma \ref{HS} we conclude
$\PiNS\circ\varrho_V^2\simeq 2\PiNS$.
In restriction to $\CKG^+$
this reads $\PiNS^+\circ\varrho_V^2\oplus
\PiNS^-\circ\varrho_V^2\simeq 
2(\PiNS^+\oplus\PiNS^-)$. 
Our previous results admit
only $\PiNS^+\circ\varrho_V^2\simeq
\PiNS^-\circ\varrho_V^2$ and hence we find
$\sigma\times\sigma=1+\rmv$. Summarizing we
rediscover the WZW fusion rules.
\begin{theorem} 
The DHR sector product reproduces the fusion rules
(\ref{toteven})  and (\ref{odd}).
\end{theorem}

\section{Remarks}
We conclude this chapter with some general remarks
on the presented analysis.

\subsection{The Chiral Ising Model}
Although the analysis of the $\son$ WZW models
requires that $N\ge 7$ our analysis with fermions
also works if one formally sets $N=1$. In this case no
current algebra appears; the chiral algebra (i.e.\ the
unbounded observable algebra) consists just of a
Virasoro algebra with central charge $c=\h$,
  \[ [ L_m,L_n ] = (m-n) L_{m+n} + \Frac{1}{24}\,
  m(m^2-1)\,\del m{-n} ; \]
then only one species of fermions is present so that
  \[ L_m = - \half \sumrZp 
  (r- \sfrac m2) \normord{b_r b_{m-r}} \]
in the Neveu-Schwarz sector, respectively
  \[ L_m = - \half \sumnZ 
  (n - \sfrac m2) \normord{b_n b_{m-n}}
  + \del m0 \,\Frac{1}{16} \, \bfe \]
in the Ramond sector.
Under the action of the Virasoro algebra the 
Neveu-Schwarz sector splits into two different modules
with conformal weights $0$ and $\h$ (the ``basic''
and the ``vector module'') and the Ramond sector splits
into two equivalent modules with conformal weight
$1/16$ (the ``spinor module'' $\sigma$).
Thus the $N=1$ case is just what is called the
chiral Ising model which was first investigated in 
the algebraic framework in \cite{MS1}. Indeed the
analysis using localized endomorphisms of even
CAR algebras reproduces the fusion rules (\ref{odd})
which are called the Ising fusion rules.
This analysis has been carried out in
\cite{leci} and is now included in our more
general setting.

\subsection{Discussion and Outlook}
The main result of this chapter is the proof of the
WZW fusion rules in terms of the DHR sector product.
We believe that this result is remarkable for two reasons.
Firstly, non-trivial CFT models could be incorporated
mathematically rigorously in the DHR framework, and
expected correspondences between CFT and AQFT could be
established. Secondly, the proof is completely independent
of the methods that are conventionally used in CFT
to derive fusion rules. This is noteworthy since methods 
like operator product expansions are not all under
sufficient mathematical control. However, our results are
based on the fact that the $\zet_2$-invariant fermion algebras
are identified to be the bounded observable algebras of
the level $1$ $\son$ WZW models. Therefore difficulties
arise when one tries to generalize this program to
other models. As we will see in the next chapter,
already at level $2$ there is no longer a gauge invariant
subalgebra of the fermion algebra that can be identified as
bounded WZW observable algebra. Also to $\sun$ WZW models
(at arbitrary level) we cannot directly translate this
program.

In order to incorporate other WZW models into the DHR
framework, Wassermann's loop group approach may be the
more suitable one. However, it would be desirable to find
the relevant localized endomorphisms of the operator
algebras in the vacuum representation so that the fusion
rules could be directly derived in terms of the DHR sector
product instead of using Connes fusion.

\chapter{$\son$ Wess-Zumino-Witten Models at Level 2}
In this chapter we tackle the problem to identify the
$\sonh_2$ highest weight modules appearing in the doubled
Neveu-Schwarz sector $\HNSh=\HNS\otimes\HNS$, the ``big
Fock space'', where $\HNS$ is the
Neveu-Schwarz sector of the level $1$ theory. We discuss
the realization of $\sonh\rtimes\Vir$ in $\HNSh$.  Crucial for
our construction is the application of the DHR theory to
a fermionic field algebra acting in the big Fock space; we
introduce the DHR gauge group $\Oz$ which leaves the chiral
algebra invariant. The decomposition of the big Fock space
into the sectors of the gauge invariant fermion algebra
turns out to be helpful for the construction of the
simultaneous highest weight vectors of $\sonh_2$ and the
coset Virasoro algebra $\Vir\coset$. A detailed analysis of
the characters ends up with the complete decomposition
of the big Fock space into tensor products of irreducible
$\sonh_2$ and $\Vir\coset$ highest weight modules.
This analysis is based on \cite{hlwzw}.

\section{The Doubled Neveu-Schwarz Sector}
In this section we introduce the big Fock space $\HNSh$
and the ``doubled'' fermion algebra acting on it. In view of
the fact that the generators of $\sonh\rtimes\Vir$ as well
as those of the Virasoro algebra $\Vir\coset$ of the coset
theory $(\sonh_1\oplus\sonh_1)/\sonh_2$ are 
both invariant under the gauge group $\Oz$
we decompose the big Fock space into the sectors
of the gauge invariant fermion algebra. According to the
results of the DHR theory, the representation theory of the
gauge group determines this decomposition. 

\subsection{The Doubled CAR Algebra}
We are interested in the theory that is obtained by doubling
the Neveu-Schwarz fermions of the type described in Chapter 3.
Thus in addition to the $\son$  index $i$ the fermion modes
will now be labelled by a ``flavor'' index
$q=1,2$. To describe this theory, we define
  \[  \KKh=\KK\oplus\KK\,,\quad \GGh=\GG\oplus\GG
  \quad\mbox{and}\quad
  \PNSh=\PNS\oplus\PNS\,,  \]
or, alternatively,
  \[  \KKh=\KK\otimes\complex^2,\quad \GGh=\GG\otimes\GG_2
  \quad\mbox{and}\quad
  \PNSh=\PNS\otimes \one_2\,,  \]
where $\GG_2$ denotes the canonical complex conjugation in
$\complex^2$. Further, for any $f\iN\KK$ we define the elements
  \[  B^q(f)=B(f\otimes v^q)\,, \qquad q=1,2\,,\]
of $\CKGh$, where $v^q$ denote the canonical unit vectors of $\complex^2$.
We denote by $(\HNSh,\PiNSh,\ONSh)$ the GNS representation
associated to the Fock state $\omega_{\PNSh}$ of $\CKGh$. 
We then define the Fourier modes
  \be  \b iqr = \PiNSh (B^q(e_r^i)) \labl{biqr}
for $i=\onetoN$, $q=1,2$ and $r\iN{\mathbb Z}+\frac12$. 
The Fourier modes $\b iqr$ generate a CAR algebra with relations
  \[  \{\b ipr, \b jqs \} \, = \, \del pq \, \del ij
  \, \del {r+s}0 \, {\bf 1}  \]
and $(\b iqr)^*=\b iq{-r}$. The modes $\b iqr$ with positive
index $r$ act as annihilation operators in $\HNSh$, i.e.\ for all $q=1,2$
and all $i=\onetoN$ we have  
  \[  \b iqr\, \ONSh =0 \qquad 
  {\rm for}\ r\in{\mathbb N}_0 +\onehalf \,. \]

\subsection{Realization of $\sonh\rtimes\Vir$ at Level 2}
Given the fermion modes \erf{biqr}, one defines again
normal ordering
  \[ \normord{\b ipr \b jqs} = \left\{ \begin{array}{rl}
  \b ipr \b jqs & \quad r<0 \\ - \b jqs \b ipr & \quad
  r>0 \eear \right. , \qquad r,s \in \zet + \half \,, \]
sums over their normal-ordered bilinears
  \[  \BX ijqm=\half\sum_{r\in\zet+1/2} 
  \normord{\b iqr\b jq{m-r}}\,, \qquad q=1,2  \]
and current operators
  \be \Jm ij=\I \,\sumq \llb \BX ijqm - \BX jiqm \lrb  \,, \labl J
for $i,j=\onetoN$. Quite analogous to the situation in $\HNS$,
all the unbounded expressions we introduce here possess an
invariant dense domain $\HNShf\subset\HNSh$ spanned by finite
energy vectors. One checks by direct computation that
  \[  [\Jm ij,\b kqr]=\I \,(\del jk\b iq{r+m}-\del ik\b jq{r+m}) 
  \,,  \]
and 
  \be  \bearl [\J ijm,\J kln]=\I \,( \del jk \J il{m+n} 
  + \del il \J jk{m+n} - \del jl \J ik{m+n} - \del ik \J jl{m+n} ) + \\[.7em]
  \qquad\qquad\qquad +   2 \, m \, \del m{-n} \, 
  ( \del ik \del jl - \del il \del jk ) \,. \eear \labl{JJ}
According to \erf{JJ} (compare also \erf{ibf}), the $\Jm ij$ with $i<j$ 
provide a basis for the affine \lie\ $\sonh$ at fixed value $\kv=2$ 
of the level. That the level of $\sonh$ has the value 2 is of course 
a consequence of the summation over two species of fermions in \erf J;
while for a single fermion we obtained  the \lie\ $\sonh$ at level 1 we
now observe that the $\J ijm$ correspond via second quantization to
operators $\ttau ijm \otimes \one_2$ on $\KKh$, so that the Schwinger 
term (\ref{cP}) is doubled now.

Recall the Chevalley basis of the affine \lie\ $\sonh_2$;
the \csa\ generators are $\HH j= \Jo{2j-1}{2j}$
for $j=\onetol$, and the Chevalley generators $\EE j\pm$  are given by
  \[  \bearl
  \EE j\pm=\pm\Jot j{j+1}\pm\mp \qquad {\rm for}\ j=\onetolme\,, \nline7 
  \EE0\pm= \pm\Jt{\pm1}12\mp\mp \,, \qquad
  \EE\el\pm = \left\{ \bearll \pm\Jot{\el-1}\el\pm\pm & \forzl \,, \nline7
       \pm J_0(\ttpm \el) & \forzle\,, \eear\right.  \eear \]
where 
  \[  \bearl
  \Jmtee ij= \half(\eps\Jm{2i}{2j-1}+\eta\Jm{2i-1}{2j}) + \halfi
  (\Jm{2i-1}{2j-1}-\eps\eta\Jm{2i}{2j}) \,, \nline7 
  \Jmte j= -\fsqz (\eps\Jm{2j-1}{2\el+1} - \I \Jm{2j}{2\el+1})  
  \eear \]
for $i,j=\onetol$ and $\eps,\eta=\pm1$.

The generators of the associated Virasoro \alg, i.e.\ the 
Laurent modes of the stress energy tensor of the \wzwt, 
will be denoted by $L_m$. In our particular case, Sugawara's
formula reads
  \[ L_m = \Frac1{2N} \sum_{1\le i<j\le N} 
  \sum_{n\in\zet} \normord{\J ijn \J ij{m-n}} \]
where the normal ordering of the current operators is defined by
  \[ \normord{\J ijm \J ijn} = \left\{ \bearll
  \J ijm \J ijn & \qquad m<0 \\ \J ijn \J ijm &
  \qquad m\ge 0\,. \eear \right. \]
This Virasoro algebra has central charge $c=N-1$. 
Also, we denote by $\LNS m$ the Laurent components of the canonical 
stress energy tensor of the fermion theory in the big Fock space
(i.e. the Sugawara operators associated to the semisimple
Lie algebra $\son\oplus\son$, compare Chapter 1, Subsection 1.2.3),
  \be  \LNS m = L_m^{(1)} + L_m^{(2)} 
  \qquad \mbox{with}\quad L_m^{(q)} 
  = -\half \sum_{i=1}^N \sum_{r\in\zet+1/2} 
  \! (r-\Frac{m}{2})\normord{\b iqr\b iq{m-r}} \,. \labl{LNS}
Thus in particular 
  \[ L_0^{(q)} = \sum_{i=1}^N \sum_{r\in\natnumo+1/2} \! 
  r\, \b iq{-r}\b iqr \,. \]
Note that the $\LNS m$ correspond via second quantization to
operators $\lambda_m\otimes\one_2$ on $\KKh$ but not the
$L_m$. Although the $\LNS m$ satisfy the same commutation
relations with the current operators as the Sugawara operators
$L_m$ do,
  \[ [ \LNS m ,\J ijn ] = [ L_m ,\J ijn ] =
  -n \, \J ij{m+n} \,, \]
they generate a Virasoro algebra with central charge $c\ns=N$.
This implies that the coset Virasoro operators 
  \[  L\coset_m=\LNS m - L_m  \]
commute with the current operators $\J ijn$ and
generate the coset Virasoro algebra $\Vir\coset$ with
central charge $c\coset=c\ns-c=1$.

\subsection{DHR Theory with Gauge Group $\Oz$}
The group $\Oz$ is generated by $\GLZ$ matrices
  \[ \sgamt = \left( \begin{array}{rr} \cost & -\sint \\
  \sint & \cost \end{array} \right), \quad t\iN\reals\,,\qquad
  \seta = \left( \begin{array}{rr} 1 & 0 \\ 0 & -1 
  \end{array} \right). \]
Correspondingly, we define Bogoliubov 
operators in $\QKGh$,
  \[ U(\sgamt) = \bfe \otimes \sgamt \,, \qquad
  U(\seta) = \bfe \otimes \seta \]
acting on $\KKh=\KK\otimes\complex^2$; they induce Bogoliubov
automorphisms $\rsgamt$, $\rseta$, respectively. These 
automorphisms fulfill
  \[ \bearl \rsgamt (B^1(f)) = \cost B^1(f) - \sint B^2(f) \,, \\[.3em]
  \rsgamt (B^2(f)) = \sint B^1(f) + \cost B^2(f) \eear \]
and
  \[  \rseta (B^1(f))= B^1(f)\,,\qquad \rseta 
  (B^2(f))= - B^2(f)\,. \]
The invariance of the Fock state $\omega_{\PNSh}$ reads now
  \[ \omega_{\PNSh} \circ \rsgamt = \omega_{\PNSh}
  = \omega_{\PNSh} \circ \rseta \,\,, \]
and hence there is a unitary (strongly continuous) 
representation $Q$ of $O(2)$ by certain implementers
$Q(\sgamt)\equiv Q_{\PNSh} (U(\sgamt))$ and 
$Q(\seta)\equiv Q_{\PNSh} (U(\seta))$
in $\mathfrak{B}(\HNSh)$ which satisfy
  \[  Q(\sgamt)\, \ONSh = \ONSh = Q(\seta)\, \ONSh \,, \]
and the action of $\rsgamt$ and $\rseta$ extends 
to $\mathfrak{B}(\HNSh)$. 

The inequivalent \findim\ \irrep s of $\Oz$  are the following.
Besides the identity $\rpo$ with $\rpo(\cdot)=1$ and another \onedim\ \rep\
$\rpj$ with
  \be  \rpj(\sgamt)=1\,, \qquad \rpj(\seta)=-1  \,, \labl{repJ}
there are only \twodim\ \rep s $\rp m$ with $m=1,2,...$\,;
their \rep\ matrices are
  \be  \rp m(\sgamt)= \left( \begin{array}{cc} 
  \E^{\I mt}&0\\0&\E^{-\I mt} \end{array} \right),
  \qquad \rp m(\seta)= \left( \begin{array}{cc} 0&1\\1&0
  \end{array} \right).  \labl{repm}
The tensor product decompositions of these \rep s read
  \be  \bearl  \rpj\times\rpj=\rpo\,, 
  \qquad\quad \rpj\times\rp m=\rp m\,,  \nline8
  \rp m\times\rp n = \rp{|m-n|} + \rp{m+n} 
  \qquad{\rm for}\ m\ne n\,, \nline6
  \rp n\times\rp n = \rpo + \rpj 
  + \rp{2n} \,.  \eear \labl{tp}
Field and observable algebras of the fermion theory are described as follows.
Choose a point $\zeta\iN S^1$ on the circle and denote by $\Jz$ the
set of those open intervals $I\subseT S^1$ whose closures do not contain
$\zeta$. For $I\iN\Jz$ let $\KK (I)$
be the subspace of functions having support in $I$. Correspondingly,
define $\KKh (I)=\KK (I)\otimes\complex^2$. The local field 
algebras $\fF (I)$ are then defined to be the von Neumann algebras 
  \[ \fF (I) = \PiNSh (\CKGIh)''\,, \]
and the global field algebra $\fFg$ is the $C^*$-algebra that is defined
as the norm closure of the union of the local algebras,
  \[ \fFg= \overline{\bigcup_{I\in\Jz} \fF (I)}\,. \]
The group $\Oz$  acts on the field algebra as a gauge group in the sense
of Doplicher, Haag and Roberts \cite{DHR1}. This a subgroup of the
automorphism group of $\fF (I)$ \resp\ $\fFg$ such that the observables
are precisely the gauge invariant fields (compare Chapter 1,
Subsection 1.1.1). Therefore the local 
observable algebras $\fA (I)$ and the global (or quasi-local)
observable algebra $\fAg$ are defined as $\Oz$-invariant
part of the field algebras,
  \[ \fA (I) = \fF (I) \cap Q(\Oz)' \]
and
  \[ \fAg= \overline{\bigcup_{I\in\Jz} \fA (I)}\,. \]
At level 2 the algebra $\fAg$ does not coincide with the
observable algebra $\AW$ of bounded operators which is
associated to the WZW theory. Indeed we will see that 
each irreducible $\fAg$-sector is highly reducible 
under the action of the observable algebra $\AW$. Nevertheless,
owing to $\AW\subset\fAg$ the \rep\ theory of $\fA$ turns out
to be crucial for our analysis of the
decomposition of the big Fock space into tensor products of
\hwm s of the level 2 chiral algebra and of the coset Virasoro algebra. 

For the construction of the \hwv s 
within the $\fAg$-sectors it is convenient to work with 
the unbounded operators of $\sonh$ (instead of the bounded
elements of $\AW$) and of the Virasoro \alg\ that is
associated to $\sonhz$ by the Sugawara formula.

The Bogoliubov automorphisms act as rotations on the flavor
index $q$ of the fermions. As a consequence, they leave 
expressions of the form
  \[ \sum_{q=1}^2 \, B^q(f)\, B^q(g)\,\qquad (f,g \in \KK) \]
invariant. In particular, it can be easily read off their
definition that the current operators $\J ijm$ remain
invariant under the action of the gauge group $\Oz$. Similarly,
owing to the summation on $q$ in the bilinear expression
\erf{LNS}, the Virasoro generators $\LNS m$ are $\Oz$-invariant,
too. This implies that the coset Virasoro operators $L\coset_m$
are gauge invariant as well. Therefore neither the current operators
of $\sonh_2$ nor the elements of $\Vir\coset$ make transitions
between the sectors of $\fAg$ (and hence in particular 
$\AW\subset\fAg$). For the decomposition of the big
Fock space $\HNSh$ into their (highest weight) modules it may 
thus be helpful to decompose $\HNSh$ first into the sectors of 
$\fAg$. Employing the results of \cite{DHR1,DRu},
we arrive at
  \[ \HNSh = \Hho \oplus \Hhj \oplus
  \bigoplus_{m=1}^\infty (\Hhm \otimes H_{[m]}) \,. \]
Here $\Hho,\,\Hhj$ and $\Hhm$ carry mutually inequivalent
irreducible representations of $\fAg$; vectors 
in $\Hho,\Hhj$ transform according to the two inequivalent \onedim\ \irrep s
$\rpo$ and $\rpj$ of the gauge group $\Oz$, respectively, and
the $H_{[m]}\simeq\complex^2$ carry the inequivalent \twodim\ irreducible
$\Oz$-representations $\rp m$. Later we will also use the notation
  \[  \Hhm\otimes H_{[m]}=\Hhm^+\oplus\Hhm^- \,, \]
where by definition, $Q(\sgamt)$ acts on $\Hhm^\pm$ as
multiplication with $\E^{\pm\I mt}$.

At level 1 we have $\HNS=\Hheo\oplus\Hhev$, and hence 
at level 2 we can write
  \be \HNSh = (\Hheo\otimes\Hheo) \oplus (\Hheo\otimes\Hhev)
  \oplus (\Hhev\otimes\Hheo) \oplus (\Hhev\otimes\Hhev) \,.  \labl{HNSh}
The four summands in this decomposition can be characterized as the common 
eigenspaces \wrt the ``fermion flips'' $Q(\sgam_\pi \seta)$ and $Q(\seta)$,
namely those associated to the pairs $(1,1)$, $(1,-1)$,
$(-1,1)$ and $(-1,-1)$ of eigenvalues, \resp. By comparison with the
action (\ref{repJ}) and (\ref{repm}) of $\Oz$  on the $\fA$ sectors,
it follows that we can decompose the tensor 
products appearing in \erf{HNSh} as
  \be \bearl
  \Hheo\otimes\Hheo = 
  \Hho \oplus \dstyle\bigoplus_{n=1}^\infty 
  \Hh {[2n]} \,, \qquad   \Hhev\otimes\Hhev = 
  \Hhj \oplus \dstyle\bigoplus_{n=1}^\infty \Hh {[2n]} 
  \,,\\{}\\[-.8em] \qquad \qquad
  \Hheo\otimes\Hhev = \dstyle\bigoplus_{n=0}^\infty 
  \Hh {[2n+1]} = \Hhev\otimes\Hheo \,.  \eear \labl{Hdec}
Later we will employ the \rep\ theory of the gauge group $\Oz$, and in
particular the decomposition \erf{Hdec}, to obtain also simple formulae 
for the characters of the level $2$ modules in the big Fock space.
As further input, we will need some information about 
the relevant coset \cfts.

\section{Highest Weight Vectors}
Recall that a \hwv\ $\Phila$ of $\sonhz$ with \hw\ 
$\Lambda$ is characterized by the
following properties. Firstly, it is annihilated by the step 
operators associated to the horizontal positive roots,
i.e.~for $1\le i<j\le\el$ and $\eps=\pm1$ one has
  \[  \Jot ij+\eps\,\Phila = 0\,, \qquad \mbox{and also } \quad
  J_0 (\ttp k) \, \Phila = 0\;\ \forzle \,;\]
secondly, it is also annihilated by the step 
operators with positive grade,
i.e.~for $m>0$, $i,j=\onetol$ and $\eps,\eta=\pm1$ it satisfies
  \[  \Jmt ij\eps\eta\,\Phila = 0\,,  \qquad \mbox{and also } \quad
  J_m (\tte k) \, \Phila = 0\;\ \forzle \,; \]
(note that the above conditions are equivalent to the
requirement $\EE j+\,\Phila = 0$, $j=\otol$)
and thirdly, $\Phila$ is an eigenvector of the \csa,
  \[ \HH k \, \Phila = \Lambda^k \, \Phila  \]
for $k=\onetol$. 

We will exploit the decomposition of $\HNSh$ into irreducible $\fA$ sectors 
to identify the \hwv s of $\sonhz$. Indeed, in each sector $\Hho$, $\Hhj$ and
$\Hhm^\pm$ we find distinguished states which are \hwv s for both 
$\sonhz$ and the coset Virasoro algebra. 

\subsection{The Combinations $\xpm jr$ and $\xbpm jr$}
For the construction of the simultaneous highest weight vectors of
$\sonh_2$ and $\Vir\coset$ it is convenient to introduce new
creation and annihilation operators in terms of linear
combinations of the $\b iqr$. We define
  \[   \xpm jr = \fsqz (\cp jr \pm \I\cbp jr) \,, \qquad
  \xbpm jr =   \fsqz (\cm jr \pm \I\cbm jr) \,, \]
for $j=\onetol$, where
  \[  \cpm jr = \fsqz (\b{2j}1r\pm\I\b{2j-1}1r) \,, \qquad 
  \cbpm jr = \fsqz (\b{2j}2r\pm\I\b{2j-1}2r) \,, \]
and also, $\forzle$,
  \[  \xbpm{\el+1}r = \fsqz ( \b{2\el+1}1r \pm
  \I \b{2\el+1}2r ) \,. \]
Further, we set
  \[  \bearll  \Xpm jr=\xpm jr\,\xpm{j-1}r\cdots \xpm1r \qquad
  & {\rm for}\ j=\onetol\,, \nline7 
  \Xbpm jr=\xbpm{j+1}r\,\xbpm{j+2}r\cdots \xbpm\el r \quad
  & {\rm for}\ j=\otolme \,, \eear \]
and $\Xbpm \el r=\bfe$.
By direct calculation, we obtain
  \be  [\HH j,\xpm kr]=\del jk\,\xpm kr \,, \qquad [\HH j,\xbpm kr]=
  -\del jk\,\xbpm kr \,,   \labl{Hx}
for all $j,k=\onetol$, and similarly, $\forzle$,
  \be  [\HH j , \xbpm {\el+1}r] = 0  \labl{Hxl}
for all $j=\onetol$. To find also the commutators of the 
fermion modes with the raising operators
$\EE j+$, we first compute
  \[  [\Jmtee ij,\cpm kr] = \half\eps\,(\eta\mp1)\,\del jk\,\ceps i{m+r}
  - \half\eta\,(\eps\mp1)\,\del ik\,\ceta j{m+r}  \,. \]
Analogous relations hold for $[\Jmtee ij,\cbpm kr]$. 
When $N=2\el+1$ we have in addition the relation
  $[ \Jmtee ij , \b {2\el+1}qr ] = 0$
and 
  \[ \bearl
  [\Jmttp j,\cpm kr] = \mp \,\del jk \, \b{2\el+1}1{m+r} \,, \qquad \
  [\Jmttm j,\cpm kr] = 0 \,, \\[.5em]
  [\Jmttpm j,\b {2\el+1}1r] = - \, \cpm j{m+r} \,,  \eear  \]
and similar relations for $[\Jmttp j,\cbpm kr]$,
$[\Jmttm j,\cbpm kr]$ and $[\Jmttpm j,\b{2\el+1}2r]$.
From these results we learn that
  \[  \bearl
  [\EE 0+,\xpm kr] = \del k2\, \xbpm 1{r+1} 
  - \del k1\, \xbpm 2{r+1}\,, \qquad 
  [ \EE 0+ , \xbpm kr ] = 0\,, \\{}\\[-.5em]
  [\EE j+,\xpm kr] = -\del k{j+1}\, \xpm jr\,, \quad 
  [\EE j+,\xbpm kr]= \del kj\, \xbpm {j+1}r \quad{\rm for}\ j=\onetolme \,, 
  \\{}\\[-.6em]
  [\EE \el+,\xpm kr] = 0\,, \quad [\EE \el+,\xbpm kr] = \left\{ \bearll
  \del k\el\, \xpm {\el-1}r-\del k{\el-1}\, \xpm\el r & \forzl\,, \nline8
  \del k\el\, \xbpm{\el+1}r-\del k{\el+1}\, \xpm\el r & \forzle \,. 
  \eear\right.  \eear \]
Taking into account that $(\xpm jr)^2 = (\xbpm jr)^2 = 0$ and $(\xpm jr)^* = 
\xbmp j{-r}$, these relations imply that
  \be  \bearl [\EE j+,\Xpm kr] = 0 \qquad {\rm for}\,\, j=\onetol \,, \\[.5em]
  [ \EE j+,\Xbpm kr ] = 0  \qquad {\rm for}\,\, j=\onetolme\,.\eear \labl{EX}
For $j=\el$ we have instead
  \be \bearl
  [\EE\el+,\Xbpm kr]\cdot\Xpm\el r = 0 \qquad \forzl\,, \\[.7em]
  [\EE\el+,\Xbpm kr]\cdot\xbpm{\el+1}r = 0 \,, \qquad 
  [\EE\el+,\xbpm{\el+1}r]\cdot \Xpm\el r = 0 
  \quad {\rm for}\ N=2\el+1 \,. 
  \eear \labl{ElXb}
Finally, for $j=0$ we find
  \be  [\EE0+,\Xbpm kr] = 0 \,, \qquad
  [\EE0+,\Xpm kr] \cdot \Xbpm0{r+1} = 0  \,. \labl{E0X}

\subsection{Simultaneous Highest Weight Vectors of 
$\sonhz$ and $\Vir\coset$}
Now we are in a position to define a lot of vectors which
will be proven to be \hws s for both, the affine \lie\ $\sonh_2$
and the coset Virasoro algebra.
\begin{definition}
For $n=0,1,2,...$ we set
  \be  \Omnpm{j}n = \Xpm j\mnh\,\Oonpm n \,,\qquad 
  {\rm for} \,\,j=\onetol \,, \labl{omnlj}
as well as
  \be  \Ombnpm jn = \left\{ \bearll \Xbpm j\mnh \Xpm\el\mnh \,
  \Oonpm n & \forzl  \,,   \, j=\onetolme \,,
  \nline6    \Xbpm j\mnh \xbpm{\el+1}\mnh \Xpm\el\mnh \,
  \Oonpm n & \forzle \,, \, j=\onetol \,,  \eear\right. \labl{ombmlj}   
where 
  \be \Oonpm{n+1} = \left\{\bearll \Xbpm0\mnh \Xpm\el\mnh\,
  \Oonpm n & \forzl
  \,, \nline6    \Xbpm 0\mnh \xbpm{\el+1}\mnh \Xpm\el\mnh \,
  \Oonpm n & \forzle \,, \eear\right. \labl{Ombmlj}   
recursively, and
  \be  \Oonpm 0 \equiv \Omo = \ONSh  \,. \ee
Further, we set
  \be \bearl
  \pm \Ovo \equiv \Ovvo = \x 1\mh \y 1\mh\,\Omo  \,, \\[.8em]
  \Ovnpm n = \xpm 1\mnh \xmp 1\nmh\,\Oonpm n \,, \quad
  n=1,2,...\,, \eear \labl{omnv}
and, for $N=2\el$,
  \be  \Osnpm n = \Omnpm\el n \,,  \qquad
  \Ocnpm n = \xbpm \el\mnh \xbmp\el\nh\,\Osnpm{n} \,.\labl{oi}
\lablth{hws}
\end{definition}
Let (compare Tables \ref{t1} and \ref{t2})
  \[  \lj j = \left\{ \begin{array}{cl}
  \Lj j & {\rm for}\ j=\onetolmz\ \;{\rm or}\;\ j=\el-1,\;N=2\el+1 \,,\nline7
  \Lj{\el-1}+\Lj\el & {\rm for}\ j=\el-1,\;N=2\el \,,\nline7 
  2\Lj\el           & {\rm for}\ j=\el,\;N=2\el+1 \,, \eear \right. \]
with the fundamental weights $\Lj i$ as defined in 
Chapter 2, Subsection 2.2.1, while $\lo=0$, $\lv=2\Lj1$,
and, $\forzl$, $\ls=2\Lj\el$, $\lc=2\Lj{\el-1}$.
We now claim
\begin{theorem}
For $n=0,1,2,...$ the vectors of Definition \ref{hws}
$\Oonpm n$, $\Ovnpm n$,
$\Omnpm jn$ and $\Ombnpm jn$, $j=\onetolme$, are \hws s of
$\sonhz$ with \hw s $\lo$, $\lv$, $\lj j$ and $\lj j$, respectively;
for $N=2\el$ the vectors $\Osnpm n$ and $\Ocnpm n$ are \hws s
with \hw s $\ls$ and $\lc$, respectively, and for $N=2\el+1$
the vectors $\Omnpm \el n$ and $\Ombnpm \ell n$ are \hws s
with \hw s $\lj \el$.
\lablth{hwst}
\end{theorem}
\bproof Firstly, we have to show that all these vectors are
annihilated by $\EE j+$ for $j=\otol$. This 
can easily be checked by inserting 
the results \erf{EX}\,--\,\erf{E0X} for the commutators between the step 
operators $\EE j+$ and the operators $\Xpm kr$, $\Xbpm kr$ into
the definitions of these states. The least trivial case occurs 
for $\EE 0+$, where one employs the first of the
identities \erf{E0X}; one then has to commute $\xbpm1\h$ and $\xbpm2\h$,
to the right and use $\xbpm1\h\Omo=0=\xbpm2\h\Omo$ when $n=0$, 
while for $n>0$ one also must employ the second identity in \erf{E0X}.
Secondly, we have to show that the states defined in Definition
\ref{hws} are eigenvectors of all \csa\ generators
$\HH k$ ($k=\onetol$). More precisely, from the commutation
relations \erf{Hx} and \erf{Hxl} it follows rather 
directly that 
  \[  \bearll
  \HH k\,\Omnpm jn= (\lj j)^k\,\Omnpm jn 
  & {\rm for}\ j=\onetol\,,
  \\[.5em] \HH k\,\Ombnpm jn = (\lj j)^k\,\Ombnpm jn   
  & {\rm for}\ j=\onetolme \eear \hsp{4.4} \]
and
  \[  \bearll  
  \HH k\,\Oonpm n =(\lo)^k\,\Oonpm n \,, \hsp{2.2} &
  \HH k\,\Ovnpm n =(\lv)^k\,\Ovnpm n \,, \nline7 
  \HH k\,\Osnpm n =(\ls)^k\,\Osnpm n \,, &
  \HH k\,\Ocnpm n =(\lc)^k\,\Ocnpm n \,, \eear \]
the theorem is proven. \eproof
We further claim
\begin{theorem}
For $n=0,1,2,...$ the vectors of Definition \ref{hws}
$\Oonpm n$, $\Ovnpm n$ and $\Omnpm jn$, $\Ombnpm jn$,
$j=\onetolme$, are \hws s of the
coset Virasoro algebra $\Vir\coset$ with coset conformal weights
  \be \Delc \circ = \Delc \rmv = \Frac{n^2N}2 \,, \labl{ccwa}
and
  \be \bearll\Delc j = \Frac1{2N} (nN+j)^2 \,,\qquad & 
  j=\onetolme\,, \nline6 
  \Delcb j = \Frac1{2N} ((n+1)N-j)^2 \,, & j=\onetolme\,, 
  \eear \hsp{.9} \labl{ccwj}
respectively; for $N=2\el$ the vectors $\Osnpm n$ and $\Ocnpm n$ 
are \hws s with coset conformal weights
  \be \Delc \rms = \Delc \rmc = \Frac1{2N} (nN+\el)^2\,,  \labl{ccws}
respectively, and for $N=2\el+1$ the vectors $\Omnpm \el n$,
$\Ombnpm \ell n$ are \hws s with coset conformal weights
  \be \Delc \el = \Frac1{2N} (nN+\el)^2 \,,\qquad 
  \Delcb \el = \Frac1{2N} ((n+1)N-\el)^2 \,, \labl{ccwb}
respectively.
\end{theorem}
\bproof As a consequence of Theorem \ref{hwst} the vectors
(\ref{hws}) are highest weight states of the Sugawara
Virasoro algebra. Hence we have to show that $\LNS m$ with 
$m>0$ annihilates these states, which is a consequence of 
  \[ [ \LNS m ,\xpm jr ] = - (r+\Frac{m}2) \,\xpm j{r+m} \,,\qquad
  [ \LNS m ,\xbpm jr ] = - (r+\Frac{m}2) \,\xbpm j{r+m} \,.\]
In particular we have
  \be [\LNS 0,\xpm ir ] = -r \, \xpm ir \,, \qquad 
  [\LNS 0,\xbpm ir ] = -r \, \xbpm ir \,. \labl{Lx}
From these relations we also deduce that
  \[ \LNS 0 \, \Omnpm jn = \Delns j \, \Omnpm jn \,, \]
with conformal weights
  \[ \Delns j = \big[ \half + \Frac32 + \ldots + \lLb n-\half \lRb
  \big] N + \lLb n+\half \lRb j = \Frac{n^2N}2 + \lLb n+\half \lRb j \]
for $j=\onetol$. Similarly,
  \[ \LNS 0 \, \Ombnpm jn = \Delnsb j \, \Ombnpm jn \,, \qquad
  \Delnsb j = \Frac{(n+1)^2N}2 - \lLb n+\half \lRb j \,,\]
for $j=\onetol$. Also, for the sectors labelled by $\circ,\,\rmv,\,\rms$
and $\rmc$ we find
  \[ \Delns \circ = \Frac{n^2N}2 \,,\qquad
  \Delns \rmv = \Frac{n^2N}2+1\,,\qquad
  \Delns \rms = \Delns \rmc = \Delns \el \,. \]
Furthermore, the conformal weights of the vectors \erf{omnlj}\,--\,\erf{oi}
\wrtt Virasoro algebra of the level $2$ \wzwt\
follow immediately from the $\son$-weights $\Lambda$ by 
the Sugawara formula for the Virasoro generator $L_0$.
This yields the conformal weights that were already listed in the
Tables \ref{t1} and \ref{t2}. By comparison of these conformal 
dimensions with the ones obtained above, the proof is 
completed. \eproof

Since the affine \lie\ $\sonhz$  and the coset Virasoro \alg\ commute,
it follows immediately that further \hwv s of $\sonhz$  are 
obtained when acting with the lowering operators of the coset 
Virasoro \alg\ on the vectors
\erf{omnlj}\,--\,\erf{oi}. For example,
applying the coset Virasoro operator $L_{-1}^{\rm c}$ to the 
\hwv\ $\Omp 1$ we get the \hwv\ (computed for the case $N=2 \el$)
  \[  L_{-1}^{\rm c} \Omp 1 = \Frac1N\, \LLb \x1{-3/2} \Omo 
  + \sum_{k=1}^{\el}
  (\xb k\mh \y k\mh - \yb k\mh \x k\mh) \Omp 1 \LRb  \]
of $\sonhz$. 

Also note that by construction the tensor product module, and hence each of
its submodules, is unitary. Thus in particular the \hwm s that are obtained
by acting with arbitrary polynomials in the lowering operators $\EE j-$ 
on the \hwv s are unitary, and hence are fully reducible.

\subsection{$\Oz$-Transformation Properties}
There is an interesting association of the $\sonh_2$
highest weight modules appearing in $\HNSh$ to the
sectors of $\fAg$ labelled by the spectrum of the gauge
group $\Oz$. This becomes apparent from the 
$\Oz$-transformation properties of the highest weight
vectors of Definition \ref{hws}.

For the Fourier modes $\cpm jr$ and $\cbpm jr$ 
the action of $\rsgamt$, $t\iN\reals$,
and of $\rseta$ read (recall that the action of these
Bogoliubov automorphisms extends to $\mathfrak{B}(\HNSh)$)
  \[  \bearll
  \rsgamt(\cpm jr)=\cost \cpm jr - \sint \cbpm jr \,,\quad &
  \rseta(\cpm jr)=\cpm jr\,, \nline5
  \rsgamt(\cbpm jr)=\sint \cpm jr + \cost \cbpm jr \,, &
  \rseta(\cbpm jr)=-\cbpm jr  \,, \eear \]
so that the combinations $\xpm jr$ transform as
  \[  \rsgamt(\xpm jr) = \E^{\pm \I t}\,\xpm jr \,,\quad
  \rseta(\xpm jr) =\xmp jr \,.  \hsp1 \]
Analogously,
  \[  \rsgamt(\xbpm jr) = \E^{\pm\I t}\,\xbpm jr \,,\quad
  \rseta(\xbpm jr) =\xbmp jr\,. \hsp1 \]
Hence the combinations $\Xpm jr$ transform as
  \[  \rsgamt(\Xpm jr) = \E^{\pm\I jt}\,\Xpm jr \,,\quad
  \rseta(\Xpm jr) =\Xmp jr\,, \hsp1 \]
and analogously,
  \[  \rsgamt(\Xbpm jr) = \E^{\pm\I(\el-j)t}\,\Xbpm jr \,,\quad
  \rseta(\Xbpm jr) =\Xbmp jr  \,. \hsp1 \]
The vacuum $\Omo$ is $\Oz$-invariant. We then deduce the following
transformations for the vectors of Definition \ref{hws}. For all
$n=0,1,2,...\,$ we have
  \[  Q(\sgamt)\,\Omnpm jn=\E^{\pm\I(nN+j)t}\,\Omnpm jn\,,\qquad\hsp{.9}
  Q(\seta)\,\Omnpm jn=\Omnmp jn  \]
for $j=\onetol$, and
  \[  Q(\sgamt)\,\Ombnpm jn=\E^{\pm\I((n+1)N-j)t}\,\Ombnpm jn\,,\qquad
  Q(\seta)\,\Ombnpm jn=\Ombnmp jn  \]
for $j=\onetol$. Also
  \[  \bearll
  Q(\sgamt)\,\Oonpm n=\E^{\pm\I nNt} \,\Oonpm n\,,\qquad\hsp{.5}
  & Q(\seta)\,\Oonpm n=\Oonmp n \,, \\[.5em]
  Q(\sgamt)\,\Ovnpm n=\E^{\pm\I nNt} \,\Ovnpm n\,,\qquad\hsp{.5}
  & Q(\seta)\,\Ovnpm n=\Ovnmp n \,, \\[.5em]
  Q(\sgamt)\,\Osnpm n=\E^{\pm\I(nN+\el)t}\,\Osnpm n\,,
  & Q(\seta)\,\Osnpm n=\Osnmp n \,, \\[.5em]
  Q(\sgamt)\,\Ocnpm n=\E^{\pm\I(nN+\el)t}\,\Ocnpm n\,,
  & Q(\seta)\,\Ocnpm n=\Ocnmp n \,. \eear  \hsp{.8} \]
We remark that the \hws s $\Ovnpm n$ and $\Oonpm n$, $n=1,2,\ldots$\,,
and for even $N$ also $\Ocnpm n$ and $\Osnpm n$, $n=0,1,2,\ldots$\,,
are connected by $\Oz$-invariant fermion bilinears, i.e.\ by
elements of the intermediate algebra $\fAg$. Explicitly, we have 
  \[  \Ovnpm n = a^n_{\rm v}\, \Oonpm n, \qquad
  a^n_{\rm v} = - (\y1\nmh \x1\mnh + \x1\nmh \y1\mnh)\,, \]
for $n=1,2,\ldots$, and 
  \[  \Ocnpm n = a^n_{\rm c}\, \Osnpm n, \qquad
  a^n_{\rm c} = - (\yb\el\nh \xb\el\mnh + \xb\el\nh \yb\el\mnh) \]
for $n=0,1,2,\ldots$.

\section{Characters}
Owing to the inclusion $\AW\subset\fAg$, the irreducible sectors of
the gauge invariant fermion algebra $\fAg$ constitute modules of the 
observable \alg\ $\AW$ of the \wzwt, which however are typically reducible.
To determine the decomposition of the \irmod s of the intermediate \alg\ 
$\fAg$ into \irmod s of $\AW$ we analyze their characters and combine the
result with the knowledge about the characters of the coset theory.

In the following calculations we directly use the argument 
$q=\exp (2\pi\I\tau)$ instead of the upper complex half plane 
variable $\tau$; so it is always understood that $|q|<1$. Moreover,
we neglect the additional term $-c/24$ in the definition
(\ref{defchar}) which is conventionally added due to simpler
transformation properties with respect to the 
modular group. For our purposes, this modification is not
needed and would only cause several confusing prefactors. Thus
we define the character $\chi (q)$ of a module simply as the
trace of $q^{L_0}$.

\subsection{$c=1$ Orbifolds}
Via the coset construction \cite{goko2}, one associates to any embedding of
untwisted \aff s that is induced by an embedding of their \hsa s another
\cft, called the coset theory. Here the relevant embedding is that of
$\sonhz$  into $\sonhe\oplus\sonhe$; the branching rules of this embedding
are just the tensor product decompositions of $\sonhe$-modules
(compare Chapter 1, Subsection 1.2.3).

The Virasoro algebra of the coset theory is easily
obtained as the difference of the Sugawara constructions of the 
Virasoro algebras of the \aff s. In contrast, the determination of the 
field contents of the coset theory is in general a difficult task (see e.g.\ 
\cite{scya6,fusS4}). But in the case of our interest, the coset theory has
conformal central charge $c=1$, and the classification of (unitary) $c=1$
\cfts\ is well known. In fact, one finds 
(compare e.g.\ \cite{scya5}) that it is a so-called rational
$c=1$ orbifold theory, which can be obtained from the $c=1$ theory of a
free boson compactified on a circle by restriction to the invariants \wrt
a $\zet_2$-symmetry. These \cft\ models have been investigated in \cite{dvvv};
for our purposes we need only the following information.

The rational $c=1$ $\zet_2$-orbifolds are labelled by a non-negative integer 
$M$. The theory at a given value of $M$ has $M+7$ sectors; they are listed in
the following table.
  \begin{table}[ptbh]
  \caption{Sectors of the $c=1$ $\zet_2$-orbifolds}\label{t3}
  \begin{center}
  \begin{tabular}{|c|c|c|} \hline &&\\[-.9em]
  field         & $\Delta$         & \qdi    \\ \hline\hline &&\\[-.9em]
  $\rmo$        & 0                & 1       \\ &&\\[-.9em]
  $\rmv$        & 1                & 1       \\ &&\\[-.8em]
  $\rms$, $\rmc$& $\Frac M4$       & 1       \\ &&\\[-.8em]%\hline &&\\[-.9em]
  $j\ \ \iN\{\onetomme\}$ &$\Frac{j^2}{4M}$ &2 \\ &&\\[-.8em]\hline&&\\[-.9em]
  $\sigma,\,\tau$ & $\Frac1{16}$ & $\sqrt M$   \\ &&\\[-.8em]
  $\sigma',\,\tau'$ & $\Frac9{16}$ & $\sqrt M$ \\[-.8em]&&\\ \hline
  \end{tabular}   \end{center}
  \end{table}
Here we have again separated the fields which correspond to the 
(doubled) \NS sector $\HNSh$ from the fields 
$\sigma,\,\tau,\, \sigma',\,\tau'$ which involve the Ramond sector;
the latter are known as ``twist fields'' of the orbifold theory.

The characters of the fields in the big Fock space are given by
  \be  \chiCj  (q) = \Frac1{\vi(q)}\, \psim j(q) \labl{chiCj}
for $j=\onetom$, where it is understood that
  \[  \chiCs  (q) = \chiCc  (q) = \half \chiCm \,, \]
and by
  \[   \chiCo  (q) = \Frac1{2\vi(q)}\,\llb \psim 0(q) 
  + \psiemq \lrb\,, \quad \chiCv  (q) = \Frac1{2\vi(q)}\,
  \llb \psim 0(q) - \psiemq \lrb\,. \]
Here the functions $\psim j$ are the infinite sums
  \[  \psim j(q)= \sum_{m\in\zet} q_{}^{(j+2mM)^2/4M} \,. \] 
One has \cite[p.\,240]{Kac}
  \be  \psiemq = \summZ(-1)^m\,q^{m^2} = \frac{(\vi(q))^2}{\vi(q^2)} \,. 
  \labl{76}
It follows in particular that
  \be   \chiCo(q) - \chiCv(q) = \frac{\vi(q)}{\vi(q^2)} \,, \labl{chiCovm} 
and
  \be   \chiCo(q) + \chiCv(q) = \frac{\psim 0(q)}{\vi(q)} \,. \labl{chiCovp}

Note that the spectrum of \wzwts\ for even and odd $N$, displayed in Tables
\ref{T1} -- \ref{t2}, is rather similar. However, to obtain the spectrum of
the coset theory also the structure of the conjugacy classes of $\son$-modules
plays an important \role, and these are rather different for even and odd 
$N$.\footnote{Also, for odd $N$ in the Ramond sector an 
additional complication 
arises, namely a so-called fixed point resolution is required
\cite{scya5,fusS4}.}
As a consequence it depends on whether $N$ is even or odd which $c=1$ orbifold
one obtains as the coset theory. Namely, for $N=2\el$ one finds $M=N/2=\el$, 
while $M=2N$ for $N=2\el+1$.

The decomposition of the products of level one characters looks as follows.
For $N=2\el$ we have
  \be \bearl 
  [\chieo ]^2 = \chicol \, \chizo  + \chicvl \, \chizv  
    + \dstyle\sum_{2\le j \le \el \atop j \,\,{\rm even}} \chicjl 
  \, \chizj \,,   \nline7 
  [\chiev ]^2 = \chicol \, \chizv  + \chicvl \, \chizo
    + \dstyle\sum_{2\le j \le \el \atop j \,\,{\rm even}} \chicjl 
  \, \chizj\,,   \nline7 
  \chieo \, \chiev  
  = \dstyle\sum_{1\le j \le \el \atop j \,\,{\rm odd}} \chicjl \, 
  \chizj  \,, \eear \labl{anse}
where it is understood that
  \be  \chizl (q) \equiv \chizs (q) + \chizc (q) \,.  \hsp{6.9} \labl{chizs}
For $N=2\el+1$, the tensor product decomposition reads instead
  \be  \bearl
  [\chieo]^2 = \chicozN \, \chizo + \chicvzN\, \chizv + 
  \dstyle\sum_{2\le j \le \el \atop j \,\,{\rm even}}\! \chicjzN {2j}  
  \, \chizj  + \dstyle\sum_{1\le j \le \el \atop j \,\,{\rm odd}}\! 
  \chicjzN{2N-2j}  \, \chizj  \,, \nline7 
  [\chiev ]^2 = \chicozN \, \chizv + \chicvzN\, \chizo + 
  \dstyle\sum_{2\le j \le \el \atop j \,\,{\rm even}}\! \chicjzN {2j}  
  \, \chizj  + \dstyle\sum_{1\le j \le \el \atop j \,\,{\rm odd}}\! 
  \chicjzN{2N-2j}  \, \chizj  \,, \nline7 
  \chieo\, \chiev = \chicjzN {2N}  \,\llb \chizo  + \chizv  \lrb
  \, + \dstyle\sum_{2\le j \le \el \atop j \,\,{\rm even}}\!
  \chicjzN{2N-2j}\, \chizj  + \dstyle\sum_{1\le j\le\el\atop j\,\,
  {\rm odd}}\! \chicjzN {2j}    \, \chizj  \,.  \eear \labl{anso}
It is worth noting that these formulae can be proven without too much
effort, whereas in general it is a difficult task to write down such 
tensor product decompositions.
Tools which are always available are the matching
of conformal dimensions modulo integers as well as conjugacy class
selection rules, which imply \cite{scya6} so-called field identifications.
In the present case, we can e.g.\ use 
the fact that the sum of conformal weights
$\Delta\kz_j=j(N-j)/2N$ and $\Delta\Coset_k=k^2/4M$ is (for generic $N$)
a half-integer only if $k=j\sqrt{2M/N}$ or $k=(N-j)\sqrt{2M/N}$. Also,
there is a conjugacy class selection rule which implies that the tensor
product of modules in the \NS sector yields only modules which are again in
the \NS sector, and the corresponding field identification tells us e.g.\ that
the branching function $b^{{\rm c};\scriptstyle M}_{\rmv,\rmv;\rmv}(q)$
coincides with $b^{{\rm c};\scriptstyle M}_{\circ,\circ;\circ}(q)=\chiCoM(q)$.

As it turns out, we are even in the fortunate situation that together with the
known classification of unitary $c=1$ \cfts, these informations
already determine the tensor product decompositions almost completely. 
In particular, the value of $M$ of the $c=1$ orbifold is determined uniquely, 
and one can prove that there are not
any further field identifications besides 
the ones implied by conjugacy class selection rules. 
Hence (\ref{anse}) and (\ref{anso}) can be viewed as a
well-founded ansatz, and the remaining ambiguities can be 
resolved by checking various consistency relations 
which follow from the arguments that we will give
in Subsections 4.3.3 and 4.3.4 below. Another possibility 
to deduce \erf{anse} and \erf{anso} is to
employ the conformal embedding of $\sonhz$  into $\unh$ 
at level one \cite{scya5},
which corresponds to regarding the real fermions as real and imaginary parts
of complex-valued fermions.

\subsection{Characters for the Sectors of $\fA$}
The characters of submodules of the space $\HNSh$, i.e.\ the trace of
$q^{L_0}$ over the modules, can be obtained as follows.
Let $\Po, \PJ$ and $\Ppm m$ denote the projections onto $\Hho, \Hhj$
and $\Hhm^\pm$ for $m\iN {\mathbb N}$, respectively.
Then the \rep\ matrices $Q(\sgamt)$ and $Q(\seta\sgamt)$ of $\Oz$
decompose into projectors as
  \[  Q(\sgamt)=\Po+\PJ+\sum_{m=1}^\infty \llb 
  \E^{\I mt}\Pp m+\E^{-\I mt}\Pm m \lrb  \hsp{5.1} \]
and
  \[  Q(\seta\sgamt) = \Po - \PJ +\sum_{m=1}^\infty \llb \E^{\I mt} 
  Q(\seta) \Pp m   +\E^{-\I mt}Q(\seta) \Pm m \lrb \,.  \]
It follows in particular that the projectors can be written as
  \[  \bearl  \Po = \IntT \llb Q(\sgamt)+Q(\seta\sgamt) \lrb \,, \qquad 
  \PJ = \IntT \llb Q(\sgamt)-Q(\seta\sgamt) \lrb \,, \\{}\\[-.4em] 
  \qquad \qquad \Ppm m = \Intt \E^{\mp\I mt} \, Q(\sgamt)
  \qquad{\rm for}\ m\iN {\mathbb N}
  \,.  \eear \]  
For the irreducible $\fA$-sectors in $\HNSh$, the $\Oz$-transformation
properties of the $\xpm ir$ and $\xbpm ir$ together with the action 
of \LNS 0 (compare \erf{Lx}) imply the following. First,
  \[  \bearl  \chio(q) \equiv \trNS \Po\,\qlo = \\[.5em]
  \quad = \dstyle\intT \llb \prod_{m=0}^\infty 
  (1+\eit\qmh)^N_{}(1+\emit\qmh)^N_{} 
  + \dstyle\prod_{m=0}^\infty (1-\qzme)^N_{} \lrb  \,. \eear \]
This can be rewritten as
   \[  \chio(q) = \intT \LLb \frac{\wi(q;-\eit q^{1/2})}{\vi(q)} \LRb^N_{}
   + \Frac12\, \LLb \frac{\vi(q)}{\vi(q^2)} \LRb^N_{} \,, \]
where $\vi$ is Euler's product function \erf{Euler} and
  \[  \wi(q;z)= \prod_{n=1}^\infty \lLb (1-q^n)(1-q^nz^{-1})
  (1-q^{n-1}z)   \lRb \,. \]
Using also the identity \cite[p.\,240]{Kac}
  \[  \wi(q;z) = \sum_{n\in\zet} (-1)^n\,q^{n(n-1)/2}z^n\,, \]
we finally arrive at 
   \be  \chio(q) = \frac{\sN 0}{2\,(\vi(q))^N}
   + \frac{(\vi(q))^N}{2\,(\vi(q^2))^N} \,, \labl{cnso}
where we introduced the functions
   \be \sN m = \!\! \summN m q_{}^{(m_1^2+m_2^2+...+m_N^2)/2}
   \;\equiv  \sumMN m q_{}^{\vecm^2/2} \, \labl S
for $m\iN\zet$. 

Analogously, we find
   \be  \chiJ(q) \equiv \trNS \PJ\,\qlo = \frac{\sN 0}{2\,(\vi(q))^N}
   - \frac{(\vi(q))^N}{2\,(\vi(q^2))^N} \, \labl{cnsJ}
and
  \be  \chim m(q) \equiv \trNS \Ppm m\,\qlo
   \!\!  = \displaystyle \intt  \E^{\mp\I mt} 
   \LLb \frac{\wi(q;-\eit q^{1/2})}{\vi(q)} \LRb^N_{}\, 
   = \frac{\sN m}{(\vi(q))^N} \labl{chimNS}
for $m\iN\natnum$.
(Note that the latter result does not depend on whether $\Pp m$ or $\Pm m$
is used, since $\sN m = \sN {-m}$.)

Expressing the integer $m$ either as $ m=nN+j$ or as $m=(n+1)N-j$ with
$1\le j\le\el$, by shifting the summation indices we obtain the relation
  $\sN {nN+j} = q^{nj+n^2N/2}_{}\, \sN j$.
Hence we have
  \be  \chim{nN+j}(q) = q^{nj+n^2N/2}_{}\, \chim j(q) \,; \labl{n+}
in the same manner we obtain
  \be  \chim{(n+1)N-j}(q) = q^{-(n+1)j+(n+1)^2N/2}_{}\, 
  \chim{j}(q) \,.\labl{n-}
For $j=0$ we have instead
  \[  \chim{nN}(q) = q^{n^2N/2}_{}\, [\chio(q)+\chiJ(q)]  \]
for all $n>0$.
  
\subsection{$\sonhz$ Characters for Even $N$}
When we use the information about the \hwv s \wrt the affine 
\lie\ $\sonh$ at level 2 that we obtained above,
we can derive the characters of the \ihwm s of $\sonhz$  
by comparing the decomposition \erf{Hdec} with the decompositions 
\erf{anse} and \erf{anso}. We first consider the case $N=2\el$. 

By comparison of \erf{Hdec} with \erf{anse} we find 
  \[ \bearll
  \chicjl (q)\, \chizj (q) \!\!&= \chim j (q) + \chim {N-j} (q)
    + \chim {N+j} (q) + \chim {2N-j} (q) + \ldots \\[.3em]
  &\equiv \dstyle\sumni 0 [\chim {nN+j} (q) + \chim {(n+1)N-j} (q)] 
  \eear \]
for even $j$. Using \erf{n+} and \erf{n-}, this becomes
  \[ \bearll  \chicjl (q)\, \chizj (q) \!\!&
  = \chim j (q) \dstyle\sumnZ q^{nj+n^2N/2} \\[.3em]
  &= q^{-j^2/2N}\, \psinh j (q)\,\chim j (q) 
   = q^{-j^2/2N}\, \psinh j (q)\, \Frac{\sN j}{(\vi(q))^N} 
  \,. \eear \]
Analogously, with \erf{anse} we obtain the same result for odd $j$.
By inserting the coset characters $\chicjl$ \erf{chiCj} we then get
  \be \chizj (q) = q^{-j^2/2N}\, \Frac{\sN j}{(\vi(q))^{N-1}}
  \,.  \labl{chizj}
For $j=\el$ one has to read this result with \erf{chizs}, which
means that our result only describes the sum of the irreducible
characters $\chizs$ and $\chizc$. By comparison with \erf{chimNS},
we may also rewrite the result in the form 
  \be \bearll \chim {nN+j} (q) \!\!&= \Frac{q^{(nN+j)^2/2N}}{\vi(q)} \, 
  \chizj (q)\,,  \\{}\\[-.5em]
  \chim {(n+1)N-j} (q) \!\!&=   \Frac{q^{((n+1)N-j)^2/2N}}{\vi(q)} 
  \, \chizj (q) \eear \labl{zvir}
for $j=\onetol$.

Comparing \erf{Hdec} again with \erf{anse}, we also find
  \be \hsp{-.7} \bearl
  \chicol(q)\, \chizo(q) + \chicvl(q)\, \chizv(q) =\\[.7em]
  \qquad = \chio(q)  + \dstyle\sumni 1 \chim {nN} (q) \\[.4em]
  \qquad = [\chio(q) + \chiJ(q)]\, \LLb \half 
    + \half\psinh0 (q) \LRb - \chiJ(q) \\[.7em]
  \qquad = \half[\chio(q)-\chiJ(q)]+\half\psinh0 (q) \,
  [\chio(q)+\chiJ(q)] \\[.6em]
  \qquad = \Frac{(\vi(q))^N}{2\,(\vi(q^2))^N} +
  \psinh0(q) \, \Frac{\sN 0}{2\,(\vi(q))^N} \eear \labl{coo}
and
  \be \bearll \chicol(q)\, \chizv(q) + \chicvl(q)\, \chizo (q) \!\!\!
  &= \chiJ (q)  + \dstyle\sumni 1 \chim {nN} (q) \nline8
  &= - \Frac{(\vi(q))^N}{2\,(\vi(q^2))^N} +
    \psinh0(q) \, \Frac{\sN 0}{2\,(\vi(q))^N}\,. \eear \hsp4 \labl{cov}
Subtraction of \erf{cov} from \erf{coo} yields 
  \[  [\chicol(q) - \chicvl(q)]\cdot[\chizo(q) - \chizv(q)]   
  = \LLb \Frac{\vi(q)}{\vi(q^2)} \LRb^N  \equiv 
  \chio (q) - \chiJ (q)   \,,  \]
so that by inserting \erf{chiCovm} we obtain 
  \be  \chizo(q) - \chizv(q) = 
  \LLb \Frac{\vi (q)}{\vi (q^2)} \LRb^{N-1}  \,. \labl{dif}
Analogously, by adding \erf{coo} and \erf{cov} we get
  \[   [\chicol(q) + \chicvl(q)]\cdot[\chizo(q) + \chizv(q)]  
  = \psinh0(q)\, \frac{\sN 0}{(\vi(q))^N} \,, \]
and hence inserting \erf{chiCovp} we obtain
  \be \chizo (q) + \chizv (q) = \frac{\sN 0}{(\vi(q))^{N-1}} \,. \labl{sum}
In summary, we have derived that
  \be  \bearll 
  \chizo(q) \!\!\!&= \half \Llb \Frac{\sN 0}{(\vi(q))^{N-1}} + 
     \LLb \Frac{\vi (q)}{\vi (q^2)} \LRb^{N-1} \Lrb \\{}\\[-.7em]
  &\equiv \Frac1{2\,(\vi(q))^{N-1}} \,\llb \sN 0 + (\psiemq)^{N-1} \lrb\,,
  \\{}\\[-.5em]
  \chizv(q) \!\!\!&= \half \Llb \Frac{\sN 0}{(\vi(q))^{N-1}} - 
     \LLb \Frac{\vi (q)}{\vi (q^2)} \LRb^{N-1} \Lrb \\{}\\[-.7em]
  &\equiv \Frac1{2\,(\vi(q))^{N-1}}\,\llb \sN 0 - (\psiemq)^{N-1} \lrb\,.
  \eear \labl{both}
Further, comparison with \erf{cnso} and \erf{cnsJ} yields
  \be \chio (q) + \chiJ (q) = \frac1{\vi(q)} \,
  [ \chizo (q) + \chizv (q) ] \,, \labl{cOJ}
while comparison with \erf{chimNS} and \erf{n+} shows that
  \be \chim {nN} (q)  = \frac{q^{n^2N/2}}{\vi(q)} \,
  [ \chizo (q) + \chizv (q) ] \,. \labl{cNSnN} 

\subsection{$\sonhz$ Characters for Odd $N$}
Now we consider the case $N=2\el+1$. From \erf{Hdec} and \erf{anso} we find 
  \[  \hsp{-1.3} \bearll
  \chicjzN {2j} (q)\, \chizj(q) \!\!\!&= \chim j (q) + \chim {2N-j} (q)
    + \chim{2N+j}(q) + \chim{4N-j}(q) +  \ldots \nline7
  &\equiv \dstyle\sumni 0 [\chim {2nN+j} (q) + \chim {2(n+1)N-j} (q)] \nline7
  &= \chim{j}(q) \dstyle\sumnZ q^{2nj+2n^2N}   
   = q^{-j^2/2N}\,\psinz {2j}(q)\, \chim{j}(q) \eear \]
for $j$ even, and 
  \[  \hsp{-.3} \bearll
  \chicjzN {2N-2j} (q)\, \chizj(q) \!\!\!
  &= \dstyle\sumni 0 [\chim{(2n+1)N+j}(q) + \chim{(2n+1)N-j}(q)] \\[.6em]
  &= \chim j (q) \dstyle\sumnZ q^{-(2n+1)j+(2n+1)^2N/2} \nline7
  &= q^{-j+N/2} \chim j (q) \, \dstyle\sumnZ q^{2n(N-j)+2n^2N} \\[.3em]
  &= q^{-j^2/2N} \, \psinz {2N-2j} (q) \, \chim j (q) \eear \]
for $j$ odd.
By inserting the coset characters \erf{chiCj} we then arrive once again at
the formulae \erf{chizj} and \erf{zvir} for $j=\onetol$. 

In the same manner we find
  \[  \hsp{-1.2} \bearl 
  \chicozN(q)\, \chizo(q) + \chicvzN(q)\, \chizv(q) =\\[.3em]
  \qquad = \chio(q)  + \dstyle\sumni 1 \chim {2nN} (q) \nline7
  \qquad = [\chio(q)+\chiJ(q)]\, \LLb \half
  +\half \dstyle\sumnZ q^{2n^2N} \LRb -\chiJ(q) \nline7
  \qquad = \Frac{(\vi(q))^N}{2\,(\vi(q^2))^N} +
  \psinz 0(q) \, \Frac{\sN 0}{2\,(\vi(q))^N} \eear \] 
and
  \[  \bearll  \chicozN(q)\, \chizv(q) + \chicvzN(q)\, \chizo(q) \!\!&
   = \chiJ (q)  + \dstyle\sumni 1 \chim {2nN} (q) \nline7 
  &= - \Frac{(\vi(q))^N}{2\,(\vi(q^2))^N} +
  \psinz 0(q) \, \Frac{\sN 0}{2\,(\vi(q))^N} \,. \eear \] 
Thus we also obtain again the relations \erf{dif} and \erf{sum} for
$\chizo$ and $\chizv$, and hence also \erf{both} and \erf{cNSnN}.

\section{Decomposition of the Tensor Product}
The comparison of the $\son_2$ characters with characters of
the $\fAg$-sectors will now yield the complete decomposition of
the big Fock space into tensor products of irreducible
$\sonh_2$-modules and $\Vir\coset$-modules.

\subsection{Evaluation of the Character Formulae}
Let us now summarize some of our results on the tensor product decompositions. 
To this end we first note that $q^\dh/\vi(q)$ is precisely the character 
of the Verma module $M(c,\dh)$ of the Virasoro \alg. For central charge
$c=1$ the Verma module $M(c,\dh)$ is irreducible as
long as $4\dh\neq m^2$ for $m\iN\zet$; otherwise there exist null
states. The characters of the irreducible modules
$V(1,\dh)$ of the $c=1$ Virasoro algebra are then given by
  \[ \chivir\dh(q) = \left\{ \bearll (\vi(q))^{-1}\,
  [q^{m^2/4}-q^{(m+2)^2/4}] \quad & \mbox{if}\
  \dh=\frac{m^2}4\ {\rm with}\ m\iN\zet\,, \\{}\\[-.8em]
  (\vi(q))^{-1}\,q^\dh & \mbox{otherwise.} \eear \right. \]
Thus for $\dh=m^2/4$ with $m\iN\zet$ the Verma module
character can be decomposed as follows:
  \[ \frac{q^{m^2/4}}{\vi(q)} = \frac1{\vi(q)} 
  \sum_{k=0}^\infty \,\LLb q^{(m+2k)^2/4} - q^{(m+2k+2)^2/4} 
  \LRb = \sum_{k=0}^\infty \chivir {(m+2k)^2/4}(q) \,. \]
Correspondingly we write 
  \be  \Reh \dh = \left\{ \bearll \dstyle\bigoplus_{k=0}^\infty 
  V(1,\Frac{(m+2k)^2}4) \qquad & \mbox{if}\ \dh=\Frac{m^2}4\
  \mbox{with}\ m\in\zet\,, \nline6
  V(1,\dh) & \mbox{otherwise.} \eear \right. \labl{deR}

Using also the formulae \erf{ccwa} and \erf{ccwb} for the coset 
conformal weights, we can summarize our results of Section 4.3
by the following description of the big Fock space $\HNSh$. 
Recalling the decomposition 
  \[ \HNSh = \Hho \oplus \Hhj \oplus
  \bigoplus_{m=1}^\infty (\Hhm \otimes \complex^2)  \]
of $\HNSh$ into $\fAg$-sectors, we can express the splitting of $\HNSh$
into tensor products of the Virasoro modules \erf{deR}
and the irreducible highest weight modules of $\sonhz$  (that is, $\Uo,\,\Uv,\,
\Uj$, and also $\Us$ and $\Uc$ when $N=2\el$) as follows.
Our results show 
\begin{theorem}
For the $\fAg$-sectors $\Hhm$, $m=1,2,...\,$, we have 
  \be \Hhh {nN} = \left[ \Uo\oplus\Uv \right] \,\otimes\,  
  \Reh{\Delc\circ} \,, \labl{decHN}
for $n=1,2,...\,$, as well as
  \be \begin{array}{r}
  \Hhh {nN+j} = \Uj \,\otimes\, \Reh{\Delc j}\,, \nline4 
  \Hhh {(n+1)N-j} = \Uj \,\otimes\, \Reh{\Delcb j}\,\, \eear \labl{decHj}
for $n=0,1,...\,$ and $j=\onetolme$. When $N=2\el+1$, \erf{decHj} also holds 
for $j=\el$, while for $j=\el$ and $N=2\el$ we have
  \be \Hhh {nN+\el} = \left[ \Us \oplus \Uc \right]  \,\otimes\, 
  \Reh{\Delc\rms}  \labl{decHsc}
for $n=0,1,...\,$. The modules $\Reh\dh$ appearing in these 
decompositions are all irreducible as long as $\sqrt{2N}\notin\natnum$.
Otherwise we can write $N=2K^2$ with $K\iN\natnum$, and then the modules
$\Reh{\Delc\circ}$ and $\Reh{\Delc j}$, $\Reh{\Delcb j}$
with $j=mK$, $m=1,2,...\,$ and $j\le\el$, split up as in \erf{deR}.
\end{theorem}

Besides the coset Virasoro generators,
the chiral symmetry algebra of the orbifold coset theory contains further
operators \cite{dvvv}. The observation above implies in particular that 
when acting on $\fAg$-sectors other than $\Hho$ and $\Hhj$,
for $\sqrt{2N}\notin\natnum$ all these additional generators make transitions 
between the sectors of the gauge invariant fermion \alg\ $\fAg$;
for $N=2K^2$ ($K\iN\natnum$) the additional generators generically still 
make transitions, except that they can map sectors with $j=mK$ to themselves.
It follows in particular that we can distinguish between elements of the
coset Virasoro \alg\ and elements of the full coset chiral \alg\ which are
not contained in the coset Virasoro \alg\ by acting with them on suitable
$\fAg$-sectors.

\subsection{The Sectors $\Hho$ and $\Hhj$}
It still remains to analyze the decomposition of the $\fAg$-sectors 
$\Hho$ and $\Hhj$ explicitly. From \erf{cOJ} we conclude that
  \be \Hho \oplus \Hhj 
  = \LLb \Uo\oplus\Uv \LRb \,\otimes\,  \Reh 0 \,. \labl{decHOJ}
Now $\Reh 0$ is always reducible, independent of the particular value
of the integer $N$. We first claim 
\begin{lemma}
The characters $\chio$ and $\chiJ$ decompose as follows:
  \be \bearl 
  \chio = 
  \chizo \cdot \dstyle\sum_{k=0}^\infty \chivir {(2k)^2}  + 
  \chizv \cdot \dstyle\sum_{k=0}^\infty \chivir {(2k+1)^2} 
  \,, \\ {} \\ [-.5em]
  \chiJ = \chizo \cdot \dstyle\sum_{k=0}^\infty \chivir {(2k+1)^2}
  + \chizv \cdot \dstyle\sum_{k=0}^\infty \chivir {(2k)^2} \,. 
  \eear \ee
\end{lemma}
\bproof
We compute
  \beaa 
  \chio (q) \!\! & = & \Frac{\sN 0}{2(\vi(q))^N} +
  \Frac{(\vi(q))^{N-2}}{2(\vi(q^2))^{N-1}}
  \dstyle\sum_{k\in\zet} (-1)^k q^{k^2} \\ %\nline7 
  & \equiv & \Frac{\sN 0}{2(\vi(q))^N} +
  \Frac{(\vi(q))^{N-2}}{2(\vi(q^2))^{N-1}}
  \dstyle\sum_{k=0}^\infty \llb q^{(2k)^2} - 2 q^{(2k+1)^2}
  + q^{(2k+2)^2} \lrb \\ %\nline7 
  & = & \chizo (q) \Frac1{\vi(q)} \dstyle\sum_{k=0}^\infty 
  \llb q^{(2k)^2} -  q^{(2k+1)^2} \lrb + \\ %\nline6
  && \qquad\qquad +
  \chizv (q) \Frac1{\vi(q)} \dstyle\sum_{k=0}^\infty 
  \llb q^{(2k+1)^2} -  q^{(2k+2)^2} \lrb \nline7 
  & \equiv & \chizo (q) \cdot \dstyle\sum_{k=0}^\infty 
  \chivir {(2k)^2} (q) + 
  \chizv (q) \cdot \dstyle\sum_{k=0}^\infty \chivir {(2k+1)^2} (q)
  \eeaa 
(in the first line we used \erf{76}), and analogously for $\chiJ$.
\eproof
Hence we arrive at
\begin{theorem}
For the $\fAg$-sectors $\Hho$ and $\Hhj$ we have
  \be \bearl
  \Hho = \Hhzo \otimes \dstyle\bigoplus_{k=0}^\infty V(1,(2k)^2)
  \,\,\,\oplus\,\,\, \Hhzv \otimes 
  \dstyle\bigoplus_{k=0}^\infty V(1,(2k+1)^2) \,,\\{}\\[-.6em]
  \Hhj = \Hhzo \otimes \dstyle\bigoplus_{k=0}^\infty V(1,(2k+1)^2)
  \,\,\,\oplus\,\,\, \Hhzv \otimes 
  \dstyle\bigoplus_{k=0}^\infty V(1,(2k)^2) \,. \eear \ee
\end{theorem}
It follows that besides $\OmfOo 0 \equiv\Omo$ and $\OmfJv 0\equiv\Ovvo$, 
there must exist further simultaneous \hwv s of $\sonhz$ and the coset 
Virasoro algebra, namely, for $k=0,1,2,...\,$, \hwv s
$\OmfOo {2k+2},\,\OmfOv {2k+1} \iN\Hho$ and 
$\OmfJo {2k+1},\,\OmfJv {2k+2} \iN\Hhj$,
with $\sonhz$-weights $\lo$, $\lv$, $\lo$, $\lv$, \resp, and with coset
conformal weights $(2k+2)^2$, $(2k+1)^2$, $(2k+1)^2$, $(2k+2)^2$, \resp. 
Those vectors with unit coset conformal weight have a relatively simple form.
Define 
  \[  Z_\mh = \left\{ \bearll  \dstyle\sum_{k=1}^\el
  \lLb \xb k\mh \y k\mh - \yb k\mh \x k\mh \lRb & \forzl\,,\nline7
  \dstyle\sum_{k=1}^\el\!\lLb \xb k\mh \y k\mh - 
  \yb k\mh \x k\mh \lRb   + \xb {\el+1}\mh \yb {\el+1}\mh & 
  \forzle\,. \eear \right. \]
Then
  \[ \OmfJo 1 = Z_\mh \, \Omo \]
as well as
  \[ \OmfOv 1 = \llb \x 1\mh \y 1\mh Z_\mh  + \x 1{-3/2} 
  \y 1\mh + \y 1{-3/2} \x 1\mh \lrb \, \Omo \,.  \]
In contrast, the \hwv s with larger coset conformal weight are more difficult
to identify.

\subsection{A Comparison of Algebras}
From gauge invariance and also from the decomposition of
$\HNSh$ one can deduce a lot of information about inclusions
of the algebras of bounded operators that are associated to
several Lie algebras acting in the big Fock space. By
$\Acos$ and $\Avirc$ we denote the $C^*$-algebras associated to
the full coset chiral algebra
$\Cos=(\sonh_1\oplus\sonh_1) / \sonh_2$
and its Virasoro subalgebra $\Vir\coset$, respectively.
Clearly we have $\Acos\subset\AW'$. It follows
by gauge invariance of $\Vir\coset$ that
  \[ \Avirc \subset \fAg \,. \]
Since the full coset algebra $\Cos$ involves different
$\fAg$-sectors we have in contrast
  \[ \Acos \not\subset \fAg \,, \]
i.e.\ not the whole operator content of the coset chiral
algebra is $\Oz$ invariant. Furthermore, we have by gauge
invariance
  \[ \AW \cup \Avirc \subset \fAg \,. \]
However, this is a proper inclusion since at least
the multiplicity space $\Reh 0$ in (\ref{decHOJ})
is reducible. In other words, there is a gauge invariant
operator content in $\Acos$ besides $\Avirc$ which may be
enlarged if $\sqrt{2N}\in\zet$.

\section{Remarks}
We conclude this chapter with some general remarks on our analysis.

\subsection{Remarks on the Characters}
Our idea to employ the \rep\ theory of the gauge group $\Oz$  allowed us to 
deduce simple formulae for the characters of the (\NS sector) \ihwm s of 
$\sonh$  at level $2$. They are given by the 
expressions \erf{chizj} for $\chizj$ and \erf{both} for $\chizo$ and $\chizv$.
Note that, not surprisingly, these results have a simple functional
dependence on the integer $N$, even though the details of 
their derivation (involving e.g.\ the relation with the orbifold
coset theory) depend quite non-trivially on whether $N$ is even or odd.

Our results for these characters are not new. In \cite{scya5}, the conformal 
embedding of $\sonhz$  into $\unh$ at level $1$ was employed to identify
(sums of) $\sonhz$  characters with characters of $\sunhe$. Indeed, the
restricted summation over the lattice vector $\vecm\iN\zet^N$ in the
formula \erf S for $\sN m$ precisely corresponds to the summation over
the appropriately shifted root lattice of $\sun$. 

With the help of the
conformal embedding only the linear combination $\chizo+\chizv$ of the
irreducible characters $\chizo$ and $\chizv$ is obtained, which is just
the level $1$ vacuum character of $\sunh$. However, the orthogonal linear 
combination $\chizo-\chizv$ is known as well; it has been obtained in 
\cite[p.\,233]{kawa} by making use of the theory of modular forms.

\subsection{A Homomorphism of Fusion Rings}
In the previous section we were able to identify the $\sonhz$ 
\hwm s within the sectors of the intermediate \alg\ $\fAg$ 
which are governed by the gauge group $\Oz$.
Our results amount to the following assignment $\rho$ of
the $\Oz$-representations to the WZW sectors:
  \[ \bearl
  \rho (\rpo) = 1 \,,\qquad \rho (\rpj) = \rmv \,,\\{}\\[-.4em]
  \rho (\rp {(n+1)N}) = 1 + \rmv \,,\\{}\\[-.4em]
  \rho (\rp {nN+j}) = \rho (\rp {(n+1)N-j}) = \pfj j 
  \qquad {\rm for}\ j=\onetolme \,,\\{}\\[-.6em]
  \rho (\rp {nN+\el}) = \rho (\rp {(n+1)N-\el}) =
  \left\{ \begin{array}{cl} \rms + \rmc  & \forzl\,, \\[.3em]
  \pfj \el & \forzle\,, \eear \right. \eear \]
for $n=0,1,2,...$
(Note that in the case of $\rpo$ and $\rpj$, the action of $\rho$ does not
directly correspond to the decomposition of the $\fAg$-sectors into $\sonhz$  
sectors.)

The multiplication rules of the \rep\ ring \Ro\ of $\Oz$  are given by the
relations \erf{tp}. The level $2$ WZW sectors generate a fusion ring, too,
which we denote by \Rw. The ring \Rw\ has a fusion subring \Rn\ which is 
generated by those primary fields which appear in the 
big Fock space $\HNSh$. The fusion 
rules, i.e.\ the structure constants of \Rw, can be computed with the help of 
the Kac\hy Walton and Verlinde formulae (see e.g.\ \cite{fuva3}).

Inspection shows that \Rn\ is in fact isomorphic to the \rep\ ring of the 
dihedral group \Dn. Now for any $N$ the group \Dn\ is a finite subgroup of
$\Oz$. As a consequence, the mapping $\rho$ actually constitutes a fusion ring
{\em homomorphism\/} from the \rep\ ring \Ro\ of $\Oz$  to the fusion 
subring \Rn\ of \Rw. (It is also easily checked that for odd $N$ the
homomorphism $\rho$ is surjective, while 
for even $N$ the image does not contain
the linear combination $\rms-\rmc$.) This observation explains to a certain 
extent why, in spite of the fact that the WZW observable
algebra $\AW$ is much smaller than the $\Oz$-invariant \alg\ 
$\fAg$, the group $\Oz$  nevertheless
provides a substitute for the gauge group in the DHR sense. 
But even in view of this relationship it is still surprising 
how closely the WZW superselection structure follows the 
representation theory of $\Oz$.

One may speculate that the presence of the homomorphism $\rho$ indicates 
that the gauge group $\Oz$  is in fact part of the full (as yet unknown)
quantum symmetry of the \wzwt\ that fully takes over the \role\ of the DHR
gauge group. This is possible because all sectors in the \NS part of the
\wzwt\ have integral quantum dimension. This is however a rather
special situation as in rational \cft\ sectors with integral quantum 
dimension are actually extremely rare.

\subsection{Discussion and Outlook}
It would be desirable to incorporate also the twisted sectors
$\sigma,\sigma',\tau,\tau'$ ($N=2\el$) respectively
$\sigma,\sigma'$ ($N=2\el+1$) in our analysis. These modules
appear in the tensor products that contain also the Ramond
sector of the level $1$ theory. In order to avoid severe
technical difficulties we did not treat Ramond fermions here.
More precisely, we expect the twisted sectors to be realized
in the tensor product $\HNS\otimes\HR$. When one tries to
incorporate this space in our analysis several unsolved 
problems arise. The level $2$ currents acting in $\HNS\otimes\HR$
are of the form $J^{\rm NS}+J^{\rm R}$ where each summand acts
non-trivially on the corresponding tensor factor. Since
$J^{\rm NS}$ is constructed from Neveu-Schwarz fermions and
$J^{\rm R}$ from Ramond fermions the $\Oz$-invariance is less
obvious. However, there is an argument coming from the framework
of bounded operators that states that $\Oz$-invariance is
just hidden here: Although this is not yet proven, local
normality of the local algebras of bounded operators 
associated to the WZW model is expected to hold also for
the twisted sectors. Hence the local algebras in any sectors
are isomorphic so that $\Oz$-invariance is given implicitly
from the vacuum sector. Unfortunately the decomposition of
$\HNS\otimes\HR$ into sectors of the gauge invariant
fermion algebra cannot simply be provided as in $\HNSh$
since the associated state $\omega_{\PNS}\otimes\omega_{\SR}$
is neither pure nor gauge invariant. Moreover, the
explicit formulae for the highest weight vectors in $\HR$
at level $1$ are already much more complicated as those in
$\HNS$. Therefore we believe that one needs some new ideas
to treat the twisted sectors as well.

Perhaps a more hopeful task is the generalization of the
analysis to higher levels $\kv$. Then one has to investigate
the $\kv$-fold tensor product $\HNSh=\HNS^{\otimes\kv}$ which
arises from the Fock representation that is associated to
the basis projection $\PNS\otimes\one_{\kv}$ of
$\KK\otimes\bbC^{\kv}$. The level $\kv$ current operators
are then invariant under the gauge group $\Ok$. Owing
to the more complex representation theory of the group
$\Ok$ the DHR decomposition of the big Fock space $\HNSh$
into sectors of the gauge invariant fermion algebra $\fAg$ 
will become more complicated. Moreover, at higher level
most of the sectors have non-integral quantum dimensions. Since
one cannot expect that such sectors possess a simple assignment
to the $\fAg$-sectors as it is realized in the fusion ring
homomorphism at level $2$, the identification of the simultaneous
highest weight vectors of $\sonh_{\kv}$ and the Virasoro algebra
of the coset theory $(\sonh_1^{\oplus\kv})/ \sonh_{\kv}$ may
be more involved. Further complications arise since the central
charge of the coset Virasoro algebra then depends on $N$,
namely it is given by
  \[ c\coset = \frac{N \kv (\kv-1)}{2(N+\kv-2)} \,. \]

The most hopeful generalization is possibly the
application of our ideas to $\sun$ WZW models.
Fortunately, no Ramond fermions are needed there; all the
level $1$ $\sunh$ unitary highest weight modules are
realized in one and the same Fock space even though
with an infinite multiplicity. At level $\kv$, the
DHR gauge group that appears is given by $\Uk$. 
It will be interesting to study the relationship between 
the \rep\ ring of $\Uk$ and the WZW fusion 
ring in these cases where most of the sectors have
non-integral quantum dimension.

\pagebreak
\section*{Acknowledgment}
I would like to thank Prof.\ K.\ Fredenhagen for many helpful
discussions, friendly atmosphere and constant support during
these investigations. Further I am grateful to Dr.\ J.\ Fuchs
for the engaged and instructive collaboration. Thanks are also
due to Dr.\ K.-H.\ Rehren for several helpful discussions.
I would like to thank C.\ Binnenhei, Dr.\ J. Fuchs and W.\ Kunhardt
for a careful reading of (parts of) the manuscript. Financial support
of the Deutsche Forschungsgemeinschaft is gratefully acknowledged.

%%%%%%%%%%%%%%%%%%%%%%%%% bibliography %%%%%%%%%%%%%%%%%%%%%%%%%%%%%%%%%%%%
\newcommand\biba[7]   {\bibitem{#1} {\sc #2:} {\sl #3.} {\rm #4} {\bf #5}
                      { (#6) } {#7}} %\hspace*{\fill} {\small\tt {#1}}}
\newcommand\bibb[4]   {\bibitem{#1} {\sc #2:} {\it #3.} {\rm #4}}
                      %\hspace*{\fill} %{\small\tt {#1}}}
\newcommand\bibp[4]   {\bibitem{#1} {\sc #2:} {\sl #3.} {\rm Preprint #4}}
                      %\hspace*{\fill} %{\small\tt {#1}}}
\newcommand\bibx[4]   {\bibitem{#1} {\sc #2:} {\sl #3} {\rm #4}}
                      %\hspace*{\fill} %{\small\tt {#1}}}
\def\AAM              {Acta Appl.\ Math.}
\def\CMP              {Com\-mun.\ Math.\ Phys.}
\def\JMP              {J.\ Math.\ Phys.}
\def\LMP              {Lett.\ Math.\ Phys.}
\def\RMP              {Rev.\ Math.\ Phys.}
\def\npbp             {Nucl.\ Phys.\ B (Proc.\ Suppl.)}
\def\nupb             {Nucl.\ Phys.\ B}
\def\adma             {Adv.\ Math.}
\def\coma             {Con\-temp.\ Math.}
\def\ijmp             {Int.\ J.\ Mod.\ Phys.\ A}
\def\FdP              {Fortschr.\ Phys.}
\def\PLB              {Phys.\ Lett.\ B}

\end{document}